# Theoretical methods for the calculation of Bragg curves and 3D distributions of proton beams

**W. Ulmer[1,2] and E. Matsinos[3]**

[1]Klinikum Frankfurt/Oder, Germany, [2]Max-Planck-Instute of Physics, Göttingen Germany. [3]ETH Zürich, Switzerland

E-mail: Waldemar.Ulmer@gmx.net

## *Abstract*

The well-known Bragg-Kleeman rule $R_{CSDA} = A \cdot E_0^p$ has become a pioneer work in radiation physics of charged particles and is still a useful tool to estimate the range $R_{CSDA}$ of approximately monoenergetic protons with initial energy $E_0$ in a homogeneous medium. The rule is based on the continuous-slowing-down-approximation (CSDA). It results from a generalized (nonrelativistic) Langevin equation and a modification of the phenomenological friction term. The complete integration of this equation provides information about the residual energy E(z) and dE(z)/dz at each position z ($0 \leq z \leq R_{CSDA}$). A relativistic extension of the generalized Langevin equation yields the formula $R_{CSDA} = A \cdot (E_0 + E_0^2/2M \cdot c^2)^p$. The initial energy of therapeutic protons satisfies $E_0 << 2M \cdot c^2$ ($M \cdot c^2$ = 938.276 MeV), which enables us to consider the relativistic contributions as correction terms. Besides this phenomenological starting-point, a complete integration of the Bethe-Bloch equation (BBE) is developed, which also provides the determination of $R_{CSDA}$, E(z) and dE(z)/dz and uses only those parameters given by the BBE itself (i.e., without further empirical parameters like modification of friction). The results obtained in the context of the aforementioned methods are compared with Monte-Carlo calculations (GEANT4); this Monte-Carlo code is also used with regard to further topics such as lateral scatter, nuclear interactions, and buildup effects. In the framework of the CSDA, the energy transfer from protons to environmental atomic electrons does not account for local fluctuations. Based on statistical quantum mechanics, an analysis of the Gaussian convolution and the Landau-Vavilov distribution function is carried out to describe these fluctuations. The Landau tail is derived as Hermite polynomial corrections of a Gaussian convolution. It is experimentally confirmed that proton Bragg curves with $E_0 \geq 120$ MeV show a buildup, which increases with the proton energy. This buildup is explained by a theoretical analysis of impinging proton beamlets. In order to obtain a complete dose calculation model for proton treatment planning, some further aspects have to be accounted for: the decrease of the fluence of the primary protons due to nuclear interactions, the transport of released secondary protons, the dose contribution of heavy recoil nuclei, the inclusion of lateral scatter of the primary and secondary protons based on Molière's multiple-scatter theory, and the scatter contributions of collimators. This study also presents some results which go beyond proton dose calculation models; namely, the application of the relativistic generalization of the Bragg-Kleeman rule to electrons, the influence of detectors to the profiles of narrow photon beams, and the application of deconvolution kernels to scatter problems in image processing.

## *Introduction*

The connection between the initial energy $E_0$ of a projectile particle (e.g., proton, electron, α-particle) and the range $R_{CSDA}$ in a (homogenous) medium (e.g., water) by continuous energy loss, which is referred to as stopping power, represents an important challenge for many questions in dosimetry. Thus, CSDA only considers the energy transfer from the projectile to the medium by continuous damping of the particle motion; the problem of statistical fluctuations (energy straggling) of the residual energy E(z) or its gradient dE(z)/dz at the position z are not taken into account. The phenomenon of the local fluctuation of the energy transfer (range straggling effects) requires the introduction of convolutions on the base of statistical mechanics and/or quantum statistics (Landau, Vavilov, and Gaussian distributions).

Nevertheless, the damping of particle motion (energy dissipation) by interaction with the environment has become an important subject in many disciplines of physics, as it also pertains to the domain of the thermodynamics of irreversible processes. In this study, we show that the determination of the range $R_{CSDA}$ can be treated by a generalized (nonrelativistic) Langevin equation or by the integration of the Bethe-Bloch equation (BBE). With respect to energy straggling, we will verify that a quantum statistical base implies a Gaussian convolution and some generalizations for energy straggling (Landau tail, if the relativistic effects are taken into account). The energy-range relation goes back to the Bragg-Kleeman rule:

$$R_{CSDA} = A E_0^p \qquad (1)$$

A special case of Eq. (1) is the Geiger rule with p = 1.5, which is assumed to be valid for α-particles with very low energy. For therapeutic protons ($E_0$ between 50 and 250 MeV) the power p has been determined to $p \approx 1.7 - 1.8$. Since $E_0$ is given in MeV and $R_{CSDA}$ in cm, the dimension of $A$ is cm/MeV$^p$; the power p is dimensionless. Therefore, the question arises as to the energy dependence of the power p (besides the specific dependence on the type of the projectile). Bortfeld (1997) has performed a least-squares fit of the range $R_{CSDA}$ of therapeutic protons, based on ICRU49, and obtained for p = 1.77 and A = 0.0022 cm/MeV$^p$ a standard error of the order of 2 %, if $E_0 \leq 200$ MeV. A precise knowledge of $p = p(E_0)$ requires again a comparison with solutions of the BBE or with Monte-Carlo calculations.

At recent times, the radiotherapy with protons has become a modality with increasing importance, which triggers a lot of work in the field of algorithms for treatment planning. One of the challenges in treatment planning, in general, is to find a reasonable compromise between the speed and the accuracy of an algorithm. The fastest dose calculation algorithms are based on look-up tables for depth dose and lateral distributions of spread-out Bragg peaks (SOBP) or single pristine Bragg peaks, see Hong et al. (1996), Petti (1992), Deasy (1998), Schaffner et al. (1999), Russel et al. (2000), Szymanowski and Oelfke (2002), and Ciangaru et al. (2005). The look-up tables often consist of measured data directly or of an analytical model fitted to the measurements in water. A depth scaling mechanism is always applied in order to convert the dose from the water to another medium. Some authors use higher-order corrections or different calculation approaches for the lateral distribution. Another approach is an iterative numerical calculation of the dose deposited by a proton beam along its path through a medium (see Sandison et al. (1997) and Hollmark et al. (2004)). Monte-Carlo methods can lead to the highest precision – especially in highly heterogeneous media – but, lacking sufficient speed, they cannot be applied to routine treatment planning (see Petti (1996), Tourovsky et al. (2005), and Jiang and Paganetti (2004)). The Monte-Carlo code proposed by Fippel and Soukup (2004) cannot be compared with GEANT4. It uses various empirical data and fits from ICRU49; the data are subjected to statistical methods and referred to as Monte-Carlo.

Common to most of the cited papers is that the authors do not provide a rigorous adaptation procedure of the dose model to different properties of the beamline. Some of the models might not be easily adaptable to another beamline or an intermediate range/energy at all. An exception to this is the model published by Bortfeld (1997). However, this model does not correctly describe the transport of secondary protons. The model proposed by Bortfeld can be obtained by methods elaborated in section 1.2. Furthermore, we present a theoretical explanation of the buildup effect, which is measured in higher energy proton beams and is based on secondary nuclear reaction protons as well as skew symmetric energy transfer from protons to electrons (Landau tails). The buildup is not modeled correctly by the Monte-Carlo code PTRAN, which is used by some authors for a comparison (see Carlsson et al. (1997)). Also, none of the other authors, referred to above, thoroughly discuss the origins of buildup effects.

In this work, we will deal with three calculation models, which we refer to as model M1, model M2 and model M3. The model M3 is incorporated in the commercial Treatment Planning System Eclipse™ (Varian Medical Systems Inc.). All three calculation models are built upon proton beamlets in water. Their outcome is the three-dimensional (3D) dose distribution delivered by the quasi-monoenergetic beam impinging on a water surface with no lateral extension and angular divergence (the term 'quasi-monoenergetic beam' refers to the spectrum as produced by the accelerator and beamline without any intended range modulation). The beamlet can be separated into a depth-dose component and a lateral distribution. The depth dose, obtained from a quasi-monoenergetic beam, is often called the pristine Bragg peak.

We also focus our interest to theoretical depth-dose models which contain some fitting parameters referring to the spectral distribution of impinging proton beams (measured pristine Bragg peaks) in order to incorporate beamline specific characteristics. We will show how well different measured Bragg peaks can be adapted – including buildup effects. The measurements have been obtained from different sources and will be described in more detail in the corresponding sections. The lateral component of the beamlet only includes scattering in the patient. It does not require any beamline specific configuration. The beamline specific component of the lateral penumbra is modeled through the lateral distribution of the in-air fluence, see Schaffner (2008).

In the development of the models, we have used the Monte-Carlo code GEANT4 (GEANT4 documents (2005)) to analyze the lateral scattering distributions, the Landau tails, the production of heavy recoil particles, and to obtain some numerical parameters. Many comparisons between Monte-Carlo results and an exact integration of the BBE, and additionally nuclear cross-sections have been previously published, see Ulmer (2007).

# *1. Theoretical Methods*

## 1.1 Abbreviations and definitions

β: ratio v/c of the particle velocity v and the velocity of light c

m: electron rest mass (electron rest energy $m \cdot c^2 = 0.511$ MeV)

M: proton rest mass (1836.15655·m; proton rest energy $M \cdot c^2 = 938.276$ MeV)

μ: reduced mass of the center-of-mass system 'proton - electron' with μ = m +m/M

$A_N$: mass number of a nucleus

Z: charge number of a nucleus

ρ: mass density (g/cm$^3$)

$e_0$: electric charge

In this work, we use the following definition of the standard deviation:

$$\mathrm{dev} = \frac{1}{N}\sum_{k=1}^{N} |x_k - y_k| \quad (2)$$

The variables $x_k$ and $y_k$ may either refer to a set of measured and calculated data or to a comparison of an approximate calculation (e.g., finite order of a power expansion) with an accurate one. The evaluation of some relativistic terms has to account for the well-known power expansion:

$$\left.\begin{array}{l} (1 \pm x)^r = 1 \pm rx/1! + r(r-1)x^2/2! \\ \pm r(r-1)(r-2)x^3/3! \pm .+(-1)^n r(r-1) \\ ..(r-n+1)x^n/n! \\ \quad r = \pm \frac{1}{2} \end{array}\right\} \quad (3)$$

The parameter x in Eq. (3) may refer to different substitutions (e.g., $x = p^2/m^2c^2$ resulting from the root $W = mc^2(1+p^2/m^2c^2)^{0.5}$ of the energy-momentum relation $W^2 = p^2c^2 + m^2c^4$).

## 1.2 Phenomenological treatment of energy-range relations of protons (Bragg-Kleeman rule and its generalization)

### *1.2.1 The energy-range relation and a modification of the Langevin-equation (classical equation of motion with energy dissipation by friction) – special case for the calculation model M1*

In thermodynamics of irreversible processes, the subsequent Langevin Eq. (4) and the related solution (5)

have been investigated in detail with respect to the energy dissipation to the environment during the motion of a projectile with velocity v(t):

$$M \cdot dv/dt = -\gamma \cdot v \quad (4)$$

The integration of Eq. (4) is yields:

$$\left.\begin{array}{l} v(t) = v_0 \exp(-\gamma t/M) \Rightarrow E(t) = E_0 \exp(-2\gamma t/M) \\ z = R_{CSDA} - \frac{v_0 M}{\gamma} \exp(-\gamma t/M) \end{array}\right\} (5)$$

Only in Eqs. (4 – 6), M may be identified with the mass of a macroscopic ‘particle’. Special cases are the Stokes law of a spherical body with $\gamma = 6\pi\eta r$ in a fluid and, in analogous sense, Ohm’s law of the electric resistance. A unique feature is the irreversible heat production in the environment by damping of the particle motion. The range $R_{CSDA}$ of a (macroscopic) particle starting at z = 0 and at time t = 0 with $v = v_0$ (or energy $E_0$) results directly from Eq. (5):

$$R_{CSDA} = \frac{v_0 M}{\gamma} = \frac{\sqrt{2 M E_0}}{\gamma} \quad (6)$$

The stopping of a particle according to Eqs. (5 – 6) is described by a continuous-slowing-down motion, which is always assumed for macroscopic processes. From a microscopic view, the damping constant $\gamma$ is the result of numerous collisions with environmental atoms/molecules. The question arises, whether the energy loss of a projectile, as a proton or an $\alpha$-particle, can be established in a similar way. With respect to Formula (1) and Eq. (4), we consider the following modification (the power q may be arbitrary, but q = −1 leading to Eq. (4) must be excluded):

$$\left.\begin{array}{l} M \cdot dv/dt = -\delta / v^q \Rightarrow \\ \int v^q \cdot dv = -\int (\delta / M) dt \end{array}\right\} (7)$$

By taking account of the initial condition, i.e., t = 0 and $v = v_0$, the integration of Eq. (7) yields:

$$v^{q+1} = -\frac{\delta t (q+1)}{M} + v_0^{q+1} \quad (8)$$

In order to perform a further integration, the $(q+1)^{th}$ root on both sides of Eq. (8) has to be taken:

$$
\left.\begin{array}{l}
z = \left\{ \begin{array}{l} R_{CSDA} + \int \left( v_0^{q+1} - \frac{\delta t (q+1)}{M} \right)^{\frac{1}{q+1}} dt \\ = R_{CSDA} - \frac{M}{\delta (q+2)} \left( v_0^{q+1} - \frac{\delta t (q+1)}{M} \right)^{\frac{q+2}{q+1}} \end{array} \right. \\
\tau = M v_0^{q+1} / \delta (q+1)
\end{array}\right\} \quad (9)
$$

The initial condition t = 0 ⇒ z = 0 now yields:

$$
R_{CSDA} = \frac{M}{\delta (q+2)} v_0^{q+2} \quad (10)
$$

Replacing the initial velocity $v_0$ by $E_0 = Mv_0^2/2$, Eq. (10) takes the form:

$$
\left.\begin{array}{l}
R_{CSDA} = A \cdot E_0^{\,p} \\
p = 1 + q/2 \\
A = 2^{p-1} / (\delta \cdot M^{p-1})
\end{array}\right\} \quad (11)
$$

Eq. (11) is identical to the Bragg-Kleeman rule (1); the already noted Geiger rule results from the restriction q = 1 ⇒ p = 3/2. A main difference between Eqs. (6) and (10) is the time interval τ needed by the particle to reach z = $R_{CSDA}$. In the case of the Eq. (4), $\tau$ is infinite, whereas for Eq. (9) the corresponding $\tau$ is finite; for $t > \tau$ the root expressions in Eq. (9) become imaginary. The CSDA range $R_{CSDA}$ and the connection to the power p (p ≈ 1.7 – 1.8) of therapeutic protons has been subjected to numerous studies, see Evans (1962), Segrè (1964), Rahu (1980), Bortfeld (1997), and Boon (1998). According to Eq. (11), we may consider *p* and *A* as depending on $E_0$. This is not useful and, in view of a relativistic treatment, we only consider $p = p(E_0)$ and keep *A* constant. This will be done in a following section. The relation $z = z(t)$, according to Eq. (9), provides a tool to calculate E(z) and the stopping power S(z) = dE(z)/dz as a function of the position z:

$$
E(z) = A^{-1/p} (R_{CSDA} - z)^{1/p} \quad (12)
$$

$$
S(z) = dE/dz = -p^{-1} A^{-1/p} (R_{CSDA} - z)^{(1/p-1)} \quad (13)
$$

It should be mentioned that for therapeutic protons a classical/phenomenological description of the proton motion and energy loss by damping makes physical sense, since E satisfies always E << $Mc^2$ and quantum mechanical effects, like production of ‘bremsstrahlung’, are negligible. The release of secondary reaction protons via proton – nucleus interactions is also rather small and cannot invalidate Eqs. (12 – 13). This will be discussed in a subsequent section.

*1.2.2 The relativistic extension of the phenomenological stopping-power model (calculation model M1)*

There are principal differences between nonrelativistic and relativistic calculations of the stopping power. The nonrelativistic treatment consists of two steps:

1. Integration of the equation of motion, i.e., dv(t)/dt, providing either dz/dt or E(v(t)) as a function of t.
2. A further integration yields z(t), and by elimination of t we obtain E(z) and dE(z)/dz (CSDA approach).

In a relativistic approach, only $ds = \sqrt{dz^2 - c^2\, dt^2}$ is invariant. Since the energy-momentum relation also holds, there is no additional integration with respect to the momentum or the energy. The relativistic extension of the motion with damping and energy dissipation according to Eq. (5) has to take the following principles into account:

The relativistic equation of a particle motion under the action of a force $F_\mu$ (though $\mu$ refers to an arbitrary component of a four-vector, we can restrict ourselves to the z-component) reads:

$$dp_z / ds = F_z; \qquad ds = \sqrt{dz^2 - c^2 dt^2} \quad (14)$$

The relativistic energy-momentum relation, which we write as:

$$W^2 = p_z^2 c^2 + M^2 c^4 \text{ and } W = Mc^2 + E \quad (15)$$

The use of the invariant variable s accounts for the length contraction besides the relativistic mass dependence. The initial energy $E_0$ has the same meaning as previously defined, i.e., $W_0 = Mc^2 + E_0$. In analogy to Eq. (7), we consider the following damping equation (the z-component of the momentum is denoted by $p_z$):

$$dp_z / ds = -\eta / p_z^q \quad (16)$$

Since the dimensions in Eq. (16) are different from those in Eq. (7), we have replaced $\delta$ by $\eta$. The integration of Eq. (16) goes parallel to previous calculations; additionally, the condition (15) has to be satisfied. Thus we obtain:

$$s = \frac{1}{\eta(q+1)}[p_{z,0}^{q+1} - p_z^{q+1}] \quad (17)$$

Inserting the solution (17) into Eq. (15) results in:

$$\left.\begin{aligned} R_{CSDA} &= \frac{1}{(q+1)\eta}[(M^2c^4 + E_0^2 + 2Mc^2E_0)/c^2 - M^2c^2]^{(q+1)/2} \\ &= A[E_0 + E_0^2/2Mc^2]^p \\ A &= (2M)^{(q+1)/2}/(\eta + \eta q) \end{aligned}\right\} (18)$$

From Eqs. (17 - 18) it follows:

$$\left.\begin{aligned} &E + E^2/2Mc^2 = [(R_{CSDA} - s)/A]^{1/p} \Rightarrow \\ &E(s) = -Mc^2 + Mc^2\sqrt{1 + 2(R_{CSDA} - s)^{1/p}/(Mc^2A^{1/p})} \end{aligned}\right\} \quad (19)$$

$$S(s) = dE/ds = -p^{-1}A^{-1/p}(R-s)^{1/p-1}/\sqrt{1+2(R_{CSDA}-s)^{1/p}/(Mc^2A^{1/p})} \quad (20)$$

The power p according to Eq. (19) is dimensionless (p = (q+1)/2). E(s) and dE(s)/ds according to Formulas (19 – 20) can be subjected to a series expansion, if $x = 2(R_{CSDA} - s)^{1/p}/(Mc^2A^{1/p}) < 1$. This condition is equivalent to E(s) ≤ $Mc^2$ (s ≤ $R_{CSDA}$). The factor A amounts to A = 0.00259 cm/$MeV^p$. For therapeutic protons with v << c the length-contraction effect expressed by s can be omitted; therefore, we may substitute s by z, i.e., s → z. This is, however, not true for therapeutic electrons with $E_0$ ≥ 4 MeV >> $mc^2$. Only for therapeutic protons, we can write Eqs. (19 – 20) in the form:

$$E(z) = -Mc^2 + Mc^2\sqrt{1 + 2(R_{CSDA} - z)^{1/p}/(Mc^2A^{1/p})} \quad (21)$$

$$S(z) = dE/dz = -p^{-1}A^{-1/p}(R-z)^{1/p-1}/\sqrt{1+2(R_{CSDA}-z)^{1/p}/(Mc^2A^{1/p})} \quad (22)$$

As expected, the expansion of the root of Eq. (22) according to Formula (3) yields Formula (13) as lowest-order approximation. Eqs. (22, 13) show a singularity at z = $R_{CSDA}$, if p > 1, valid for therapeutic protons. In the relativistic case, this singularity is weakened due to the nonsingular relativistic corrections.

### 1.3 Integration of the Bethe-Bloch equation (calculation model M2)

Monte-Carlo codes for the computation of electronic stopping power of protons and other charged particles have to be based on the BBE:

$$\left.\begin{aligned} -dE(z)/dz &= (K/v^2)\cdot[\ln(2mv^2/E_I)-\ln(1-\beta^2)+ \\ &+ a_{shell}+a_{Barkas}+a_0 v^2+a_{Bloch}] \\ K &= (Z\rho/A_N)\cdot 8\pi q^2 e_0^4/2m \end{aligned}\right\} (23)^1$$

$E_I$ is the atomic ionization energy, weighted over all possible transition probabilities of atomic/molecular shells, and q denotes the charge number of the projectile (proton). The meaning of the correction terms $a_{shell}$, $a_{Barkas}$, $a_0$ and $a_{Bloch}$ are explained in literature, see Bethe (1930), Bloch (1933), Bethe et al. (1953), Bethe (1953), ICRU49 (1993), and Boon (1998). Since the Bloch correction $a_{Bloch}$ will be introduced in Eq. (31), we present, for completeness, the remaining correction terms according to ICRU49:

$$a_0 : \text{constant value; ICRU49} : a_0 = -1 \quad (23a)$$

$$a_{shell} = -\tfrac{C}{Z} \quad (23b)$$

$$\left.\begin{aligned} a_{Barkas} &= \sqrt{2}\cdot F_{ARB}(b/\sqrt{\kappa})/(\sqrt{Z}\cdot\kappa^{3/2}) \\ \kappa &= Z^{-1}\cdot(\beta/\alpha)^2 \end{aligned}\right\} (23c)$$

Some comments to Eqs. (23b – 23c):

The parameter C in Eq. (23b), referring to shell corrections, is determined by different models (ICRU49 and references therein). A unique parameterization of C depending on Z, $A_N$, and $E_I$ does not exist. It is therefore recommended to select C according to proper domains of validity. It must be noted that several models have been proposed to account for shell transitions. Therefore, the recommendations of ICRU49 have been applied in this work. The function $F_{ARB}$ in Eq. (23c) refers to the theory of the Barkas effect developed by Ashley, Ritchie and Brandt, see Ashley et al. (1974). The parameter α refers to Sommerfeld's fine structure constant and b to a fitting parameter. Unfortunately, b is not a unique fitting parameter; this results in an uncertainty of about 2 %. Modifications of Eq. (23c) for high-Z materials are not of interest in this work.

---

Since we consider at first a nonrelativistic approach, the term $-\ln(1-\beta^2)$ is added to the Bloch correction.

It is also possible to substitute the electron mass m by the reduced mass m $\Rightarrow$ μ. However, this leads for protons to a rather small correction (i.e., less than 0.1 %). For complex systems $E_I$ and some other contributions like $a_{shell}$ and $a_{Barkas}$ can only be approximately calculated by quantum-mechanical models (e.g., harmonic oscillator); the latter terms are often omitted and $E_I$ is treated as a fitting parameter, but different values are proposed and used (ICRU49). The restriction to the logarithmic term leads to severe problems, if either $v \rightarrow 0$ or $2mv^2/E_I \rightarrow 1$.

With regard to the integration procedure, we start with the logarithmic term of Eq. (23) and perform the substitutions:

$$v^2 = 2E/M; \quad \beta_I = 4m/ME_I; \quad E = (1/\beta_I)\exp(-u/2) \qquad (24)$$

With the help of substitution (24) (and without any correction terms), Eq. (23) leads to the integration:

$$-\int du \exp(-u)\cdot(1/u) = \tfrac{1}{2} K \cdot \beta_I^2 \cdot M \int dz \quad (25)$$

The boundary conditions of the integral (25) are:

$$\left.\begin{aligned} z = 0 &\Rightarrow E = E_0 \,(\text{or}: u = -2\ln(E_0 \cdot \beta_I)) \\ z = R_{CSDA} &\Rightarrow E = 0 \,(\text{or}: u \Rightarrow \infty) \end{aligned}\right\} (26)$$

The general solution is given by the Euler exponential integral function Ei(ξ) with P.V. = principal value:

$$\left.\begin{aligned} &\tfrac{1}{2} K \cdot M \cdot \beta_I^2 \cdot R_{CSDA} = -P.V. \int_{-\xi}^{\infty} u^{-1} \exp(-u)\, du = Ei(\xi) \\ &\xi = 2\ln(4mE_0/ME_I) \quad \text{and} \quad \xi > 0 \end{aligned}\right\} \quad (27)$$

Some details of Ei(ξ) and its power expansions can be found in Abramowitz and Stegun (1970). The critical case ξ = 0 results from $E_{critical} = ME_I/4m$ (for water with $E_I$ = 75.1 eV, the critical energy $E_{critical}$ amounts to 34.474 keV; for Pb with $E_I \approx$ 800 eV to about 0.4 MeV). Since the logarithmic term derived by Bethe implies the Born approximation, valid only if the transferred energy $E_{transfer}$ >> the energy of shell transitions, the above corrections, exempting the Bloch correction, play a significant role in the environment of the Bragg peak, and the terms $a_0$ and $a_{shell}$ remove the singularity. With respect to numerical integrations (Monte Carlo), we note that, in the environment of $E = E_{critical}$, the logarithmic term may become crucial (leading to overflows); rigorous cutoffs circumvent the problem. Therefore, the shell correction is an important feature for low proton energies. In similar fashion, we can take account of the Barkas correction. Since this correction is also important for low proton energies, it is difficult to make a quantitative distinction to the shell correction, and different models exist in the literature implying overall errors up to 2 % (ICRU49). Using the definitions/suggestions of the correction terms according to ICRU49 and the substitutions (24), we obtain:

$$\tfrac{1}{2} K \cdot M \cdot \beta_I^2 \int dz = \int du \, exp(-u) [u + 2\alpha_s + \\ + 2\alpha_{Barkas} (4 \cdot m / E_I)^{p_B} exp(p_B u / 2) + \alpha_0 (E_I / 2m) exp(-u/2)]^{-1} \quad (28)$$

A closed integration of Eq. (28) does not exist, but it can be evaluated via a procedure valid for integral operators, see Feynman (1962):

$$[A' + B']^{-1} = A'^{-1} - A'^{-2} B' + A'^{-3} B'^{2} - A'^{-4} B'^{3} + .. + (-1)^{n} A'^{-n-1} B'^{n} \quad (29)$$

A’+B’ is equated to the complete denominator on the right-hand side of Eq. (28). The (small) Barkas correction is identified with A’ and the other (more important) terms with B’:

$$\left.\begin{aligned} A' &= 2\alpha_{Barkas} (4m / E_I)^{p_B} exp(p_B u / 2) \\ B' &= u + 2\alpha_s + \alpha_0 (E_I / 2m) exp(-u/2) \end{aligned}\right\} (30)$$

The integration of Eq. (28) with the help of Relation (29) leads to standard tasks (i.e., to a series of usual exponential functions). In the following, we add the Bloch correction to the denominator of Eq. (28). In order to use the procedure (29), we define now the nonrelativistic energy $E_{nr}$ by: $E_{nr} = 0.5 \cdot Mv^2$ and write the relativistic energy expression $E_{rel}$ (the rest energy $Mc^2$ is omitted) in terms of an expansion:

$$\left.\begin{aligned} & a_{Bloch} = -(q^2\alpha^2 / \gamma^2)[1.042 - 0.8549\, q^2\alpha^2 / \gamma^2 + \\ & 0.343\, q^4\alpha^4 / \gamma^4 - . + \text{higher order terms}] \\ & \alpha = 1/137.036, \\ & \gamma^2 = 2E_{nr} / Mc^2 = 2\, exp(-u/2) / \beta Mc^2 \end{aligned}\right\} (31)$$

Relation (31) provides a sequence of exponential functions:

$$a_{Bloch} = -(\tfrac{1}{2} q^2 \alpha^2 \beta_I Mc^2 \exp(u/2))[1.042 - \tfrac{1}{2} 0.854 \cdot q^2 \alpha^2 \beta_I Mc^2 \exp(u/2) + \\ + \tfrac{1}{4} 0.343 q^4 \alpha^4 \beta_I^{\,2} M^2 c^4 \exp(u) \; - .. + \; \text{higher} - \text{order terms}] \qquad (32)$$

We add the term $a_{Bloch}$ to the term *A'* in Eq. (30); all relativistic corrections are included by taking account of Relation (32):

$$A' = 2\alpha_{Barkas}(4m/E_I)^{p_B} \exp(p_B u/2) + a_{Bloch} \qquad (33)$$

$$\tfrac{1}{2} K \cdot M \cdot \beta_I^2 \int dz = \int du \exp(-u) \cdot [A' + B']^{-1} \qquad (34)$$

The integration of Eq. (30) is carried out with the boundary conditions (24). Since these conditions are defined by logarithmic values, which have to be inserted to an exponential function series, the result yields a power expansion for $R_{CSDA}$ in terms of $E_0$:

$$R_{CSDA} = \frac{1}{\rho} \cdot \frac{A_N}{Z} \sum_{n=1}^{N} \alpha_n E_I^{pn} E_0^{\,n} \quad (N \Rightarrow \infty) \quad (35)$$

The coefficients $\alpha_n$ are determined by the integration procedure and only depend on the parameters of the BBE. For applications to therapeutic protons, i.e., $E_0 < 300$ MeV, a restriction to N = 4 provides excellent results (Fig. 1). For water, we have to take $E_I = 75.1$ eV, $Z/A_N = 10/18$, $\rho = 1$ g/cm$^3$; Formula (35) becomes:

$$R_{CSDA} = \sum_{n=1}^{N} a_n E_0^{\,n} \quad (N \Rightarrow \infty) \quad (36)$$

The values of the parameters of Formulas (35 – 36) with restriction to N = 4 are displayed in Tables 1 and 2.

**Table 1:** Parameter values for Eq. (35) if $E_0$ is in MeV, $E_I$ in eV and $R_{CSDA}$ in cm

| $\alpha_1$ | $\alpha_2$ | $\alpha_3$ | $\alpha_4$ | p1 | p2 | p3 | p4 |
|---|---|---|---|---|---|---|---|
| $6.8469 \cdot 10^{-4}$ | $2.26769 \cdot 10^{-4}$ | $-2.4610 \cdot 10^{-7}$ | $1.4275 \cdot 10^{-10}$ | 0.4002 | 0.1594 | 0.2326 | 0.3264 |

**Table 2:** Parameter values for Eq. (36), if $E_0$ is in MeV, $E_I$ in eV and $R_{CSDA}$ in cm

| $a_1$ | $a_2$ | $a_3$ | $a_4$ |
|---|---|---|---|
| $6.94656 \cdot 10^{-3}$ | $8.13116 \cdot 10^{-4}$ | $-1.21068 \cdot 10^{-6}$ | $1.053 \cdot 10^{-9}$ |

The determination of $A_N$ and Z is not a problem for atoms or molecules, where weight factors can be introduced according to the Bragg rule; for tissue heterogeneities, it is already a difficult task. Much more difficult is the accurate determination of $E_I$, which results from transition probabilities of all atomic/molecular states to the continuum (δ-electrons). Thus, according to the report ICRU49 of stopping powers of protons in different media, there are sometimes different values of $E_I$ proposed (e.g., for Pb: $E_I$ = 820 eV and $E_I$ = 779 eV). If we use the average (i.e., $E_I$ = 800.5 eV), the above formula provides a mean standard deviation of 0.27 % referred to stopping-power data in ICRU49, whereas for $E_I$ = 820 eV or $E_I$ = 779 eV we obtain 0.35 % - 0.4 %.

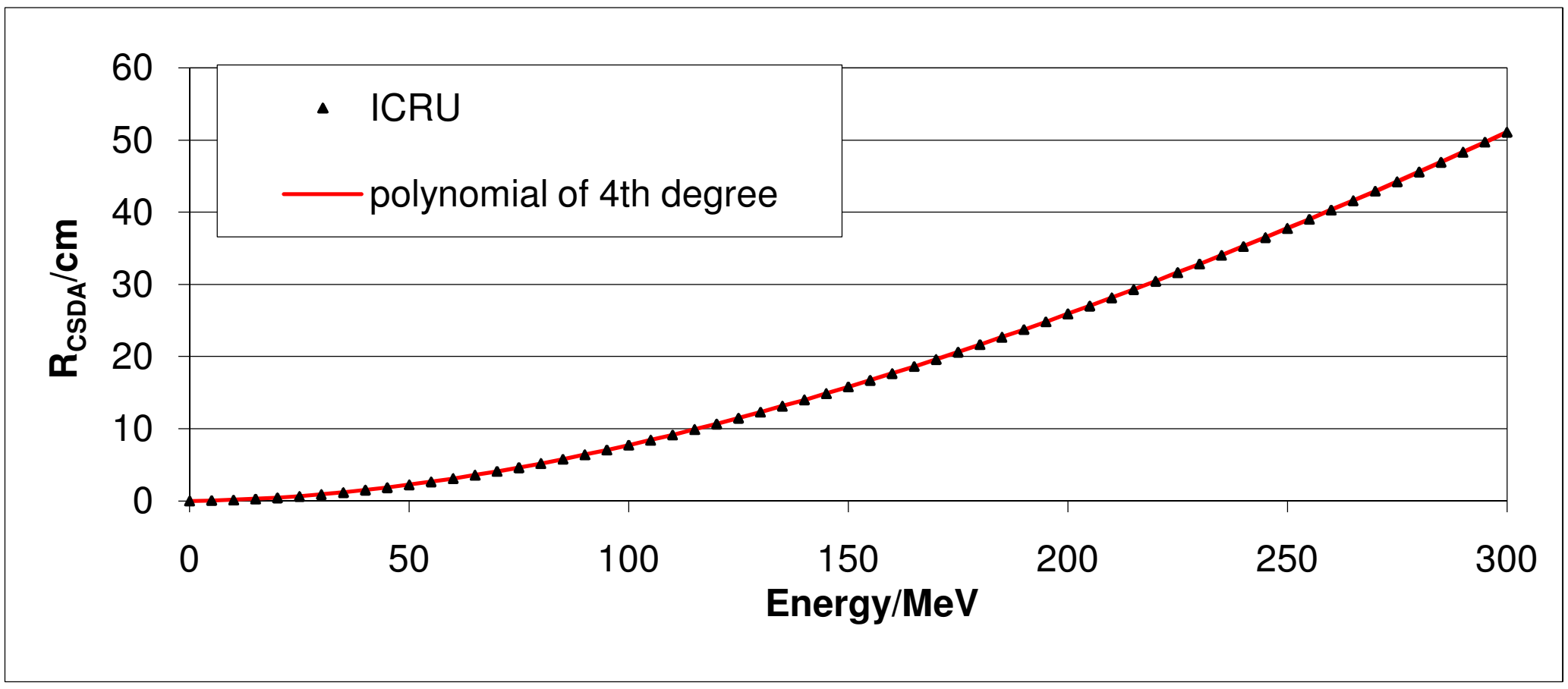

**Fig. 1:** Comparison between ICRU49 data of proton $R_{CSDA}$ range (up to 300 MeV) in water and the fourth-degree polynomial (Eq. 36). The average deviation amounts to 0.0013 MeV.

If we apply the above formula to data of other elements listed in ICRU49, the mean standard deviations also amount to about 0.2 % - 0.4 %. Instead of the usual power expansion (36), we can represent all integrals in terms of Gompertz-type functions multiplied with a single exponential function by collection of all exponential functions obtained by the expression of [A' +B']$^{-1}$ and the substitution $\beta_l E = \exp(-u/2)$. A Gompertz-function is defined by:

$$\left.\begin{aligned} &\exp(-\xi \exp(-u/2)) = 1 - \xi \exp(-u/2) + \tfrac{1}{2!}\xi^2 \exp(-u). - .. + = \\ &= 1 + \sum_{k=1}^{\infty} \tfrac{1}{k!}(-1)^k \xi^k \exp(-ku/2) \end{aligned}\right\} \quad (37)$$

By inserting the integration boundaries u = 2·ln·4m·$E_0$/(M·$E_I$), i.e., E = $E_0$ and u → ∞ (E = 0), the integration leads to a sequence of exponential functions; the power expansion (36) is replaced by:

$$R_{CSDA} = a_1 E_0 \cdot [1 + \sum_{k=1}^{N}(b_k - b_k \exp(-g_k \cdot E_0))] \quad (\lim N \Rightarrow \infty) \quad (38)$$

For therapeutic protons, the restriction to N = 2 provides the same accuracy (Fig. 2) as Relation (36); the parameters are given in Table 3 ($a_l$ is the same as in Table 2).

**Table 3:** Parameters of Formula (38); $b_1$ and $b_2$ are dimensionless; $g_1$ and $g_2$ are given in $MeV^{-1}$.

| $b_1$ | $b_2$ | $g_1$ | $g_2$ |
|---|---|---|---|
| 15.14450027 | 29.84400076 | 0.001260021 | 0.003260031 |

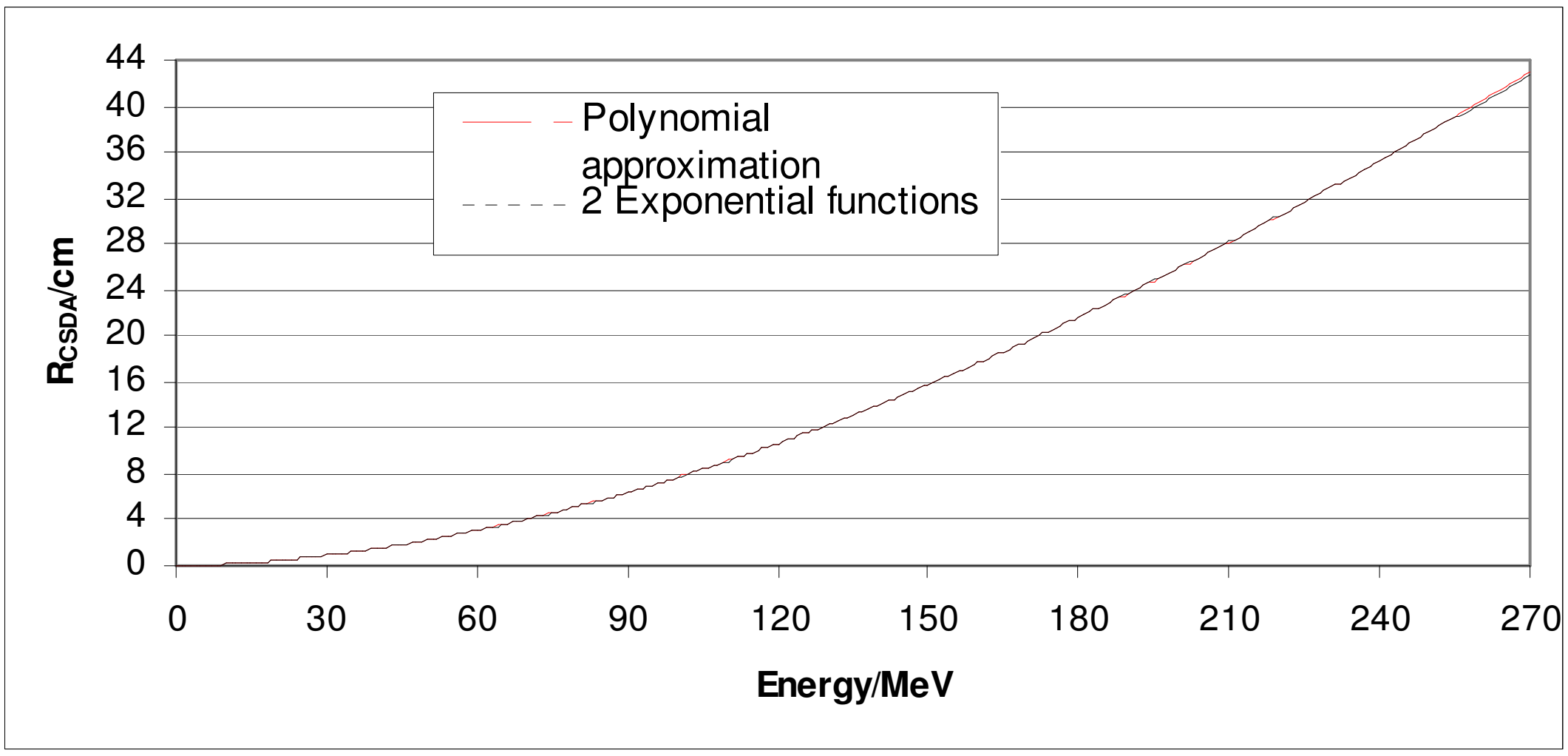

**Fig. 2:** $R_{CSDA}$ calculation - comparison between a fourth-degree polynomial (Eq. (36)) and two exponential functions (Eq. (38)).

In the following, we will verify that the latter formula provides some advantages with respect to the inversion $E_0 = E_0(R_{CSDA})$.

### *1.3.1 The Inversion problem: calculation of $E_0(R_{CSDA})$ or $E(z)$*

Formulas (36, 38) can also be used for the calculation of the residual distance $R_{CSDA} - z$, relating to the residual energy E(z); we have only to perform the substitutions $R_{CSDA} \rightarrow R_{CSDA} - z$ and $E_0 \rightarrow E(z)$ in these formulas. In various problems, the determination of $E_0$ or E(z) as a function of $R_{CSDA}$ or $R_{CSDA} - z$ is an essential task. The power expansion (36) implies again a corresponding series $E_0 = E_0(R_{CSDA})$ in terms of powers:

$$\left.\begin{array}{l} E_0 = \sum_{k=1}^{\infty} c_k R_{CSDA}^{\,k} \\ c_1 = 1/a_1, \;\; c_2 = -a_2/a_1^3, \;\; c_3 = (2a_2^2 a_1^{-1} - a_3)/a_1^4, .... \\ c_k = f(a_k, a_{k-1}, a_{k-2} ...)/a_1^{k+1} \;\; (k > 3) \end{array}\right\} \quad (39)$$

The coefficients $c_k$ are calculated by a recursive procedure; we have given the first three terms in formula (39). Due to the small value of $a_l = 6.8469 \cdot 10^{-4}$, this series is ill-posed, since there is no possibility to break off the expansion; it is divergent and the signs of the coefficients $c_k$ are alternating, see Abramowitz and Stegun (1970). The inversion procedure of Eq. (38) leads to the formula:

$$\left.\begin{aligned} E_0 &= R_{csda} \sum_{k=1}^{N} c_k \exp(-\lambda_k R_{csda}) \quad (\lim N \to \infty) \\ E(z) &= (R_{csda} - z) \sum_{k=1}^{N} c_k \exp(-\lambda_k (R_{csda} - z)) \end{aligned}\right\} \quad (40)$$

The following modification of Eq. (40), which results from some substitutions, represent the inverse formula of Eq. (35). The necessary parameters are stated in Table 4.

$$\left.\begin{aligned} c'_k &= c_k \cdot (18/10) \cdot Z \cdot \rho \cdot (75.1/E_I)^{qk} / (A_N \cdot \rho_w) \\ \lambda^{-1\prime}_k &= \lambda^{-1}_k \cdot (10 \cdot \rho_w / 18) \cdot (75.1/E_I)^{pk} \cdot A_N / (\rho \cdot Z) \\ E(z) &= (R_{CSDA} - z) \cdot \sum_{k=1}^{5} c'_k \cdot \exp[-(R_{CSDA} - z) \cdot \lambda'_k] \end{aligned}\right\} \quad (40a)$$

For therapeutic protons, a very high precision is obtained by the restriction to N = 5 (Table 4 and Fig. 4). Formula (40) is also suggested by a plot $S(R_{CSDA}) = E_0(R_{CSDA})/R_{CSDA}$ according to Eq. (38). This plot is shown in Fig. 3 and gives rise for an expansion of $S(R_{CSDA})$ in terms of exponential functions. This plot is obtained by an interchange of the plot $E_0$ versus $R_{CSDA}$ and a calculation according to Relation (38).

**Table 4:** Parameters of the inversion Formula (40) with N = 5 (dimension of $c_k$: cm/MeV, $\lambda_k$: $cm^{-1}$).

| $c_1$ | $c_2$ | $c_3$ | $c_4$ | $c_5$ | $\lambda_1^{-1}$ | $\lambda_2^{-1}$ | $\lambda_3^{-1}$ | $\lambda_4^{-1}$ | $\lambda_5^{-1}$ |
|---|---|---|---|---|---|---|---|---|---|
| 96.63872 | 25.0472 | 8.80745 | 4.19001 | 9.2732 | 0.0975 | 1.24999 | 5.7001 | 10.6501 | 106.72784 |

| $P_1$ | $P_2$ | $P_3$ | $P_4$ | $P_5$ | $q_1$ | $q_2$ | $q_3$ | $q_4$ | $q_5$ |
|---|---|---|---|---|---|---|---|---|---|
| -0.1619 | -0.0482 | -0.0778 | 0.0847 | -0.0221 | 0.4525 | 0.195 | 0.2125 | 0.06 | 0.0892 |

One way to obtain the inversion Formula (40) of Formula (38) is to find $S(R_{CSDA})$ by a sum of exponential functions with the help of a fitting procedure. Thus it turned out that the restriction to five exponential functions is absolutely sufficient and yields a very high accuracy. A more rigorous way (mathematically) has been described in the LR of Ulmer (2007).

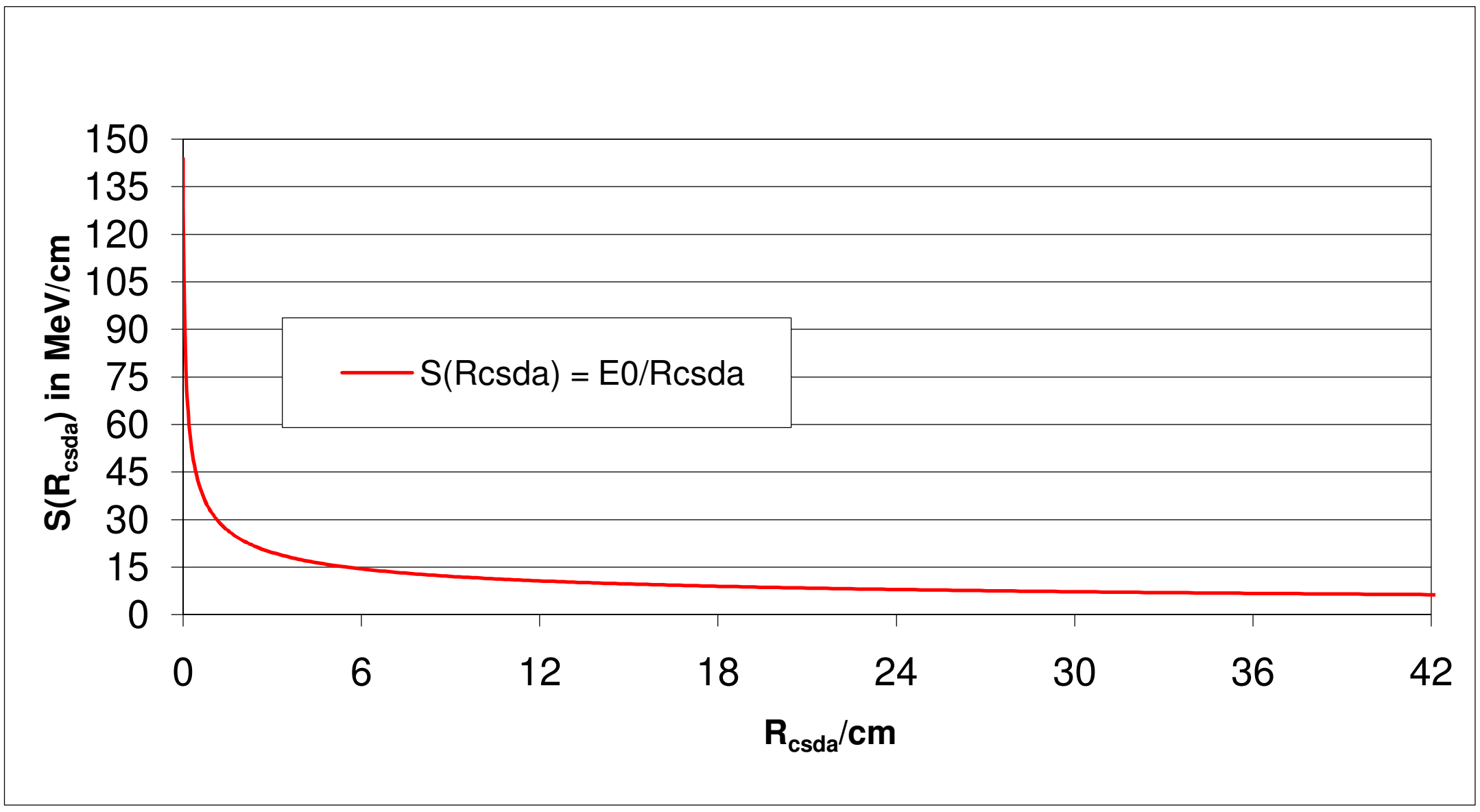

**Fig. 3:** Plot $S(R_{CSDA}) = E_0/R_{CSDA}$ provides a justification of the representation of S by exponential functions.

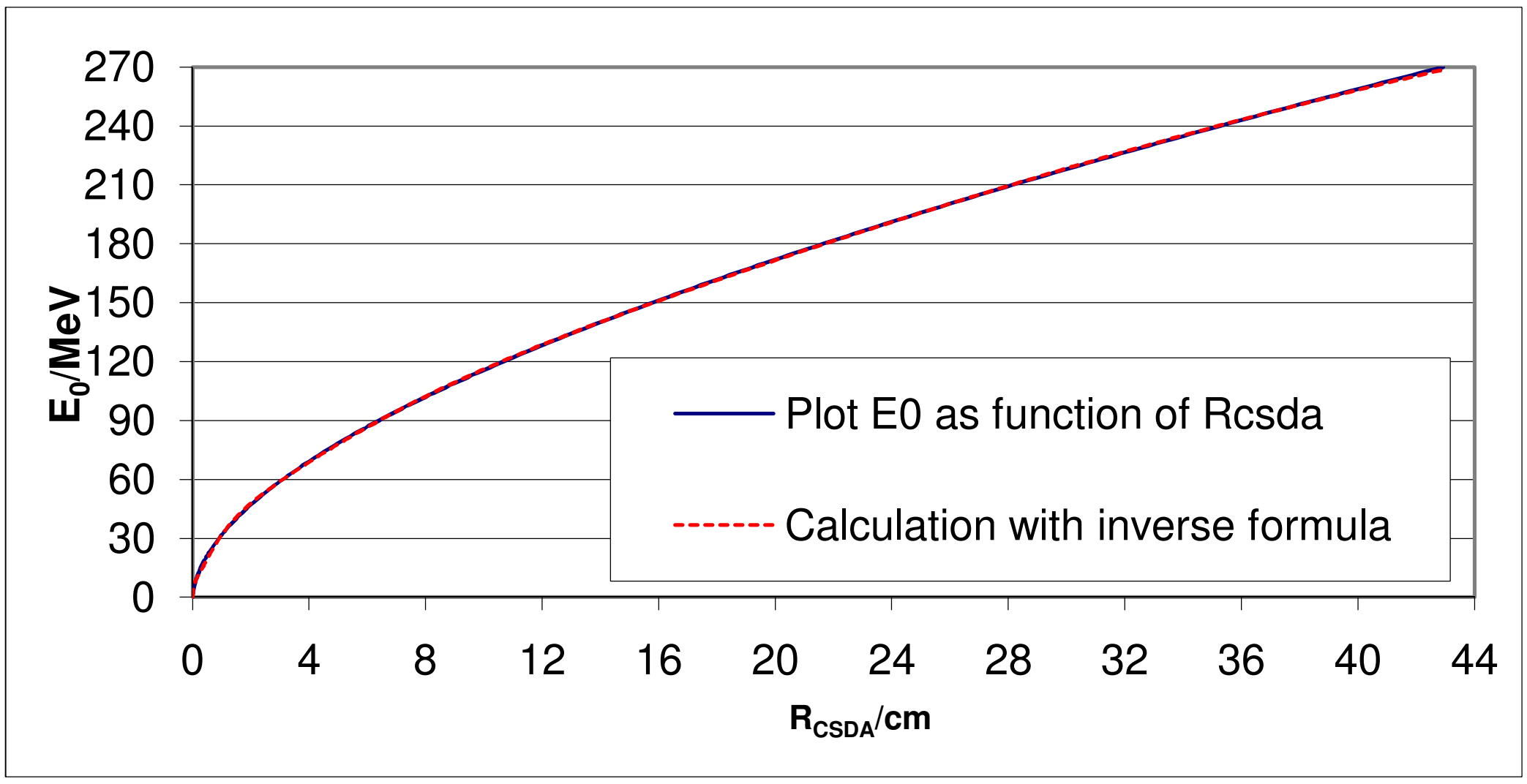

**Fig. 4:** Test of the inverse Formula (40) $E_0 = E_0(R_{CSDA})$ by five exponential functions. The mean deviation amounts to 0.11 MeV. The plot results from Fig. 1.

The residual energy E(z), appearing in Eq. (40), is the desired analytical base for all calculations of stopping power and comparisons with GEANT4. The stopping power is determined by dE(z)/dz and yields the following expression:

$$\left.\begin{aligned} S(z) &= dE(z)/dz \\ &= -E(z)/(R_{CSDA} - z) + \sum_{k=1}^{N} \lambda_k E_k(z) \ (\lim \ N \to \infty) \\ E_k(z) &= c_k (R_{CSDA} - z) \cdot \exp[-\lambda_k (R_{CSDA} - z)] \end{aligned}\right\} (41)$$

The aforementioned restriction to N = 5 is certainly extended to Eq. (41), which can be considered as a representation of the BBE in terms of the residual energy E(z). Due to the low-energy corrections ($a_0$, $a_{shell}$, $a_{Barkas}$), the energy-transfer function dE(z)/dz remains finite for all z (i.e., $0 \leq z \leq R_{CSDA}$). This is, for instance, not true for the corresponding results according to Eqs. (13 – 22) at z = $R_{CSDA}$. The calculation of E(z) and dE/dz according to Eqs. (40 – 41), referred to as LET, is presented in Fig. 5. The figure shows that, within the framework of CSDA, the LET of protons is rather small, except at the distal end of the proton track.

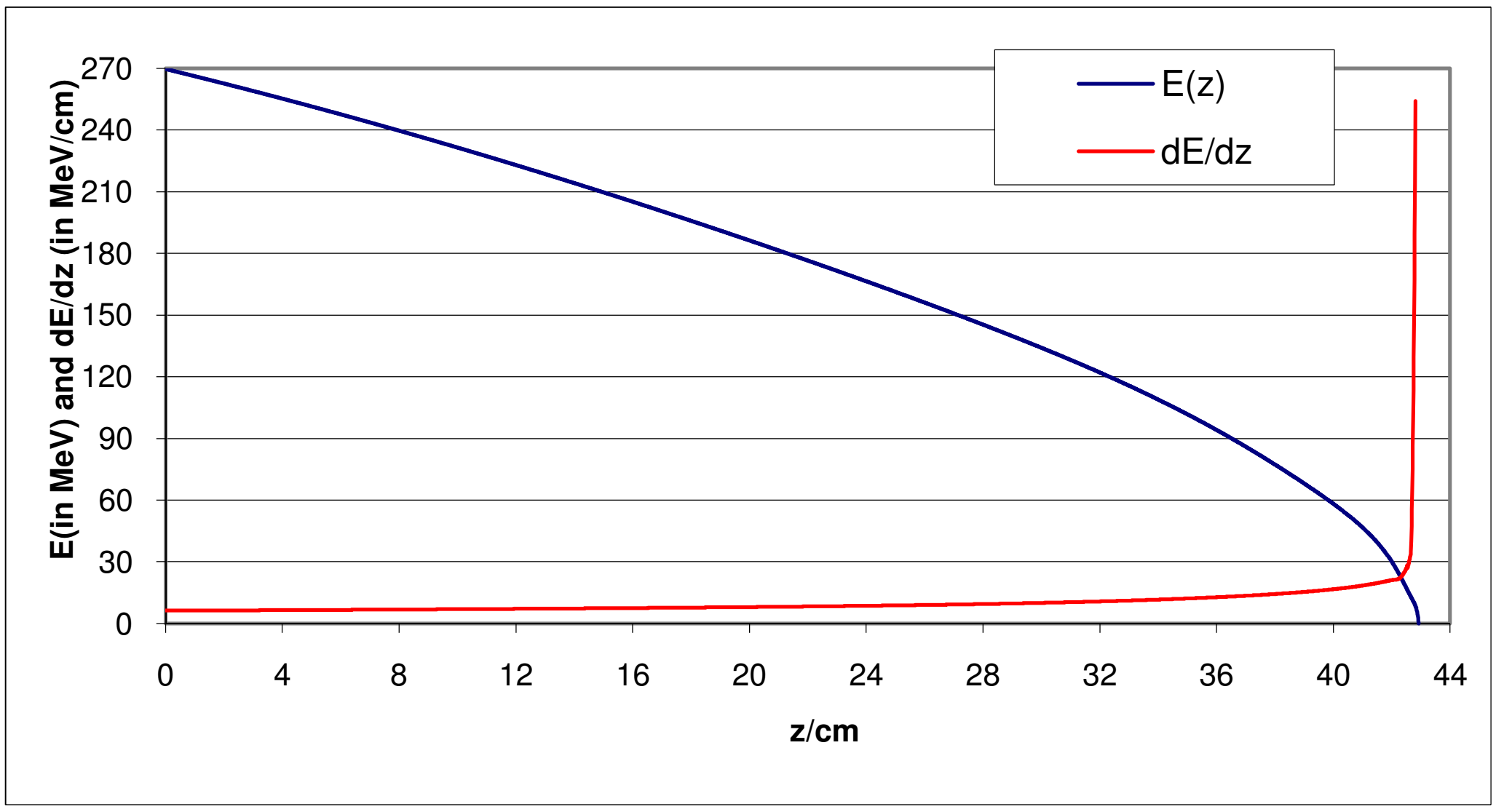

**Fig. 5:** E(z) and dE(z)/dz as a function of z (LET based on CSDA).

A change from the interacting reference medium water to any other medium can be carried out by the calculation of $R_{CSDA}$, where the substitutions have to be performed and used in Eqs. (40 – 41):

$$R_{CSDA}(medium) = R_{CSDA}(water) \cdot (Z \cdot \rho / A_N)_{water} \cdot (A_N / Z \cdot \rho)_{medium} \quad (42)$$

It is also possible to apply Eqs. (40 – 42) stepwise (e.g., voxels of CT). This procedure will not be discussed here, since it requires a correspondence between $(Z \cdot \rho / A_N)_{Medium}$ and information provided by CT, see Schneider et al. (1996).

### *1.3.2 Approximate solution of the Bethe-Bloch equation (calculation model M3)*

Eq. (41) consists of a sum of five exponential functions. In order to speed up this expression with respect to dose calculations (50 MeV: factor 2.4 up to 250 MeV: factor 4.7), we reduce the five exponential functions of S(z) = −dE(z)/dz to a single one, and introduce four more convenient (less time-consuming) functions according to the following criteria of Formula (41) leading to the functions $\varphi_1 \ldots \varphi_5$ of Eq. (43):

1. Optimization of the exponential behavior and coefficient weight, slightly depending on the initial energy $E_0$, by an envelope exponential function $\exp(-Q_p \cdot (R_{CSDA} - z))$, to provide the main contribution of the exponentially increasing part of Bragg curves ($\varphi_3$).

2. A Gaussian term ($\varphi_1$) containing a half-width $\tau_0$ with $\tau_0 \approx 10^{-5}$ cm aims at reflecting the behavior of the Bethe-Bloch function in the environment of the CSDA range, which would otherwise be singular and could not be integrated. Thus, we use $\exp(-(R_{CSDA} - z)^2/ \tau_0^2)$ instead of the δ-function (if lim $\tau_0 \to 0$); in the subsequent Eq. (44), the undefined square of a δ-function would appear instead of products with Gaussian terms resulting from $\varphi_1$. The problem of the singularity does not exist anymore in the presence of energy/range straggling as represented by the parameter τ (which will be defined in a later section). Later on, $\tau_0$ will be neglected ($\tau_0 << \tau_{straggle}$).

3. A power expansion of Eq. (43) with respect to the initial plateau and slowly increasing domain of S(z) up to the order $z^2/R_{CSCA}^2$ provides the functions $\varphi_2$, $\varphi_4$, and $\varphi_5$.

With the help of these five functions we are able to develop an accelerated algorithm:

$$\left.\begin{array}{l}
1.\ \varphi_1 = C_1(E_0)\cdot \exp(-(R_{CSDA} - z)^2/\tau_0^{\,2})\cdot \theta(R_{CSDA} - z) \\
2.\ \varphi_2 = 2\cdot C_2(E_0)\cdot \theta(R_{CSDA} - z) \\
3.\ \varphi_3 = 2\cdot C_3(E_0)\cdot \exp(-Q_p(E_0)\cdot (R_{CSDA} - z)\cdot \theta(R_{CSDA} - z)) \\
4.\ \varphi_4 = 2\cdot C_4(E_0)\cdot (z/R_{CSDA})^2\cdot \theta(R_{CSDA} - z) \\
5.\ \varphi_5 = 2\cdot C_5(E_0)\cdot (1 - z/R_{CSDA})\cdot \theta(R_{CSDA} - z)
\end{array}\right\} (43)$$

Explanations: $\theta(R_{CSDA} - z)$ is a unit step function, i.e., $\theta(R_{CSDA} - z) = 1$ (if $z \le R_{CSDA}$) and 0 (otherwise). The purpose of the unit step function is that the energy E(z) is zero for $z > R_{CSDA}$. $Q_p = \pi \cdot P_E/z_{max}$ appears in the function $\varphi_3$; $z_{max}$ will be explained at the end of this section. The parameter $P_E$ and the coefficients $C_1$, $C_2$, $C_3$, $C_4$, and $C_5$ depend linearly on $E_0$ and are determined by the variation procedure:

$$\left.\begin{array}{l}\sum_{E_0=1}^{300}\int_0^{R_{CSDA}}\left|S(z)-\sum_{k=1}^{5}\varphi_k(z,E_0)\right|^2 dz = Minimum\\ S(z)\approx\sum_{k=1}^{5}\varphi_k(z,E_0)\end{array}\right\}\quad(44)$$

It should be mentioned that the determination of the stopping-power function according to Eq. (43) agrees with an expansion of the solution functions of the BBE with Feynman propagators, but the energy dependence of the parameters had to be defined via Monte-Carlo calculations (GEANT4). In particular, the free particle propagator is also a Gaussian kernel, and $\varphi_1$ immediately results from this kernel; the remaining contributions 2 – 4 of Eq. (43) are generated via iterated integral operators. Since an analytical integration of the BBE is superior to a perturbation expansion based on propagators, we prefer to present here the latter solution method. The result of the adaptation of S(z) according to Eq. (44) is stated in Table 5. The mean standard deviation amounts to 0.7 %. It turned out that the contribution of the coefficient $C_5$ is rather negligible, since it amounts to 0.007; therefore, this contribution may be omitted ($C_5$ = 0). Eventually, the backbone of the accelerated algorithm can be restricted to the four coefficients $C_1,\ldots,C_4$ and the parameter $P_E$ related to $\varphi_3$. The coefficients $C_1,\ldots,C_4$ and $P_E$ are given by:

$$C_p = \alpha_{0,p} + \alpha_{1,p}\cdot E_0 \text{ and } P_E = \alpha_{0,5} + \alpha_{1,5}\cdot E_0 \quad (45)$$

**Table 5:** The parameters to calculate the energy dependence of the coefficients $C_p$ and $P_E$ according to Eq. (45).

| $C_p$ | $C_1$ | $C_2$ | $C_3$ | $C_4$ | $P_E$ |
|---|---|---|---|---|---|
| α0,p | 2.277463 | 0.2431 | 1.0295 | 0.4053 | 6.26751 |
| α1,p | - 0.0018473 | 0.0007 | - 0.00103 | - 0 .0007 | 0.00103 |

A determination of the parameters presented in Table 5 with GEANT4 (least-squares fit) leads to the following maximal deviations: $C_1$: +0.07 %; $C_2$ = +0.08 %; $C_3$: +0.04 %; $C_4$: –0.09 %; $P_E$: –0.08 % (corresponding to different values of $E_0$). The mean standard deviations are of the order of 0.04 % – 0.06 %. We should also note that some formulas of section 1.3 are still used in the calculation model M3 (e.g., Eqs. (36 – 37)). The parameter $z_{max}$, appearing in function 3 of the Equation system (43), is given by the definition

$$z_{max} = R_{CSDA} + \tau_{Range} \quad (46)$$

$$\tau_{Range} = R_{CSDA}\cdot(2.11791\cdot 10^{-5}\cdot E_0 + 0.9192399\cdot 10^{-7}\cdot E_0^2) \quad (47)$$

The meaning of $z_{max}$ and $\tau_{Range}$ results from adaptations of the function system (43), originally derived by Feynman propagators, to Monte-Carlo data of monoenergetic protons. In a subsequent section, we will verify that the parameter $\tau_{Range}$ will have some importance in the lateral scatter functions at the distal end.

## 1.4 Determination of the power p(E$_0$) of section 1.2 (phenomenological model M1)

The empirical Bragg-Kleeman rule (1) and its relativistic generalization (18) have been developed on the basis of phenomenological principles and classical (nonrelativistic/relativistic) equations of motion. The undefined parameters A and p can either be adapted by fits to experimental data or by comparison with $R_{CSDA}$ calculations based on the BBE. Since different numerical values for A and p have been proposed in relation to the considered energy domain $E_0$, we have performed an adaptation of these parameters based on the results (36 – 38). Formula (18) agrees with Formula (1) in the nonrelativistic limit with $E_0 \rightarrow 0$. Therefore, we have determined A by this request and permitted only a dependence of p on $E_0$. The result of this fit is given in Fig. 6; due to relativistic corrections, p turns out to be lower than the appropriate value used in the Bragg-Kleeman rule (Relation (1)). The dimensionless power p (nonrelativistic and relativistic) can be obtained from the expressions:

$$p = -5\cdot 10^{-15}\cdot E_0^6 + 5\cdot 10^{-12}\cdot E_0^5 - 2\cdot 10^{12}\cdot E_0^4 + \\ + 4\cdot 10^{-7}\cdot E_0^3 - 5\cdot 10^{-5}\cdot E_0^2 + 0.0003\cdot E_0 + 1.6577 \quad (48)$$

$$p = -4\cdot 10^{-15}\cdot E_0^6 + 4\cdot 10^{-12}\cdot E_0^5 - 2\cdot 10^{-9}\cdot E_0^4 + \\ + 4\cdot 10^{-7}\cdot E_0^3 - 5\cdot 10^{-5}\cdot E_0^2 + 0.0027\cdot E_0 + 1.6576 \quad (49)$$

Eq. (48) is the nonrelativistic determination of p(E$_0$) and Eq. (49) is valid for the relativistic case. It appears to be a simple task to invert Formulas (1 – 18) to obtain $E_0 = E_0(R_{CSDA})$, yet the energy dependence $p = p(E_0)$ prevents simple calculations; hence, iterative procedures are necessary. If one keeps A and p constant in a certain energy domain – see Bortfeld (1997) and Boon (1998), then the deviations turn out to be much higher for inverse calculations than for the original problem, i.e., $R_{CSDA}$ as a function of $E_0$. In Fig. 6, we have assumed that A = 0.00259 cm/MeV$^p$. Formulas (1, 18) are only valid for water. Therefore, the question as to a connection to Formula (35) has to be considered.

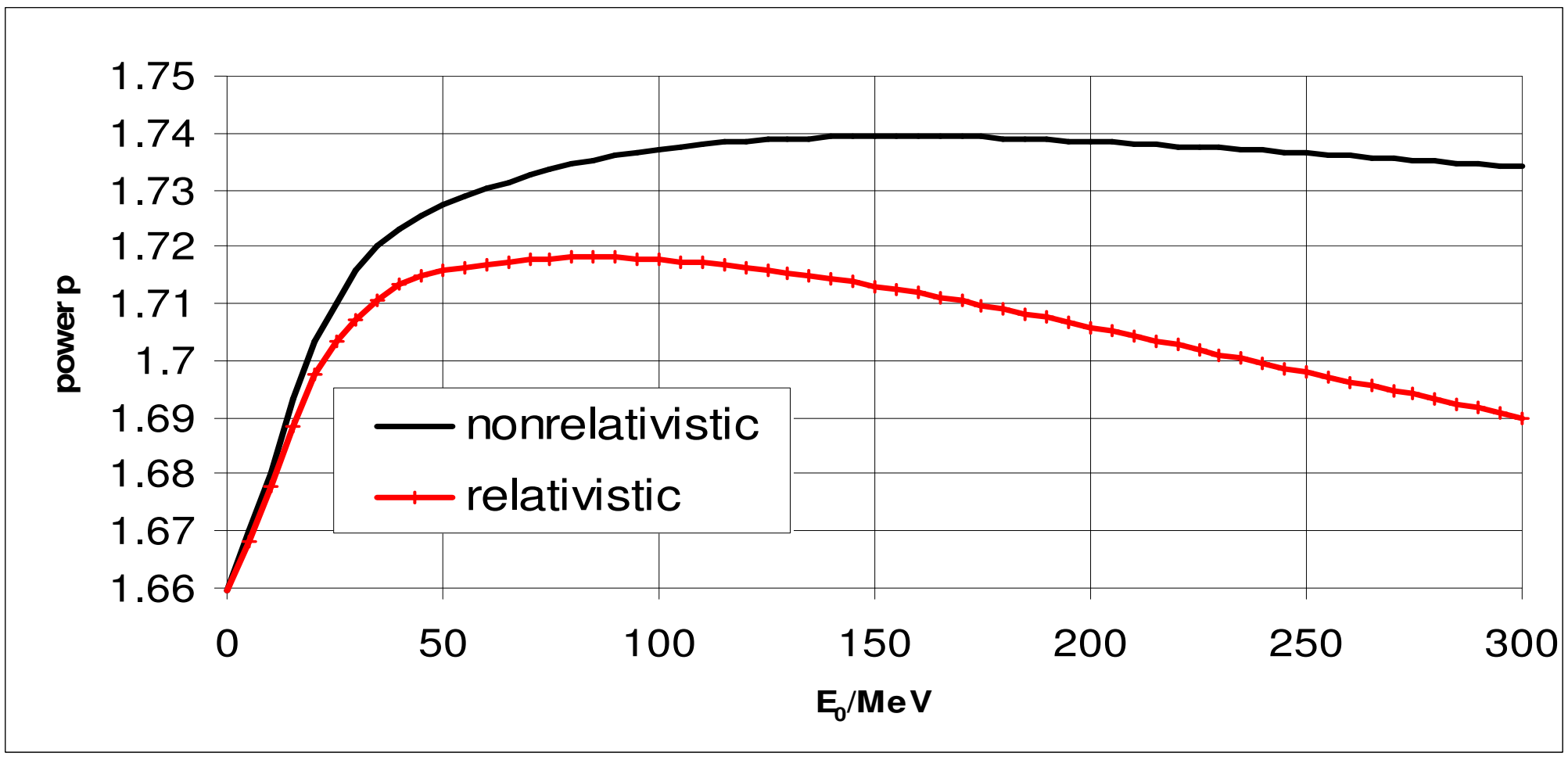

**Fig. 6:** The energy dependence of the power p in Formulas (48) and (49) is determined by Formula (36).

If we pass from water to any other (homogeneous) medium, then the Bragg rule suggests the following substitutions:

$$A \Rightarrow A \cdot \frac{Z_{water} \cdot \rho_{water} \cdot A_{N,medium}}{A_{N,water} \cdot Z_{medium} \cdot \rho_{medium}} \qquad (50)$$

A comparison of substitution (50) with Formula (35) showed that this substitution only holds, if $E_{I,water} \approx E_{I,medium}$. If $E_{I,water}$ significantly differed from $E_{I,medium}$, modifications also in the power p in Eqs. (1, 18) are required. For this purpose, we substitute the power p by $p_{water}$ and write $p_{medium}$ as the modified power:

$$\left.\begin{aligned} & p \Rightarrow p_{water} \\ & p_{medium} = p_{water} + \frac{1}{2 \cdot p_{water}} \cdot \ln(E_{I,water} / E_{I,medium}) \end{aligned}\right\} \quad (51)$$

We have performed various comparisons of Eq. (35) with Eqs. (50 – 51) for Pb, Cu, Ca, Al, and Be. The mean standard deviation amounted to 2.16 %. The maximal deviation was obtained for some ranges of protons in Pb with 3.76 %.

## 1.5 Monte-Carlo calculations and theoretical results of nuclear interactions

### *1.5.1 The BBE, fluctuations of the energy transfer, and multiple scatter*

The Monte-Carlo code GEANT4 is described in detail in the reference manual (2005). The calculation of the stopping power is based on a numerical manipulation of the Bethe equation with the Bloch corrections. The cutoff amounts to 1 MeV. Since GEANT4 is an open system, all correction terms according to the BBE in accordance with ICRU49 have been implemented. These corrections only imply the removal of singularities, if the total integration procedure is carried out analytically and not by numerical step-by-step calculations of $\Delta E/\Delta z$; these calculations depend on the actual velocity v and have to be performed for each term of the BBE separately. The scatter of protons is treated by the Molière multiple-scatter theory or (optionally) by the Lewis scatter theory. The code also contains a hadronic generator for the simulation of nuclear interaction processes. With respect to proton dose deposition, the basic theory is the BBE and a numerical fit of the Vavilov distribution function. A Vavilov distribution function takes account of the Landau tails; in the limit case of fluctuations of small transfer energies from protons to environmental electrons, a Vavilov distribution function assumes a Gaussian shape. In GEANT4, it is also possible to restrict these fluctuations to a Gaussian shape. This fact is of interest with regard to the role of the Landau tails in the initial plateau (entrance region) of Bragg curves. A theoretical analysis of Gaussian convolutions and their generalizations to account for Landau tails will be given in a later section.

### *1.5.2 Nuclear interactions of protons, release of secondary protons, and heavy recoils*

The most important aspect is the hadronic generator and the energy transport of secondary (and higher-order) particles. However, the default nuclear cross-section implemented in GEANT4 is very poor. Instead of using the default routine, which is based on data of Berger et al. (2000), we have implemented the cross-section data of $O^{16}$ of Chadwick and Young (1996). Furthermore, we have calculated this nuclear cross-section with the help of the extended nuclear shell theory containing, apart from the strong interaction spin-spin and spin-orbit couplings, the electrostatic interaction and the exchange interactions between the nucleons due to the Pauli principle (see Appendix). Thus, the wavefunctions of ground and excited states can be calculated by a perturbed $SU_3$, see Elliott (1963) and Meyer-Goeppert and Jensen (1970). The calculation of the cross-section due to energy transfer by an external proton is based on well-elaborated principles (determination of the transition probability and density of states). The result is given in Fig. 7.

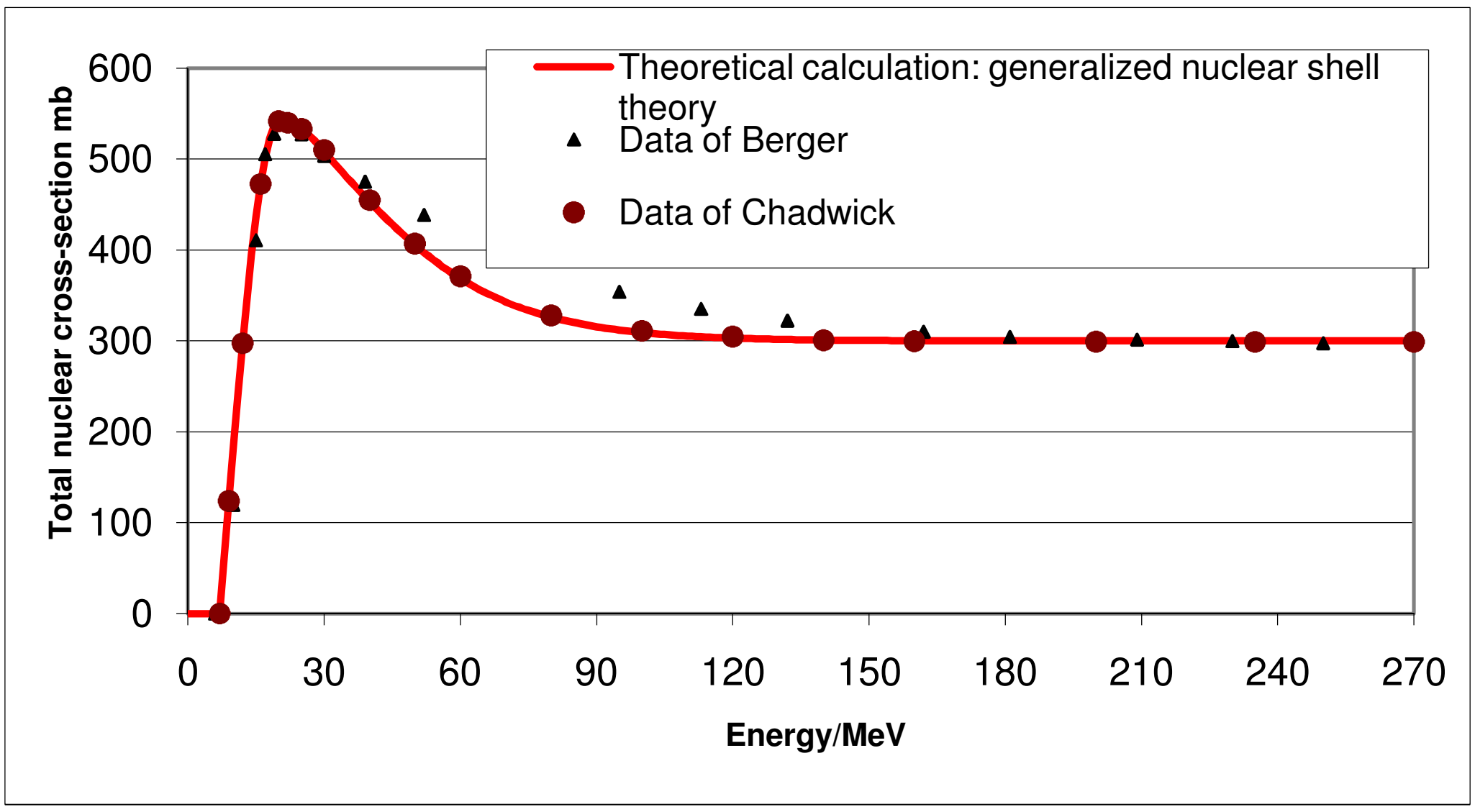

**Fig. 7:** The total nuclear cross-section of the proton – nucleus (O) interaction. By 'Data of Berger', we imply 'Data of Berger et al. (2000)'. By 'Data of Chadwick', we imply 'Data of Chadwick and Young (1996)'.

The decrease of fluence of primary protons can also be calculated from the results of this figure; we will deal with this issue in the next section. Protons with energy lower than the threshold energy $E_{Th}$ cannot surmount the potential wall of $O^{16}$. Fig. 7, which presents the total nuclear cross-section, shows that there is a threshold energy $E_{Th}$ = 7 MeV (more accurate: 6.997 MeV), which a proton should have to perform nuclear interactions with the $O^{16}$. For proton energies lower than the resonance maximum at $E_{res}$ = 20.12 MeV, the primary proton is preferably scattered by the nucleus (the secondary proton is now identical to the deflected primary one); the nucleus is excited, to undergo rotations/oscillations and emission of X-rays of very low energy (around 1 keV), leading to the release of Auger electrons. A complete classification of the total nuclear cross-section of the proton – nucleus (O) interaction for therapeutic protons is given by the following two types:

1. Potential scatter of protons by the strong interaction potential in the environment of the nucleus (note that $R_{strong}$: $\approx 1.2 \cdot A_N^{1/3} \cdot 10^{-13}$ cm determines a typical distance where the strong-interaction and the Coulomb force 'balance' one another; $A_N = 16$). For $R > R_{strong}$, only the Coulomb part is present. Potential scatter accounts for most of the protons undergoing a nuclear interaction. Resonance scattering of the incident protons at the nucleus occurs by inducing transitions between different states of the nucleus (e.g., vibrations leading to intermediate deformations, rotation bands, excited states by changing the spin multiplicity).

2. Nuclear reactions, which produce heavy recoils (see the listing 52 below). These protons are sometimes referred to as reaction protons – or 'secondary protons' by Boon (1998) and Paganetti (2002). For protons in the therapeutic energy domain, the amount of reaction protons is about 1 % - 3.5 % (see Appendix).

Within the total nuclear cross-section, the case 1 above plays the dominant role, if the residual proton energy

E is lower than 150 MeV. However, for E > 150 MeV case 2 is dominant. In first order, case 1 is described by the Breit-Wigner formula; for a detailed representation of this formula, see Segrè (1964). The Breit-Wigner formula results from the exact scatter theory and the restriction of the general *S-matrix* to 'S states', i.e., to l = 0. This restriction is only valid for light nuclei with rotational symmetry. If the number of neutrons is significantly different from the atomic number Z, the related *S-matrix* has also to account for 'higher states', i.e., for l ≠ 0. The total nuclear cross-section obtained by the Breit-Wigner formula has an elastic and inelastic part. The restrictions of the Breit-Wigner formula are insufficient in our problem; an extension to more resonances has been given by Flügge (1948). By taking account for the above-mentioned collective vibrations (oscillations of the nucleus by deformations and distortions) and rotations, we have taken into account all these degrees of freedom. Therefore, the total nuclear cross-section shows, around its maximum at $E_{res}$, a Gaussian behavior for $E > E_{Th}$, before it exponentially decreases to reach the asymptotic behavior.

It has to be added that, in all three cases, about 1 – 7 MeV of the proton energy (depending on the deflection angle) is transferred to whole nucleus to satisfy the energy and the momentum conservation in the center-of-mass system. This implies that for a neutron release, the proton energy has to be 21 – 27 MeV and not simply 20 MeV. Due to the potential barrier, the energy of the colliding proton has, at least, to be 30 MeV in order to release a secondary proton. These two processes have, however, an exception and may also occur at very low energies via exchange of mesons due to the Pauli principle, as pointed out in the discussion of listing (52).

The reaction protons (which always result from an inelastic process) are closely related to heavy recoil particles; the most probable heavy recoil elements resulting from the nuclear reactions of therapeutic protons are given by the types (1 – 5); types (6 – 7) result from types (1 – 2):

$$\left.\begin{array}{l}
1.\, p + O_8^{16} \Rightarrow n(neutron) + F_9^{16}\ (\beta^+ - decay, T_{1/2} = 22\ \sec + \gamma) \\
2.\, p + O_8^{16} \Rightarrow p + n + O_8^{15}\ (+\beta^+ - decay, T_{1/2} = 124\ \sec + \gamma) \\
3.\, p + O_8^{16} \Rightarrow p + p + N_7^{15}\ (+\beta^+ - decay,\ T_{1/2} = 10\min + \gamma) \\
4.\, p + O_8^{16} \Rightarrow \alpha + N_7^{13}\ (+\beta^+ - decay,\ T_{1/2} = 124\ \sec + \gamma) \\
5.\, p + O_8^{16} \Rightarrow d + O_8^{15}\ (+\beta^+ - decay,\ T_{1/2} = 124\ \sec + \gamma) \\
\left.\begin{array}{l}
6.\, n + O_8^{16} \Rightarrow p + N_7^{16}\ (+\beta^- - decay, T_{1/2} = 120\ \sec + \gamma) \\
7.\, n + O_8^{16} \Rightarrow p + n + N_7^{15}\ (see\ case\ 3)
\end{array}\right\}
\end{array}\right\} (52)$$

All types of $\beta^+$-decay emit one $\gamma$ quantum; its energy is around 0.6 MeV – 1 MeV. The $\beta^+$-decay of $F_9^{16}$ has a half-life of about 20 seconds, and $\gamma$ quanta are produced by collisions of positrons with environmental electrons. Some of the remaining heavy recoil fragments have half-times up to ten minutes ($N^{15}$). Since Fig. 7 refers to the total nuclear cross-section in relation to the actual (residual) proton energy, we have to add some qualitative aspects to the five different types with regard to the required proton energy: if E < 50 MeV, the type (1) is the most probable case with rapid decreasing tendency between 50 MeV < E < 60 MeV, to vanish for E > 60 MeV. Type (2) also pushes out a neutron, but the incoming proton is not absorbed; the required energy amounts, at least, to 50 MeV. Type (3) is similar, but requires, at least, about 60 MeV, with probability increasing with the energy. The release of α-particles, resulting from clusters in the nucleus, requires an energy E ≈100 MeV and the probability is increasing up to E ≈ 190 MeV; thereafter, it is decreasing rapidly, since higher-energy protons destroy these clusters by pushing out deuterons (type 5). Thus, case 5 is energetically possible for E > 60 MeV, but the significance is only increasing for E > 200 MeV. It has to be pointed out that, with regard to the nuclear reactions listed by the types (52), the exchange interactions between protons and neutrons, which result from the Pauli principle (applied to spin and isospin), play a dominant role. If the incident proton hits the nucleus, it is also included in these exchange interactions. According to the Pauli principle, a preferred interaction (resonance) of this proton with an environmental neutron occurred by virtue of the exchange of a $\pi^-$ meson (neutron n of the nucleus + p (incident) → p + n (outgoing)); this implies that a neutron of the nucleus converts to a proton and the incident proton leaves the nucleus as a neutron (reaction type (1) of the listing (52)). The inclusion of nuclear reaction processes to a generalized Breit-Wigner formula according to Flügge requires results of the extended nuclear shell theory. It is obvious that secondary protons may again release tertiary protons by additional inelastic scatter or by the types (6 – 7) of the listing (52), where released neutrons are responsible for resonance effects, e.g., type (6) represents the reversal process of type (1). The incoming (secondary) neutron is converted to a proton via $\pi^+$ exchange. The calculation procedure of the stopping power contribution of reaction protons $S_{sp,r}$ is presented in Appendix.

A particular feature of the nucleus is the lack of a central force, unlike the case of atomic electrons. Each nuclear constituent (proton, neutron) is therefore moving in the field of all remaining constituents (self-consistent field). The average kinetic energy amounts to 24 MeV. This fact implies that, due to the spatial charge distribution of the nucleus, the potential barrier with the threshold energy $E_{Th}$ is not proportional to $Z^2$; the exponent has to be somewhat lower, i.e., $Z^{\kappa}$ ($\kappa < 2$). An analysis of theoretical results and of the nuclear cross-section data of the Los Alamos library resulted in the following connection between $E_{Th}$ and Z (valid for those nuclei, for which the number of neutrons is approximately equal to that of protons):

$$\left.\begin{array}{l} E_{Th} = C_F \cdot Z^{\kappa} \cdot F(Z, A_N) \\ \kappa = 1.659 \quad (Z \geq 6) \end{array}\right\} \qquad (53)$$

$$\left.\left.\begin{array}{l} E_{Th} = C_F \cdot Z^{\kappa} \cdot F(C_6^{12}) \cdot (6 - Z) \\ \kappa = 1.659 + 0.341 \cdot (6 - Z)/5 \end{array}\right\} (Z < 6) \right\} \quad (54)$$

$$C_F = 0.222265 \qquad (55)$$

The form-factor function $F(Z, A_N)$ accounts for the total mass distribution of a nucleus; it is defined and explained in the Appendix. From Relation (54), it follows that κ converges to 2, if $Z \rightarrow 1$ (proton – proton repulsion); $F(A_N = 12.01, Z = 6) = 1.328917$. With regard to $E_{Th}$, Fig. 8 shows an application of the Formula (53). Referring to point 1 (potential scatter) and point 2 (resonance scatter), we have evaluated the Breit-Wigner formula for S and P states, and only the nuclear reaction processes producing neutrons, secondary protons, α-particles, deuterons, etc., have partially accounted for by data of the Los Alamos library. An analysis of the presented total nuclear cross-sections $Q^{tot}$ (and of some further cases) available in the Los Alamos library suggests the following adaptation model ($E_m$: the characteristic energy of the maximum value $Q^{tot}{}_{max}$ of $Q^{tot}$, $Q^{tot}{}_{as}$: the asymptotic value of $Q^{tot}$, approximately equal to the geometric cross-section; $\sigma_{res}$: the half-width of the resonance region; $\sigma_{as}$: a characteristic value used in the description before reaching the asymptotic behavior).

$$\left.\begin{aligned} A_{Th} &= exp(-(E_{Th} - E_{res})^2 / \sigma_{res}^2) \\ E_m &= E_{res} - E_{Th} \\ \sigma_{res} &= \sqrt{\pi} \cdot E_m \end{aligned}\right\} \quad (56)$$

$$\left.\begin{aligned} Q^{tot} &= Q^{tot}{}_{max} \cdot [exp(-(E - E_{res})^2 / \sigma_{res}^2) - A_{Th}] \cdot \\ &\quad \cdot (1 - A_{Th})^{-1} \qquad (\text{if } E_{Th} \le E_{res}) \\ Q^{tot} &= Q^{tot}{}_{max} \cdot exp(-(E - E_{res})^2 / 2\sigma_{res}^2) \; (\text{if } E_{res} < E < E_c) \\ E_c &= E_{res} + \sqrt{-2 \cdot ln(I_c)} \end{aligned}\right\} \quad (57)$$

$$\left.\begin{aligned} Q^{tot} &= Q^{tot}{}_c - (Q^{tot}{}_c - Q^{tot}{}_{as}) \cdot \tanh[(E - E_c)/\sigma_{as}] \; (if \quad E > E_c) \\ Q^{tot}{}_c &= I_c \cdot Q^{tot}{}_{\max} \\ \sigma_{as} &= \sigma_{res} \cdot (Q^{tot}{}_c - Q^{tot}{}_{as}) / (Q^{tot}{}_{\max} \cdot \sqrt{-2 \cdot \ln(I_c)}) \end{aligned}\right\} \quad (58)$$

The parameters $E_{res}$, $Q^{tot}{}_{max}$, $\sigma_{res}$, $I_c$ ($I_c = Q^{tot}{}_c/Q^{tot}{}_{max}$), $Q^{tot}{}_{as}$, and $\sigma_{as}$ are still not defined. A discussion on these parameters (as well as an overview of some theoretical aspects of nuclear physics, e.g., of the nuclear shell theory and extensions) will be given in Appendix). The parameters required in Eqs. (57 – 58), referring to Fig. 8, are given in Table 6.

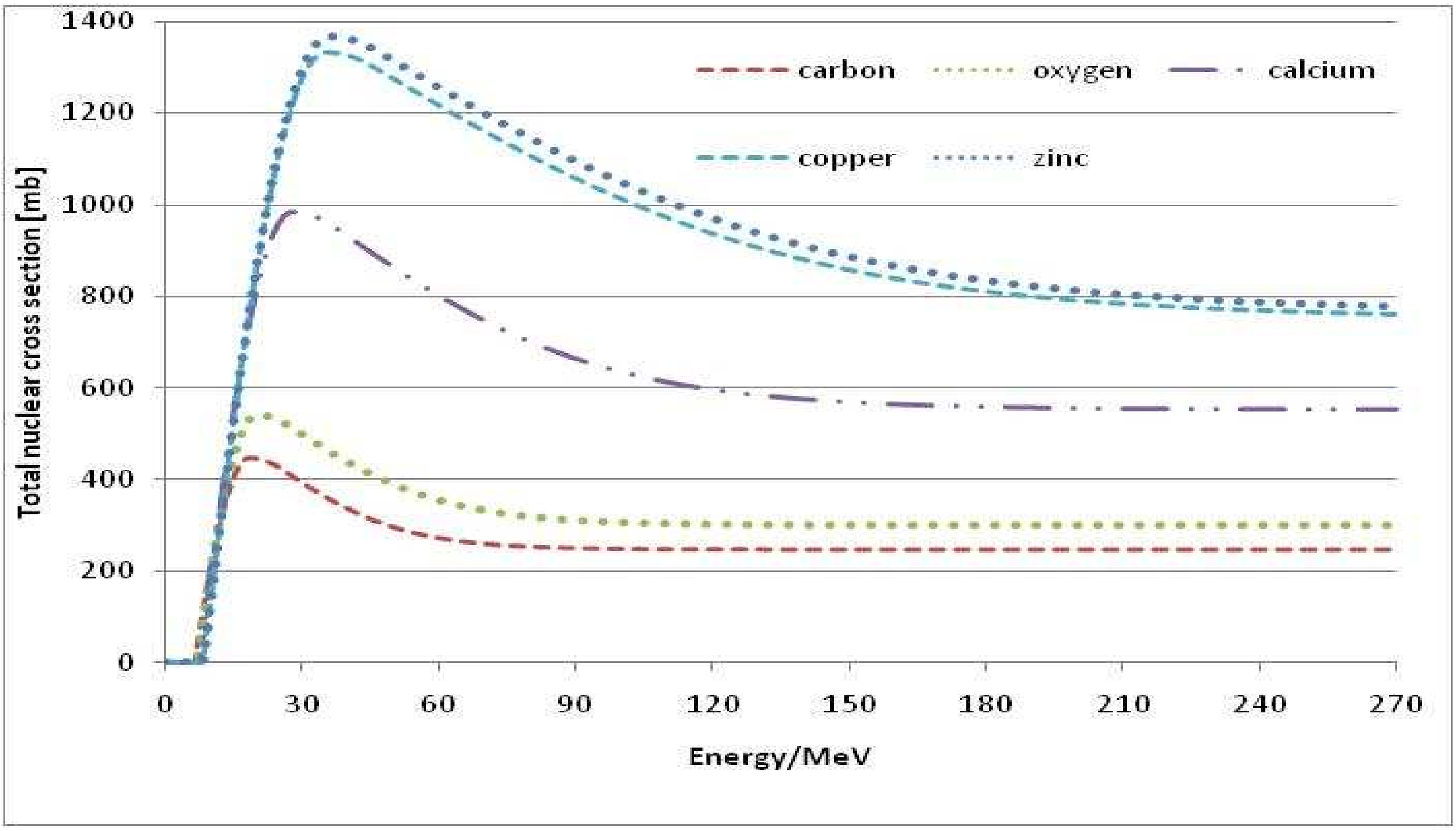

**Fig. 8:** The total nuclear cross-section for C, O, Ca, Cu, and Zn (taken from Ulmer and Schaffner (2006)).

**Table 6:** Numerical parameters for the evaluation of Fig. 8 according to Eqs. (57 – 58).

| Nucleus | $E_{Th}$/MeV | $E_{res}$/MeV | $\sigma_{res}$/MeV | $\sigma_{as}$/MeV | $Q^{tot}{}_{max}$/mb | $Q^{tot}{}_{c}$/mb | $Q^{tot}{}_{as}$/mb |
|---|---|---|---|---|---|---|---|
| C | 5.7433 | 17.5033 | 21.1985 | 27.1703 | 447.86 | 426.91 | 247.64 |
| O | 6.9999 | 20.1202 | 23.2546 | 34.1357 | 541.06 | 517.31 | 299.79 |
| Ca | 7.7096 | 25.2128 | 35.6329 | 58.4172 | 984.86 | 954.82 | 552.56 |
| Cu | 8.2911 | 33.4733 | 47.6475 | 93.2700 | 1341.94 | 1308.07 | 752.03 |
| Zn | 8.3213 | 33.9144 | 48.6416 | 96.8560 | 1365.50 | 1332.31 | 766.35 |

An inspection of Fig. 8 and of Eq. (57) indicates that two Gaussian distributions are needed between $E_{Th} \leq E \leq E_{res}$ and $E_{res} < E \leq E_c$, this being a result of the different interaction mechanisms of the proton with the nuclei. The parameter $\sigma_{as}$ describes the asymptotic behavior of $Q^{tot}$ for $E > E_c$ and is determined by the condition that, at $E = E_c$, Eqs. (57 – 58) have to be compatible (that is, the functions and their first derivatives must be equal). It is known from nuclear (and even particle) physics that inelastic cross-sections show an exponential decrease represented by the sum of some exponential functions before reaching the asymptotic region. We have verified that a single exponential function is not sufficient and that the hyperbolic-tangent (tanh) function provided more accurate results.

## 1.6 The Fluence decrease of primary protons and nuclear interactions

Using the methods described in section 1.5.2, we are able to define the fluence decrease of primary protons, the creation of secondary protons, recoil protons/neutrons, and the contribution of heavy recoil particles to the total stopping power S. The fluence decrease of primary protons can be calculated by a method of Segrè (1964):

$$\left.\begin{aligned} & d\Phi/\Phi = Z\cdot\rho\cdot(N_{Avogadro}/A_N)\cdot Q^{tot}(E)\cdot dz(E) \Rightarrow \\ & \int_{\Phi_0}^{\Phi} d\Phi/\Phi = \ln(\Phi/\Phi_0) = Z\cdot\rho\cdot(N_{Avogadro}/A_N)\cdot\int_{E_1}^{E_2} Q^{tot}(E)\cdot dE\cdot[dE/dz]^{-1} \end{aligned}\right\} \quad (59)$$

Note that $dz(E) = [dE/dz]^{-1}\cdot dE$; the boundary condition for the fluence $\Phi$ may be chosen by $\Phi = \Phi_0 = 1$ at the surface. The decrease function of primary protons, obtained via Eq. (59), is given in the subsequent sections.

*1.6.1 Primary (monoenergetic) protons $\Phi_{pp}$:*

$$\left.\begin{aligned} & \Phi_{pp} = \Phi_0\cdot[1-(\tfrac{E_0-E_{Th}}{Mc^2})^f\cdot\tfrac{z}{R_{CSDA}}]\cdot[1+erf((R_{CSDA}-z)/\sigma_{pp}))]\cdot\tfrac{1}{2} \\ & f = 1.032;\;\; E_{Th} = 7\;\; MeV \end{aligned}\right\} \quad (60)$$

It might be surprising that, in Formula (60), the error function erf($\xi$) and $\sigma_{pp}$ appear. In principle, the behavior of $\Phi_{pp}$, valid within the CSDA framework, should be a straight line as long as $E = E_{Th}$ is not yet reached. From $E < E_{Th}$ to $E = 0$, $\Phi_{pp}$ should be constant and, at $E = 0$ ($z = R_{CSDA}$), a jump to $\Phi_{pp} = 0$ is expected. However, due to energy/range straggling, proton beams can never remain monoenergetic in the sense of the CSDA. The parameter $\sigma_{pp}$ refers to the half-width of a Gaussian convolution, which introduces 'roundness' in the shape. The range of 7-MeV protons is less than 1 mm; therefore, we cannot verify whether $\Phi_{pp}$ is constant in the fluence profile of primary protons. The half-width parameter $\sigma_{PP}$ will be defined in a forthcoming section. An evaluation on the basis of Eq. (60) yields Fig. 9.

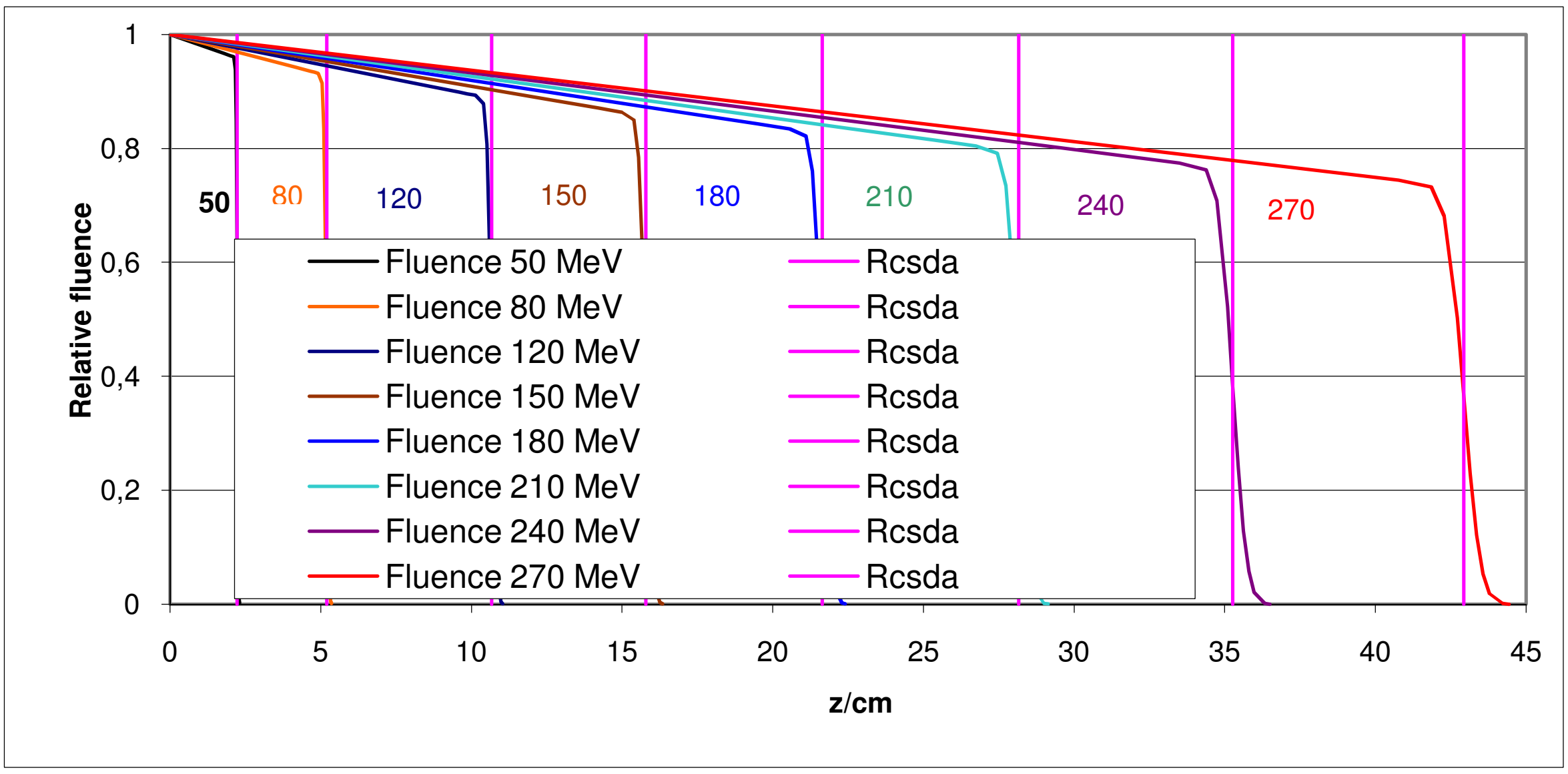

**Fig. 9:** The decrease of the fluence of the primary protons due to the nuclear interactions (O); evaluation of Fig. 8 using Gaussian convolutions with nuclear scatter and energy straggling.

Fig. 9 clearly shows that the slope of the straight lines depends on the difference $E_0 - E_{Th}$, if $E_0 \geq E_{Th}$. If $E_0 < E_{Th}$, the expression $(E_0 - E_{Th})^f$ becomes complex and we have to impose the condition $\Phi_{pp} = 1$. There have been many attempts – e.g., see Janni (1982) and Bortfeld (1997) – to fit the fluence decrease of primary protons by an approximated form, which solely depends on $R_{CSDA}$. Apart from the fact that these fits are unnecessarily complicated, they are rather inaccurate, since they do not involve the threshold energy $E_{Th}$, which decides, whether a nuclear interaction is possible at all or not.

### *1.6.2 Secondary protons $\Phi_{sp}$:*

In this section we separate the whole number of secondary protons by their origins, i.e., we differ between reaction protons $\Phi_{sp,r}$ and nonreaction protons $\Phi_{sp,n}$. Due to the complexity the contributions of reaction protons will be determined in section Appendix.

$$\left.\begin{aligned}&\Phi_{sp,n} = \Phi_0 \cdot [\upsilon' \cdot (\tfrac{E_0 - E_{Th}}{Mc^2})^f \cdot \tfrac{z}{R_{CSDA}}] \cdot [1 + erf((R_{CSDA} - z - z_{shift}(E_0))/\sigma_{sp})] \cdot \tfrac{1}{2}\\ &\upsilon' = \upsilon - 2 \cdot C_{heavy};\ \upsilon = 0.958\end{aligned}\right\} \quad (61)$$

It should be noted that $\sigma_{sp}$ is somewhat different from $\sigma_{pp}$ of Eq. (60). The argument of the error function in Eq. (61) is slightly changed by the additional $z_{shift}$, which results from an average energy loss of the secondary protons. The uncertainty intervals in the value $\upsilon$ = 0.958 are + 0.40 % and – 0.42 %. Thus, $\upsilon$ =

0.958 represents the value with the lowest mean standard deviation.

### *1.6.3 Recoil protons/neutrons $\Phi_{rp}$:*

$$\Phi_{rp} = \Phi_0 \cdot [\eta \cdot (\frac{E_0 - E_{Th}}{Mc^2})^f \cdot \frac{z}{R_{CSDA}}] \cdot [1 + erf((R_{CSDA} - z - z_{shift}) / \sigma_{rp})] \cdot \frac{1}{2} \quad (62)$$

For the recoil factor η, we use η = 0.042. The remaining parameters are the same as in Eqs. (61 – 62). The fluence decrease of primary protons, resulting from Fig. 8, is analogous to that of water (Fig. 9).

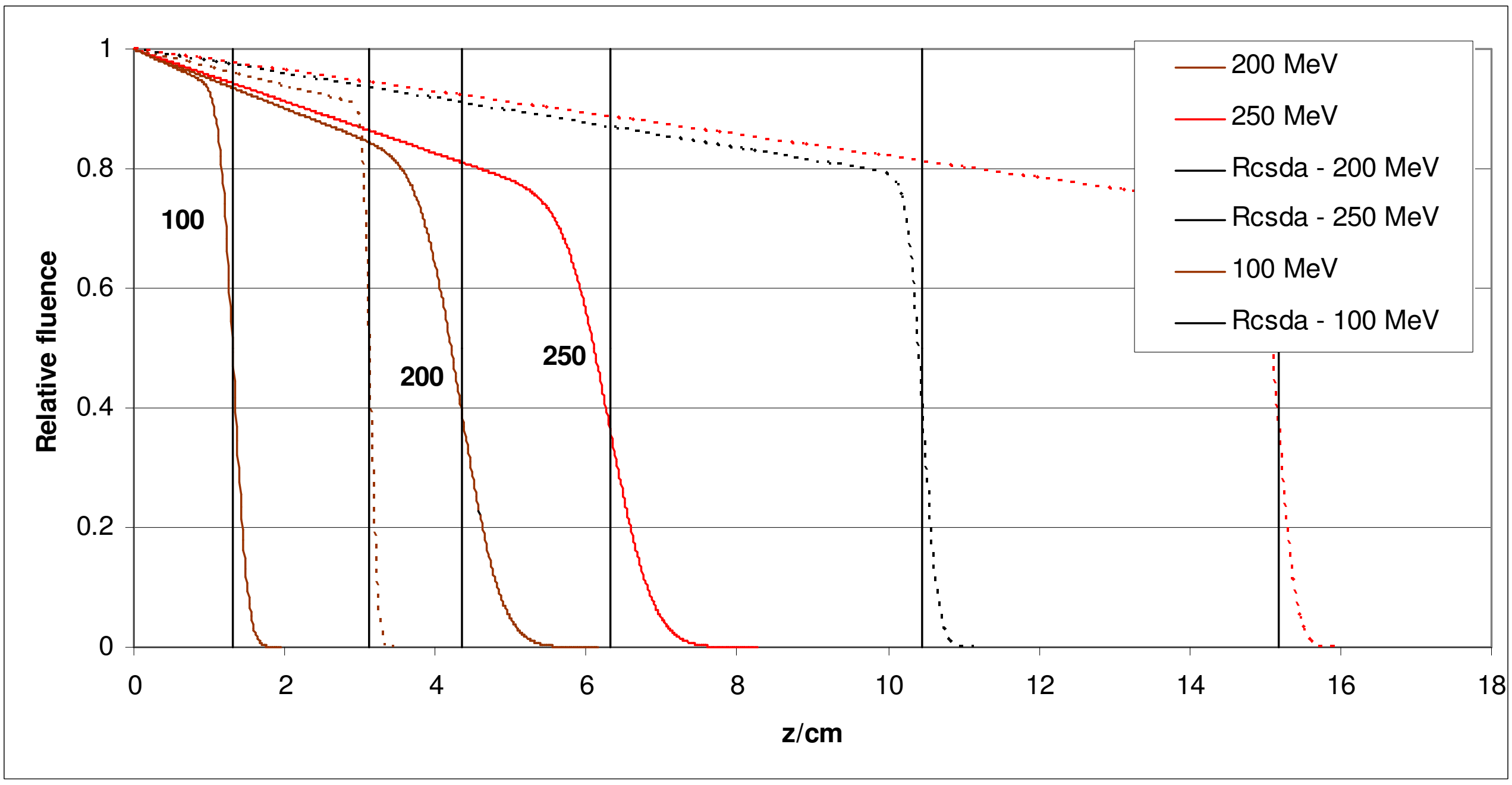

**Fig. 10:** Decrease of the fluence of 100, 200, and 250 MeV primary protons in copper (solid lines) and in calcium (dashes). The $R_{CSDA}$ ranges are stated by perpendicular lines.

A comparison of Figs. 9 and 10 shows, that the slope of the fluence decrease is steeper in the latter case, in particular for proton energies below 100 MeV. The power f =1.032, which is valid for water to compute the slope of the straight line in Eqs. (60 – 62), has to be slightly modified (f = 0.755 for copper and f = 0.86 for calcium), and the general formulas are given by:

$$f(Z, A_N) = a \cdot A_N^{-1} + b \cdot A_N^{-2/3} + c \cdot A_N^{-1/3} + d \cdot Z \cdot A_N^{-1/2} \quad (63)$$

$$\left(\frac{E_0 - E_{Th}}{M \cdot c^2}\right)^{1.032} \Rightarrow \left(\frac{E_0 - E_{Th}}{M \cdot c^2}\right)^{f} \quad (64)$$

After these modifications Eqs. (60 – 62) can be applied. Eq. (63) is closely related to the nuclear collective model (see Appendix), where the parameters *a, b, c* and *d* are interpreted in terms of different contributions to the total cross section. These parameters have been determined by this model: a = -0.087660001, b = -6.379250217, c = 5.401490050and d = - 0.054279999. There are two applications, in which Eqs. (63 - 64) are relevant: 1. Passage of protons through collimators. 2. Passage of protons through bone/metallic implants. In case 2, only a small path length has to be corrected; however, Fig. 10 shows that the fluence decrease has also to be corrected to fulfill continuity at the boundaries.

## 1.7 Convolution theory and the energy-range straggling by fluctuations of the energy transfer to environmental electrons

This section requires knowledge of some elements of advanced mathematical physics. The result of this section, in a concise form, is that a Gaussian convolution is rigorously only valid in the nonrelativistic domain of a proton track, i.e., in the environment of the Bragg peak, whereas relativistic corrections have to be included in the initial plateau. Readers who may have some difficulty with mathematical physics may directly proceed to the next Chapter 1.8.

### *1.7.1. Bohr's classical formula of energy straggling*

According to Bohr's formalism – Bethe et al. (1953), the formula for energy straggling (or fluctuation) $S_F$ is given by

$$S_F = \frac{1}{\sqrt{\pi}\,\sigma_E} \exp[\,-(E - E_{Average})^2 / \sigma_E^{\,2}\,] \qquad (65)$$

The fluctuation parameter $\sigma_E$ can be best determined using the method of Bethe et al. (1953).

$$\left.\begin{array}{l} \Delta\sigma_E^{\,2} = \Delta z \cdot \frac{1}{2} \cdot (Z/A_N) \cdot \rho \cdot f \cdot \frac{2mc^2}{1-\beta^2}(1 - \beta^2/2) \ \text{(for finite intervals } \Delta z) \\ f = 0.1535 \ \text{MeVcm}^2/\text{g} \\ d\sigma_E^{\,2}/dz = \frac{1}{2} \cdot (Z/A_N) \cdot \rho \cdot f \cdot \frac{2mc^2}{1-\beta^2}(1 - \beta^2/2) \end{array}\right\} \quad (66)$$

$\Delta\sigma_E^2$ contains as a factor the important magnitude $E_{max}$, that is, the maximum energy transfer from the proton to an environmental electron; it is given by $E_{max} = 2mv^2/(1-\beta^2)$. In a nonrelativistic approach, we get $E_{max} = 2mv^2$. $E_{max}$ can be represented in terms of the energy E, and, for the integrations to be performed, we recall the relation E = E(z) according to Eq. (40):

$$E_{max}(\text{in keV}) = \sum_{k=1}^{4} s_k \cdot E^k \quad (E \text{ in MeV}) \quad (67)$$

This formula is plotted in Fig. 11 (the nonrelativistic limit is given by the straight line $s_1 \cdot E$); the parameters $s_k$ are displayed in Table 7.

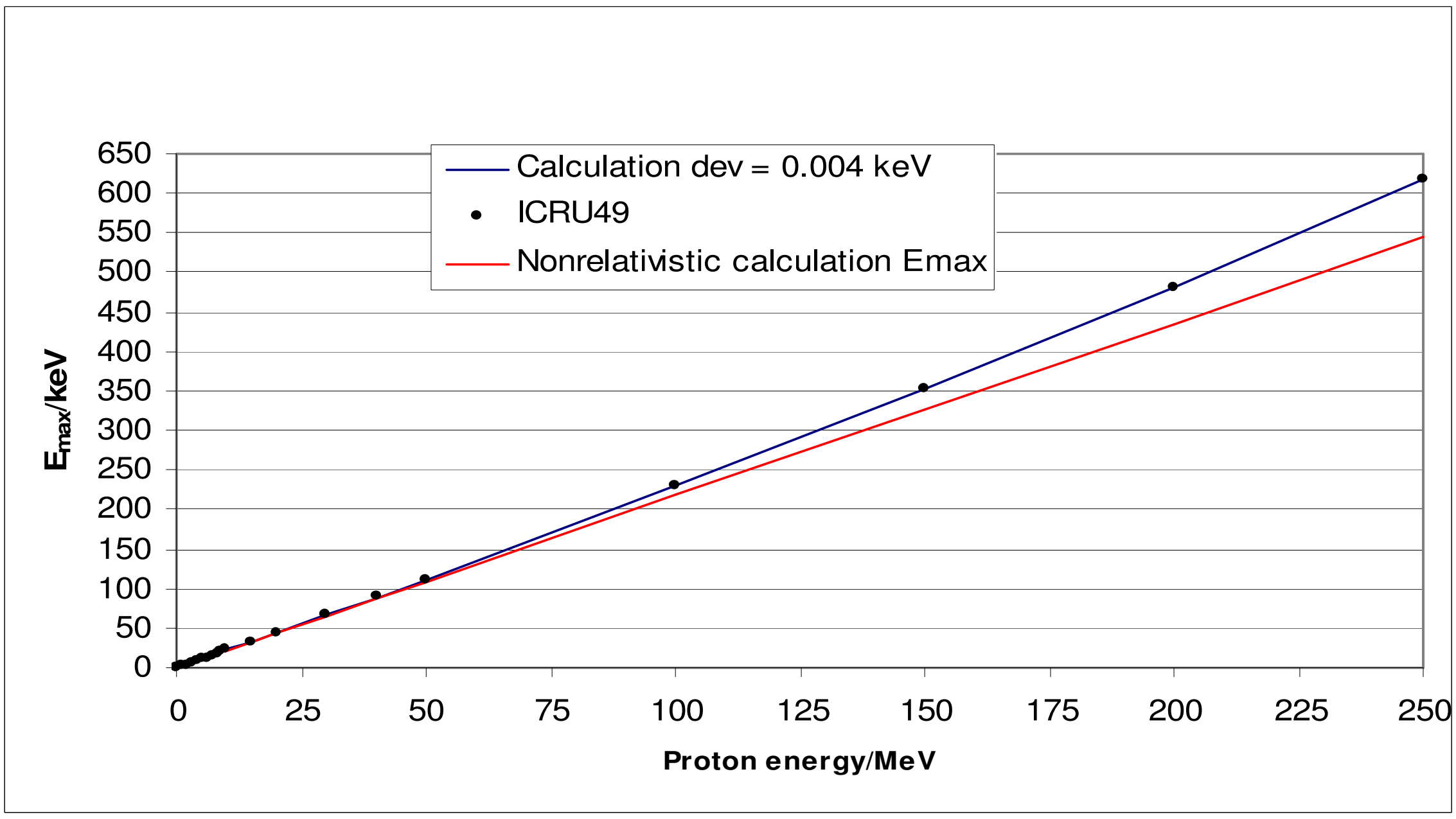

**Fig. 11:** Calculation of $E_{max}$ according to Eq. (67). The straight line refers to the nonrelativistic limit.

**Table 7.** The parameters $s_k$ for the calculation of $E_{max}$ (Formula (67)).

| $s_1$ | $s_2$ | $s_3$ | $s_4$ |
|---|---|---|---|
| 2.176519870758 | 0.001175000049 | -0.000000045000 | 0.0000000000348 |

*1.7.2 Gaussian convolution in configuration space and its operator notation*

A Gaussian distribution is assumed to account for the energy straggling, if the stopping power is given by the CSDA approach. However, this is valid only in the nonrelativistic domain, and the Gaussian convolution kernel in the configuration space must not be mixed with Bohr's relation. The convolution kernel reads as:

$$K(\sigma, u - z) = \frac{1}{\sigma\sqrt{\pi}} \exp(-(u - z)^2 / \sigma^2) \quad (68)$$

Here the half-width parameter σ is arbitrary; this quantity will be later on identified with the energy/range straggling or the lateral scatter, if Eq. (68) is appropriately modified.

In a 3D version, linear combinations of K(σ, u – x) and the inverse kernel $K^{-1}$ are also used in scatter problems of photons and electrons, see Ulmer et al. (2005). If the dose deposition of protons is calculated by the BBE or by the phenomenological Eqs. (13, 22) based on classical energy dissipation, then the energy fluctuations are usually accounted for by:

$$D(z) = \int D_{R_{CSDA}}(u) K(\sigma, u - z) du \quad (69)$$

This kernel may either be established by nonrelativistic transport theory (Boltzmann equation) or, as we prefer here, by a quantum statistical derivation, where a relation to the path-integral formulation of Neumann's density matrix will be obtained, see Feynman and Hibbs (1965).

Let φ be a distribution function and Φ a source function, mutually connected by the operator $F_H$ (operator notation of a canonical ensemble):

$$\left.\begin{array}{l} \varphi = \exp(-H / E_{ex})\Phi = F_H \Phi \\ F_H = \exp(-H / E_{ex}) \end{array}\right\} \quad (70)$$

An exchange Hamiltonian H couples the source field Φ (proton fluence) with an environmental field φ by $F_H$, due to the interaction with electrons:

$$\left.\begin{array}{l} H = -\frac{\hbar^2}{2m} d^2 / dz^2 \\ \exp(0.25\sigma^2 d^2 / dz^2)\Phi = F_H \Phi = \varphi \\ \sigma^2 = 2\hbar^2 / m E_{ex} \end{array}\right\} \quad (71)$$

It must be noted that the operator Eq. (71) was formally introduced (see Ulmer and Kaissl (2003)) to obtain a Gaussian convolution as Green's function and to derive the inverse convolution. $F_H$ may formally be expanded in the same fashion as the usual exponential function exp(ξ); ξ may either be a real or complex number. This expansion is referred to as Lie series of an operator function. Only in the thermodynamical limit (equilibrium), can we write $E_{ex} = k_B T$, where $k_B$ is the Boltzmann constant. This equation can be solved by the spectral theorem (functional analysis). According to this theorem, we have to consider the eigenvalue problem:

$$\left.\begin{aligned}
&F_H\Phi = \gamma\Phi \\
&\Phi_k = exp(-ikz)/\sqrt{2\pi} \\
&F_H\Phi_k = \gamma(k)exp(ikz)/\sqrt{2\pi} = exp(-\sigma^2k^2/4)\cdot exp(ikz)/\sqrt{2\pi} \\
&K(\sigma, u-z) = \int \Phi_k^*(z)\Phi_k(u)\gamma(k)dk = \tfrac{1}{2\pi}\int exp(-\sigma^2k^2/4)exp(ik(u-z))dk \\
&K(\sigma, u-z) = \tfrac{1}{\sigma\sqrt{\pi}}exp(-(u-z)^2/\sigma^2)
\end{aligned}\right\} (72)$$

As shown in a previous study, Green's function of the operator $A^{-1}$ is a two-point Gaussian convolution kernel as stated above, see Ulmer and Kaissl (2003). In order to avoid confusion with the proportionality factor $A$ in Eqs. (1, 11, 13, 18 – 22), we denote this operator function by $F_H$. The density-matrix formulation of quantum mechanics is based on the definition:

$$\rho(u, z) = \sum_{k=0}^{\infty} \exp(-E_k / k_B T)\psi^*(k, u)\cdot\psi(k, z) \quad (73)$$

For the statistical motion of free particles ($E_k = \hbar^2k^2/2m$), we obtain $\rho(u - z) = K(\sigma, u - z)$, if $E_{ex} = k_BT$. In the continuous case, the summation over k has to be replaced by an integral. With the help of Eq. (73), various properties (e.g., the partition function) can be calculated. According to Feynman and Hibbs (1965), we equate $\rho(u - z)$ to the path-integral kernel $K_F(k_BT, u - z)$, useful for the calculation of perturbation problems of statistical mechanics, if the formal substitution $it/\hbar \rightarrow 1/k_BT$ is carried out. Thus, the operator-function formalism according to Relations (71 – 72) can be regarded as an operator calculus of path-integral kernels. The question now is, which temperature T or exchange energy $E_{ex}$ should be used and whether $E_{ex} = k_BT$ may hold in the energy straggling of protons and, consequently, this parameter may be kept constant along the pathway. For this purpose, we consider some further properties which can easily be derived from the operator notation. It immediately follows from $F_H = \exp(0.25\sigma^2 d^2/dz^2)$ that by n-times repetition of this operator (method of iterated operators), we obtain:

$$(F_H)^n = [exp(0.25\cdot\sigma^2 d^2/dz^2)]^n = exp(0.25\cdot n\cdot\sigma^2 d^2/dz^2) \quad (74)$$

The kernel K resulting from Eq. (74) is:

$$\left.\begin{aligned}
&K(\sigma_n, u-z) = \frac{1}{\sqrt{\pi}}\frac{1}{\sigma_n}\exp(-(u-z)^2/\sigma_n^2) \\
&\sigma_n^2 = n\cdot\sigma^2
\end{aligned}\right\} (75)$$

A composite application (e.g., the energy-range straggling of polychromatic proton beams) is now given by $F_H(\sigma_1)\cdot F_H(\sigma_2)$ and yields:

$$\left.\begin{aligned} &\mathrm{F_H}(\sigma_1)\cdot \mathrm{F_H}(\sigma_2) = \exp(\ 0.25\cdot(\sigma_1^2+\sigma_2^2)\mathrm{d}^2/\mathrm{d}z^2\ ) \Rightarrow \mathrm{K}(\sigma_t, u-z) \\ &\sigma_t^2 = \sigma_1^2+\sigma_2^2 \end{aligned}\right\} \quad (76)$$

The advantage of the operator calculus of iterated kernels is the straightforward calculation of generalized kernels K, even if these kernels are not Gaussian, such in case of a Landau – Vavilov distribution kernel. The question arises as to the order of magnitude following from Eq. (71) for $\sigma^2$, which is expressed there in terms of fundamental constants. For this purpose, we partition the whole proton path from z = 0 to z = $R_{CSDA}$ in molecular intervals of water (or other) atoms/molecules. With respect to water molecules ($\rho$ = 1g/cm$^3$, $N_{Avogadro} = 6.022\cdot10^{23}$, $A_N = 18$, $M_{Mol} = 18$ g), the average distance between the centers of mass amounts to $l_A = 8.4446\cdot10^{-7}$ cm. The exchange particle is the electron, which undergoes further collisions with environmental atoms/molecules, to yield finally a locally stored energy that can be recorded by calorimetric measurements. In the case of electrons $E_{ex}$ = 1 eV, the calculated $\sigma$ amounts to $3.26\cdot10^{-7}$ cm, and for $E_{ex}$ = 0.1 eV (thermal energy) to $\sigma = 1.03\cdot10^{-6}$ cm. It is obvious that this process has to be repeated. This means that $\sigma_n^2 = n\cdot\sigma^2$ enters the convolution kernel as a parameter, which is now $\sigma^2(z)$. It should be noted that the problem of composite convolutions can easily be handled by the operator notation. If the exchange energy $E_{ex}$ would remain always constant, then we should obtain $\sigma_n \sim R_{CSDA}^{1/2}$ at the distal end. However, this is not true, as the locally stored exchange energy is increasing with decreasing proton energy. In Figs. 12 and 13, we show the 'global average: $E_{av_global}$', which results from the energy $E_0$ divided by the number of water molecules per unit length, and the 'local average: $E_{av_local}$', obtained by a subtraction method (starting with the lowest energy $E_0$ = 1 keV).

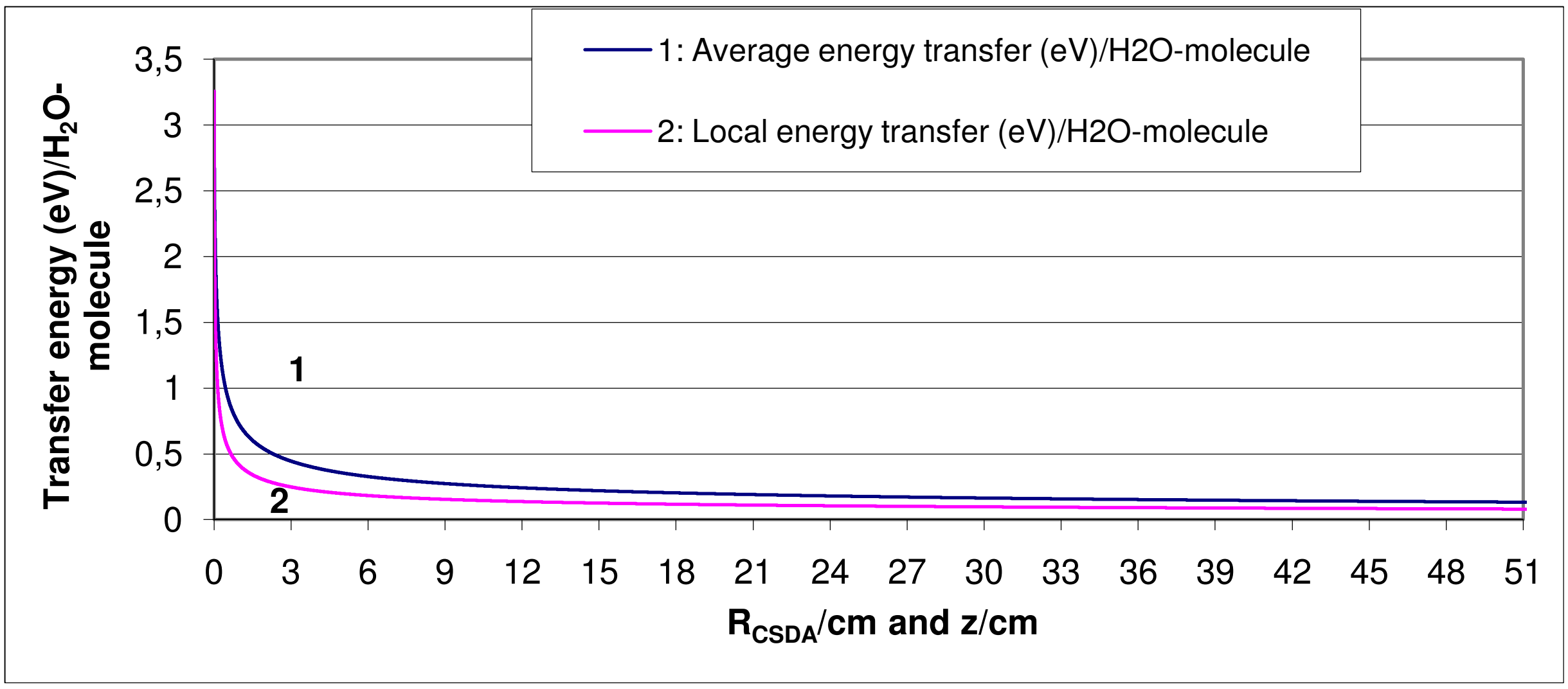

**Fig. 12:** The average transfer energy (1) and the local transfer energy (2) from proton to water molecules (CSDA approach).

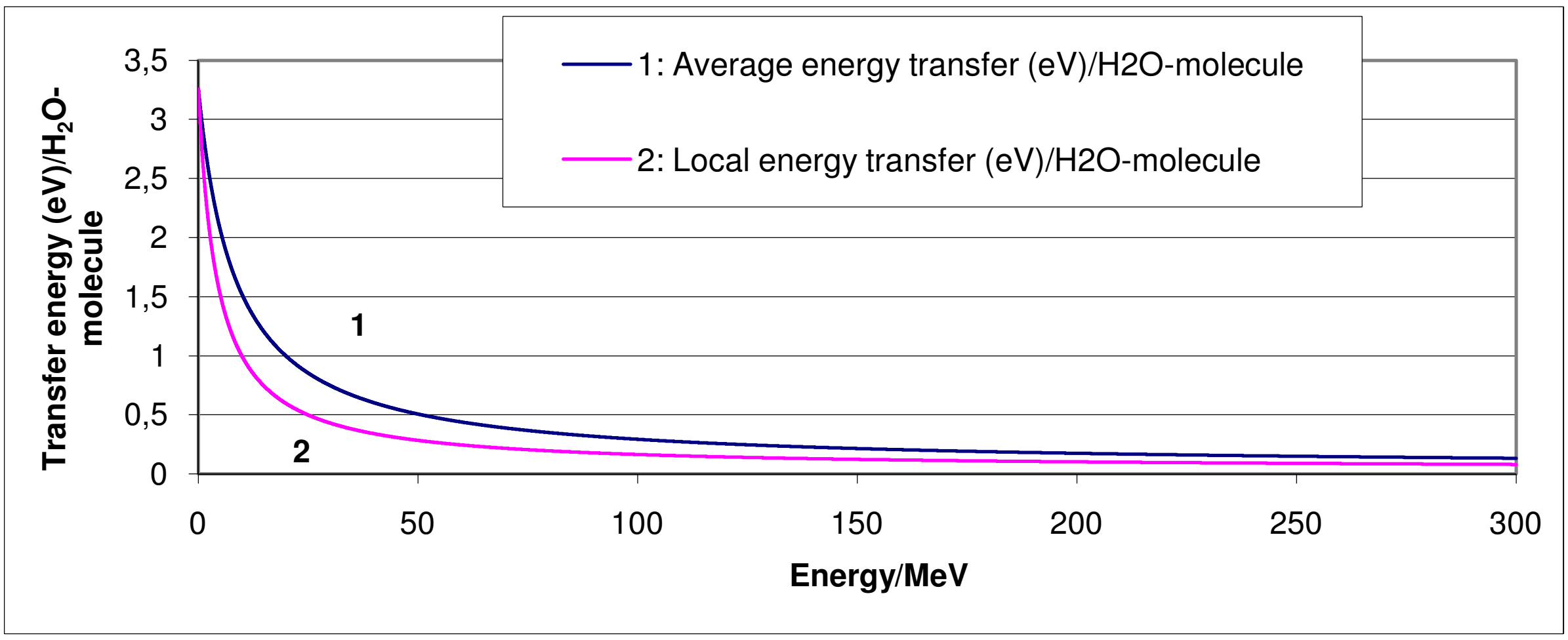

**Fig. 13**: The average transfer energy (1) and the local transfer energy (2) versus the total energy E.

If we compare this result with $E_{max}$ according to Fig. 11, we may verify that, in particular for high proton energies, a considerable amount of the proton energy is stored in δ-electrons, and, only by further collisions of these electrons along the track with the electrons of the environmental molecules, is the stored energy locally downgraded to thermal energies. (These processes comprise the starting point of calorimetric measurements.) Some specific numerical cases for exchange energies $E_{ex}$ are given in Table 8.

**Table 8:** Exchange length σ in dependence energy $E_{ex}$ according to Eq. (71).

| Energy | exchange length σ (in cm) | exchange length σ (in cm) |
|---|---|---|
| Exchange particle | proton | electron |
| 1 MeV | $0.91 \cdot 10^{-12}$ | $3.86 \cdot 10^{-11}$ |
| 1 keV | $0.24 \cdot 10^{-10}$ | $1.03 \cdot 10^{-9}$ |
| 1 eV | $0.76 \cdot 10^{-8}$ | $3.26 \cdot 10^{-7}$ |
| 0.1 eV | $0.24 \cdot 10^{-7}$ | $1.02 \cdot 10^{-6}$ |

One should note that, only for proton – electron collisions, is the exchange particle an electron, where a Boltzmann partition function can be applied. For proton – nucleon or electron – electron collisions, due to the Pauli principle, we have to make use of Fermi – Dirac statistics (see later section). The order of the exchange range lengths of Table 8 remains valid. With respect to the operator calculus of the Gaussian kernel, we finally add two items. The 3D version of $F_H$ with different σ values in the z direction and x/y plane is given by:

$$\left.\begin{array}{l} F_H = \boldsymbol{exp[-0.25\cdot(\sigma^2\cdot\partial^2/\partial z^2+\sigma_l^2(\partial^2/\partial x^2+\partial^2/\partial y^2))]} \Rightarrow \\ K(u-z, v-x, w-y) = \frac{1}{\sigma\sqrt{\pi}}\frac{1}{\pi\cdot\sigma_l^2}\boldsymbol{exp[-((u-z)^2/\sigma^2+((v-x)^2+(w-y)^2)/\sigma_l^2)]} \end{array}\right\} \quad (77)$$

Eq. (77) additionally represents the basis for multiple lateral scatter. If it is modified according to principles developed in the next section, i.e., that proton scatter is only *approximately* described by one Gaussian on the x/y plane, we are able to describe the Molière multiple scatter and energy fluctuations in a unique manner. It should be noted that $\sigma^2 + \sigma_l^2 \sim N_{Avogadro}$ holds now. The general relation $\sigma = \sigma(z)$ does not imply any constraint, since $F_H(\sigma(z))$ yields the kernel $K(\sigma(z), u - z)$.

### *1.7.3 Generalizations of the Gaussian convolution kernel*

Owing to the linearity of $F_H$, that is, the absence of terms of the form $(F_H\cdot\Phi)^2$, the solution function $\Phi_k = \exp(-i\,k\,z)/\sqrt{2\pi}$ is not the only possible one; the multiplication with an arbitrary function g(k) also solves Eq. (72):

$$\Phi_k = g(k)\,\boldsymbol{exp}(-ikz)/\sqrt{2\pi} \quad (78)$$

The power expansion (78) and the application of the spectral theorem (77) lead (for each power $k^n$) to the set:

$$f(k) = g^*(k)g(k) = a_0 + a_1 k + a_2 k^2 + \ldots + a_n k^n \quad (79)$$

$$J(n,\sigma,u-z) = \int k^n\, \boldsymbol{exp}(ik(u-z))\,\boldsymbol{exp}(-\sigma^2 k^2/4)\,dk/2\pi \quad (80)$$

We perform the substitutions:

$$\left.\begin{array}{l} k' = k - 2i(u-z)/\sigma^2 \\ \text{and} \\ \int k'^n\, \boldsymbol{exp}(-k'^2/a)\,dk' = \Gamma((n+1)/2)/(2\cdot a^{(n+1)/2}) \end{array}\right\} \quad (81)$$

With the help of Relation (81), we obtain the kernel expansion:

$$K(\sigma, u-z) = \sum_{n=0}^{N} a_n \sum_{j=0}^{n} \binom{n}{j} \Gamma((j+1)/2)(\tfrac{1+(-1)^j}{2})(\sigma^2/4)^{-(j+1)/2}(2i(u-z)/\sigma^2)^{n-j} \cdot$$

$$\cdot exp(-(u-z)^2/\sigma^2)/2\pi \qquad (82)$$

$$(lim\ N \to \infty)$$

Every finite sum running from 0 to N (N < ∞) is also a solution. In particular, N = 0 provides the familiar Gaussian kernel. A rather similar expression can be derived by a Boltzmann equation; the resulting kernel can be rewritten in terms of a Gaussian multiplied with Hermite polynomials, see Ulmer (1980):

$$\left.\begin{array}{l} K = \sum_{n=0}^{N} B_n(\sigma) \cdot (\frac{u-z}{\sigma})^n \cdot exp(-(u-z)^2/\sigma^2) \\ \quad \text{or} \\ K = \sum_{n=0}^{N} P_n(\sigma) \cdot H_n(\frac{u-z}{\sigma}) \cdot exp(-(u-z)^2/\sigma^2) \end{array}\right\} \quad (83)$$

$$(lim\ N \to \infty)$$

One should note that there is a rigorous correspondence between ordinary polynomials and Hermite polynomials. By that, we are able to express $B_l$ in terms of $P_0$, $P_1$, …, $P_l$, see Abramowitz and Stegun (1970) and Ulmer and Kaissl (2003). Although the basis of Eq. (83) is of quantum-mechanical nature, it can readily be interpreted as a general solution of the Boltzmann equation, where the two-point polynomials refer to the corresponding order of statistical moments (N may be kept finite). This provides a connection to Grad's solutions, which have been studied in radiation physics, see O'Brien (1979). An interesting special case of the expansion (83) is given by:

$$f(k) = c_g + c_e \cdot exp(\pm k/k_0) \quad (84)$$

The power series of $exp(\pm k/k_0)$ leads to the expansion (84) with fixed coefficients $a_k$; the resulting kernel K is obtained in a straightforward manner:

$$K = [c_g \cdot exp(-(u-z)^2/\sigma^2) + c_e \cdot exp(-(u-z \pm 1/k_0)^2/\sigma^2)] \frac{1}{\sigma \cdot \sqrt{\pi}} \quad (85)$$

The minus sign before the term $1/k_0$ in the previous equation implies a shift to a lower energy and reverse;

the normalization condition of the kernel K is $c_g + c_e = 1$ and other properties such as $\sigma_n^2 = n\cdot\sigma^2$ also hold, if the convolution is repeated n times. The Gaussian convolutions and their generalizations developed in this section represent Poisson distributions in the energy space, as may be verified easily by the operator $F_H\Phi = \exp(-H/E_{ex})\Phi = \exp(-E/E_{ex})\Phi$ or by the definition of the density matrix (73); only in the configuration space, do we obtain Gaussian distribution functions (and generalizations) $K(\sigma, u - z)$. A connection between $E_{ex}$ and $E_{max}$ may be established, but this is not quite satisfactory, since $E_{max}$ according to Eq.(67) and Fig. 11 are inconsistent with a nonrelativistic calculation and poor fitting methods should be avoided.

### *1.7.4 Generalization to a quantum statistical Klein-Gordon equation*

A relativistic generalization of the operator function $F_H$ is obtained by replacing the Hamiltonian (81, 82) by a stationary Klein-Gordon equation. We now write (restriction to z direction):

$$\left.\begin{aligned} F_{KG} &= \exp(-H^2/2E_{ex}^2) \\ H^2 &= -\hbar^2 c^2 d^2/dz^2 + m^2c^4 \\ \sigma^2 &= 2\hbar^2 c^2/E_{ex}^2 \end{aligned}\right\} \quad (86)$$

All previously elaborated principles can be used for the determination of the spectral distribution and the kernel K. With $F_{KG}\Phi_k = \gamma(k)\Phi_k$ and $\Phi_k = \exp(-ikz)/\sqrt{2\pi}$, we obtain the results:

$$\left.\begin{aligned} &\gamma(k) = \exp(-E(k)^2/2E_{ex}^2) \text{ and } E(k)^2 = \hbar^2c^2k^2 + m^2c^4 \\ &K = N_f\cdot\exp(-(u-z)^2/\sigma^2)\cdot\exp(-m^2c^4/E_{ex}^2) \text{ and } N_f = \tfrac{1}{\sigma\cdot\sqrt{\pi}}\cdot\exp(m^2c^4/E_{ex}^2) \end{aligned}\right\} (87)$$

With respect to Bohr's energy-straggling Formula (65) and the Landau – Vavilov theory, we now consider the modifications (86 - 87), obtained by the Klein-Gordon equation. For this purpose, we introduce the iterated operator equation:

$$\left.\begin{aligned} &[(1+\varepsilon\cdot p/mc)^n\cdot\exp(-n\cdot i\cdot p/\hbar k_0)\cdot\exp(n\cdot(p^2c^2+m^2c^4)/2E_{ex}^2)\Phi_k = \gamma(k)\Phi_k \\ &p \Rightarrow \tfrac{\hbar}{i}\cdot\tfrac{d}{dz} \text{ and } \varepsilon\text{: dimensionless factor} \end{aligned}\right\} (88)$$

The resulting spectral distribution $\gamma(k)$ and the convolution kernel K are given by:

$$\gamma(k)=\sum_{j=0}^{n}\binom{n}{j}\cdot(\varepsilon\cdot\hbar\cdot k/mc)^{j}\cdot exp[-n\cdot(\hbar^{2}\cdot c^{2}\cdot k^{2}+m^{2}c^{4})/2E_{ex}^{2}]\cdot exp(-n\cdot k/k_{0}) \quad (89)$$

$$\left.\begin{array}{l} K= N_f\cdot exp[-(u-z)^2/\sigma_n^2]\cdot exp(-n\cdot m^2c^4/E_{ex}^2)+ \\ +N_f\cdot exp[-(u-z-\frac{1}{2}\sigma_n^2/k_0)^2/\sigma_n^2]\cdot exp(-n\cdot m^2c^4/E_{ex}^2)\cdot[\sum_{j=1}^{n}P_j(\sigma_n,mc,\varepsilon)\cdot H_j(u-z-\frac{1}{2}\sigma_n^2/k_0)/\sigma_n)] \\ N_f=\frac{1}{\sigma_n\cdot\sqrt{\pi}}\cdot exp(n\cdot m^2c^4/E_{ex}^2)\text{ and }\sigma_n^2=n\cdot\sigma^2 \end{array}\right\}(90)$$

Eq. (90) can be brought into connection to Bohr's formula (65), if we neglect all powers of k and reduce the complete expression to exponential functions. We then obtain the special case $\gamma(k) = \gamma([E(k)/2E_{ex} - k/k_0]^2)$, since (by a suitable quadratic completion) the term $\exp(-nk/k_0)$ can be added to the preceding exponential function. This is not surprising, since it has been pointed out that Bohr's energy-straggling formula stands in close relation to the relativistic limit of the Bloch corrections (ICRU49). The Hermite polynomial expansion of the generalized kernel K is established by the same procedures as in the nonrelativistic case. In general, these contributions yield various complicated tails and asymmetries. Thus, for low proton energies with $E_{max} << mc^2$, all higher-order powers of $(\hbar\cdot k/mc)^j$ with j > 1 can be neglected and only the contribution $n\cdot\hbar\cdot\varepsilon\cdot k/mc$ remains. The inclusion of this term in the integration procedure finally leads to a Landau tail, since the history of energy transfers along the proton path with spectral distributions beyond the Gaussian maximum leads to shifts towards lower proton energies in the environment of $z = R_{CSDA}$. The implications of the Landau tail and energy shifts in the proton-fluence calculation, by taking account of the range straggling in the convolution procedure, will be established in a forthcoming section. Formulas (89 - 90) represent a quantum-statistical foundation of the Landau – Vavilov theory of fluctuations. For practical applications, where the initial proton energy $E_0$ does not exceed 270 MeV, a restriction to low-order corrections of a Gaussian kernel is certainly justified.

### *1.7.5 Dirac equation and Fermi-Dirac statistics*

In the following, we make use of properties, which an interested reader may find in a textbook, e.g., see Feynman (1962). The Dirac Hamiltonian $H_D$ reads as:

$$\left.\begin{array}{l} H_D=c\vec{\alpha}\vec{p}+\beta mc^2 \\ \vec{\alpha}=\begin{pmatrix}0&\vec{\sigma}\\ \vec{\sigma}&0\end{pmatrix}\ \beta=\begin{pmatrix}1&0\\0&-1\end{pmatrix} \\ H_D^2=c^2p^2+m^2c^4 \end{array}\right\}\quad(91)$$

The Dirac equation is obtained by the substitution $\vec{p} \Rightarrow \frac{\hbar}{i} \cdot \nabla$ :

$$(\beta mc^2 + \frac{\hbar c}{i} \cdot \vec{\alpha} \cdot \nabla)\psi = E_D \cdot \psi \quad (92)$$

According to Feynman (1962), $E_D$ satisfies the relation:

$$E_D = \pm mc^2 \sqrt{1 + 2 \cdot E_{Pauli} / mc^2} \quad (93)$$

$E_{Pauli}$ is the energy of the related Pauli equation, and, if we neglect spin effects, we can replace $E_{Pauli}$ by $E_{Schrödinger}$. Since the energy level, required for the creation of positrons, is distant to the available $E_{max}$ value, we also omit the minus sign in Eq. (93). With respect to the Fermi-Dirac statistics, we use the notation $\hat{H} = H_D - E_F$ ($E_F$: energy of the Fermi edge). The Fermi distribution function is:

$$f_F(\hat{H}) = f(\hat{H}) \cdot d_s(H_D) \quad (94)$$

The notation $d_s(H_D)$ refers to the density of states corresponding to the energy $E_D$ (or Hamiltonian $H_D$) and $E_F$ to the Fermi edge. By use of these definitions, the operator equation of Fermi-Dirac statistics, which will be used, assumes the shape:

$$f_F(\hat{H}) = \frac{1}{1+\exp((H_D - E_F)/E_{ex})} \cdot d_s(H_D) \quad (95)$$

Since the Relation (93) considerably simplifies a lot of calculations, we have to recall that, for the eigenvalues E of *continuous* operators H, the following property is valid:

$$\left.\begin{aligned} H\psi &= E\psi \\ f(H)\psi &= f(E)\psi \end{aligned}\right\} \quad (96)$$

The operator function f(H) may result from an iteration of H (Lie series); in Section 1.8 (energy straggling), we will intensively make use of this property. With the restriction to the z-axis and n-times repetition of the Fermi operator $f_F(\hat{H})$, we obtain:

$$\left.\begin{aligned} &\eta(k) = [mc^2\sqrt{1+\hbar^2\cdot k^2/m^2c^2} - E_F]/E_{ex} \\ &\gamma(\eta(k)) = [\tfrac{1}{2}(\eta\cdot E_{ex}+E_F)/mc^2]^n \cdot exp(-n\eta/2)\cdot sech(\eta/2)^n \end{aligned}\right\} \quad (97)$$

The general convolution kernel $K_F$ is given by:

$$K_F = \frac{1}{2\pi}\cdot\int_{-\infty}^{\infty} exp(ik(u-z))\cdot d\gamma(\eta(k)) \quad (98)$$

This integral cannot be evaluated by elementary methods, but it is possible to expand the hyperbolic-secant function $sech(\eta/2)$ in terms of a Gaussian multiplied with a specific power expansion. By that, we obtain a rigorously generalized version of the nonrelativistic/relativistic kernels, which are already investigated in sections 1.7.2 and 1.7.3. Thus, it is known from mathematical textbooks that the polynomial expansion of sech($\xi$) is given by:

$$sech(\xi) = \sum_{l=0}^{\infty} E_{2l}\cdot\xi^{2l}/(2l)! \qquad (|\xi| < \pi/2) \quad (99)$$

$E_{2n}$ are the Euler numbers; because of the very restricted convergence conditions, this expansion does not help. In view of the evaluation of the kernel $K_F$, we have derived a more promising expansion without any restriction for convergence:

$$\left.\begin{aligned} &sech(\xi) = exp(-\xi^2)\cdot\sum_{l=0}^{\infty}\alpha_{2l}\cdot\xi^{2l} \\ &\alpha_{2l} = E_{2l}/(2l)! + \sum_{l'=1}^{l}(-1)^{l'+1}\cdot\alpha_{2l-2l'}/l'! \end{aligned}\right\} \quad (100)$$

The coefficients $\alpha_{2l-2l'}$ are recursively defined. Some low-order coefficients are: $\alpha_0 = 1$, $\alpha_2 = 1/2$, $\alpha_4 = 5/4!$, $\alpha_6 = 29/6!$, $\alpha_8 = 489/8!$. It is easy to verify that, due to the Gaussian, the expansion above converges for $|\xi| \leq \infty$. Since $\eta(k)$ is given by a root, some terms are connected with difficulties, but if we expand $\eta(k)$ by the power expansion (3), then – besides the rest energy $mc^2$ – the first expansion term is the nonrelativistic contribution; thus, we may consider only Fermi statistics without relativistic terms (Dirac). The higher-order terms are now readily evaluated (up to arbitrary order). The energy distribution function $S_E$ assumes the shape:

$$S_E = N_f \cdot \exp(-(E_n(k) - E_{Average,n})^2 / 2\sigma_E(n)^2) \cdot \sum_{l=0}^{\infty} b_l(n, mc^2) \cdot (E_n(k)/2E_{ex})^l \quad (101)$$

$N_f$ is a normalization factor. Eq. (101) provides a very interesting special case: if we take l = 0, we obtain Bohr's classical formula (65) and a connection between $E_{Average}$ and $E_F$. The convolution kernel $K_F$ is determined by the general structure:

$$K_F = N_f \cdot \sum_{l=0}^{\infty} H_l((u - z - z_{shift}(l))/\sigma_n) \cdot B_l(n, mc^2) \cdot \exp(-(u - z - z_{shift}(l))^2 / 2\sigma_n^2) \quad (102)$$

Now $E_{Average}$ and $z_{shift}$ depend in any order on the Fermi edge energy. Detailed descriptions of the above terms are avoided in this paper; Formulas (101 – 102) represent only a general outline. If we recall that $\sigma_n$ is a function of z (that is, σ(z)), and take the proportionality $E_F \sim E_{Average}$ into account, we can verify that the Landau – Vavilov theory represents a specific form of a Fermi-Dirac distribution. It should also be emphasized that $E_F$ and consequently $E_{Average}$ depend on the proton track, but in a low order (that is, the proton energy is $E_0 << Mc^2$ and the kinetic energy $E_\delta$ of δ-electrons satisfies $E_\delta << 2mc^2$), the Bohr approximation formula is a good basis. One might be surprised that, in practical calculations, we can use $\sigma(R_{CSDA})$ instead of the continuously increasing σ(z). For homogeneous media, this simplification works, since the residual energy E(z) of a proton is monotonically decreasing. (For heterogeneous media, we have always to take slabs with different densities and material properties like $(Z/A_N)_{medium}$ into account.) The question arises as to when the kernel $K_F$ can be used. The first case pertains to the energy/range straggling of fast electrons as primary projectiles; to avoid unnecessary evaluations of higher order, it is necessary to check the order of $K_F$ by comparison to the measured stopping power. The second case is the calculation of transition probabilities and of the energy transfer of protons passing through nuclei; the Pauli principle for spin and isospin gives raise to exchange interactions.

## 1.8 Calculation methods for pristine Bragg curves by inclusion of energy/range straggling and Landau tails for the models M1, M2, M3

### *1.8.1 Determination of the required calculation parameters of the three models*

This section deals with the parameters required for the description of fluctuations, the energy shifts due to the nuclear interactions, and the contributions of heavy recoils to the total stopping power.

#### *1.8.1.1 Energy-range straggling and rms parameters required in convolutions*

It is obvious that, even for initially monoenergetic proton beamlets, there are fluctuations in the energy transfer from the protons to the environmental electrons. In first order, we assume that these fluctuations are symmetrical (i.e., Gaussian distributed). In particular, Fig. 5 ignores these fluctuations. It is also commonly assumed that the incident-proton spectra are also of Gaussian type. We adopt here these first-order assumptions and add, as a second-order correction, the Landau tails for the description of the complete beamline. However, these corrections are necessary, if the initial proton energy $E_0$ exceeds 100 MeV; the importance is increasing with $E_0$. We denote the rms of monoenergetic proton beamlets by $\tau_{straggle}$ and the rms of the impinging proton beam by $\tau_{in}$. The rms of the polychromatic energy spectrum $\tau$ is given by:

$$\tau^2 = \tau_{\text{straggle}}^{\;2} + \tau_{\text{in}}^{\;2} \qquad (103)$$

It is a favorite property of Gaussian convolutions that the two successive convolutions lead to the result of one convolution, performed according to Eq. (103). With regard to primary protons, we will give the results of all required convolutions in terms of $\tau$; by setting $\tau_{in} = 0$, we obtain the pristine Bragg curves of initially monoenergetic protons.

It is clear that $\tau_{straggle}(z)$ has to be a monotonically increasing function of z. However, it is usual in various calculation models to restrict $\tau_{straggle}(z)$ to $\tau_{straggle}(R_{CSDA})$. In almost all applications in therapy planning, this restriction is justified, since S(z) in the CSDA framework is a monotonically-increasing function, too, and $\tau_{straggle}(z)$ does not have an influence in the initial plateau of a pristine Bragg curve. We now explain the calculation of $\tau_{straggle}(R_{CSDA})$ according to the generalization of the Bortfeld model (model M1). If the relativistic correction terms are omitted, and p and A are appropriately determined – Bortfeld (1997), then we are in agreement with the Bortfeld model. With respect to this calculation, we are closely related to the Bortfeld model, which makes use of the Bohr approximation, i.e., $E_{Average} = 0$, if E < 10 MeV:

$$\tau_{\text{straggle}}^{\;2} = \int_0^{R_{\text{CSDA}}} (d\sigma_E^{\;2} / dz)[dE/dz]^{-2} dz \qquad (104)$$

Using the relativistic formula for E(z) and dE(z)/dz, i.e., Relations (21 – 22), and the evaluation of $d\sigma_E^2/dz$ corresponding to Bortfeld, we obtain:

$$\left.\begin{aligned} \tau_{\text{straggle}}^{\;2} &= 2 \cdot C_{\text{Bohr}} \cdot \int_0^{R_{\text{CSDA}}} [p^2 A^{1/p} \cdot (R_{\text{CSDA}} - z)^{2-2/p} + (2p^2 A^{1/p} / Mc^2) \cdot (R_{\text{CSDA}} - z)^{2-1/p}] dz \\ &= 2 \cdot C_{\text{Bohr}} \cdot [p^2 A^{2/p} \cdot (3 - 2/p)^{-1} R_{\text{CSDA}}^{\;3-2/p} + (2p^2 A^{1/p} / Mc^2) \cdot (3 - 1/p)^{-1} \cdot R_{\text{CSDA}}^{\;3-1/p}] \end{aligned}\right\} \quad (105)$$

$C_{Bohr}$ amounts to 0.087 $MeV^2/cm$. The factor 2 goes back to the definition of a Gaussian by $\exp(-z^2/\sigma^2)$ instead of $\exp(-z^2/2\sigma^2)$. If we use A = 0.0022 and p = 1.77 – Bortfeld (1997), we obtain:

$$\left.\begin{aligned}\tau_{straggle}(R_{CSDA}) &= \sqrt{2}\cdot 0.012703276\cdot R_{CSDA}{}^{0.9358} \quad (if \quad R_{CSDA} \geq 1 \ cm)\\ \tau_{straggle}(R_{CSDA}) &= \sqrt{2}\cdot 0.012703276\cdot R_{CSDA}{}^{1.763} \quad (if \quad R_{CSDA} < 1 \ cm)\end{aligned}\right\} \quad (106)$$

If the integrations according to Eq. (104) are only carried out up to a final value E(z) > 0 and with regard to Eq. (40), the two most important contributions are accounted for; we are then able to obtain a fitting formula for $\tau_{straggle}(z)$ and $Q_z = 2.887$ cm$^{-1}$:

$$\tau_{straggle}(z) = \tau_{straggle}(R_{CSDA}) \cdot \frac{e^{Q_z \cdot z} - 1}{e^{Q_z \cdot R_{CSDA}} - 1} \qquad (107)$$

Fig. 6 shows that we cannot deal with constant values of p; a good approximation in the relativistic case is:

$$\left.\begin{aligned} p &= 1.71 \quad (\text{if } E_0 \geq E_{50} = 50 \text{ MeV}) \\ p &= 1.66 + 0.05 \cdot E_0 / E_{50} \quad (\text{if } E_0 \leq E_{50}) \end{aligned}\right\} \quad (108)$$

These two restrictions can be used for the evaluation of the result given by Eq. (104). The applicability of the connection (107) is independent of the way in which $\tau_{straggle}(R_{CSDA})$ is calculated.

In Ulmer (2007), we have presented an alternative method. We decomposed the total energy straggling into a Gaussian part and a Landau tail. Since the resulting Landau tail represents a comparably small correction, we assume that $\tau_{straggle}$ for the Landau tail is the same as for the Gaussian part. On the other hand, we do not permit any approximation with respect to the evaluation of $[dE(z)/dz - E_{average}]^{-2}$ according to Formulas (40 – 41). The resulting formula is a power expansion in terms of exponential functions:

$$\left.\begin{aligned} \tau_{straggle}(z) = {} & R_{CSDA} \cdot [c_{m,1} \cdot (1 - \exp(-g_1 \cdot R_{CSDA})) + \\ & + c_{m,2} \cdot (1 - \exp(-g_2 \cdot R_{CSDA}))] \\ & - (R_{CSDA} - z) \cdot [c_{m,1} \cdot (1 - \exp(-g_1 \cdot (R_{CSDA} - z))) + \\ & + c_{m,2} \cdot (1 - \exp(-g_2 \cdot (R_{CSDA} - z)))] + 0(\text{higher order}) \end{aligned}\right\} \quad (109)$$

The higher-order terms are smaller than 0.001 and hence negligible. In view of a possible step-by-step calculation, we write $\tau_{straggle}(z)$, not simply $\tau_{straggle}(R_{CSDA})$. The dimensionless coefficients $c_{m,1}$ and $c_{m,2}$ are given by: $c_{m,1} = 0.4968$ and $c_{m,2} = 0.1662$. For the factors $g_1$ and $g_2$: $g_1 = 176.752$ cm$^{-1}$ and $g_2 = 112.384$ cm$^{-1}$. We emphasize that these values are valid for water. For other media, the coefficients $c_{m,1}$ and $c_{m,2}$ have to be

rescaled:

$$\left.\begin{aligned} c_{m,1}(\text{medium}) &= c_{m,1} \cdot \sqrt{(Z \cdot \rho / A_N)_{\text{medium}} / (Z \cdot \rho / A_N)_{\text{water}}} \\ c_{m,2}(\text{medium}) &= c_{m,2} \cdot \sqrt{(Z \cdot \rho / A_N)_{\text{medium}} / (Z \cdot \rho / A_N)_{\text{water}}} \end{aligned}\right\} \quad (110)$$

Based on Formula (22) with a rigorous account for p = p($E_0$) according to Formula (49) and without the Bohr approximation, we have also performed a numerical calculation of $\tau_{\text{straggle}}(R_{\text{CSDA}})$. The results (Fig. 14) are rather close to those of Formula (106). However, it appears that small errors, obtained (for instance) by the Bortfeld approximation, have no significant importance.

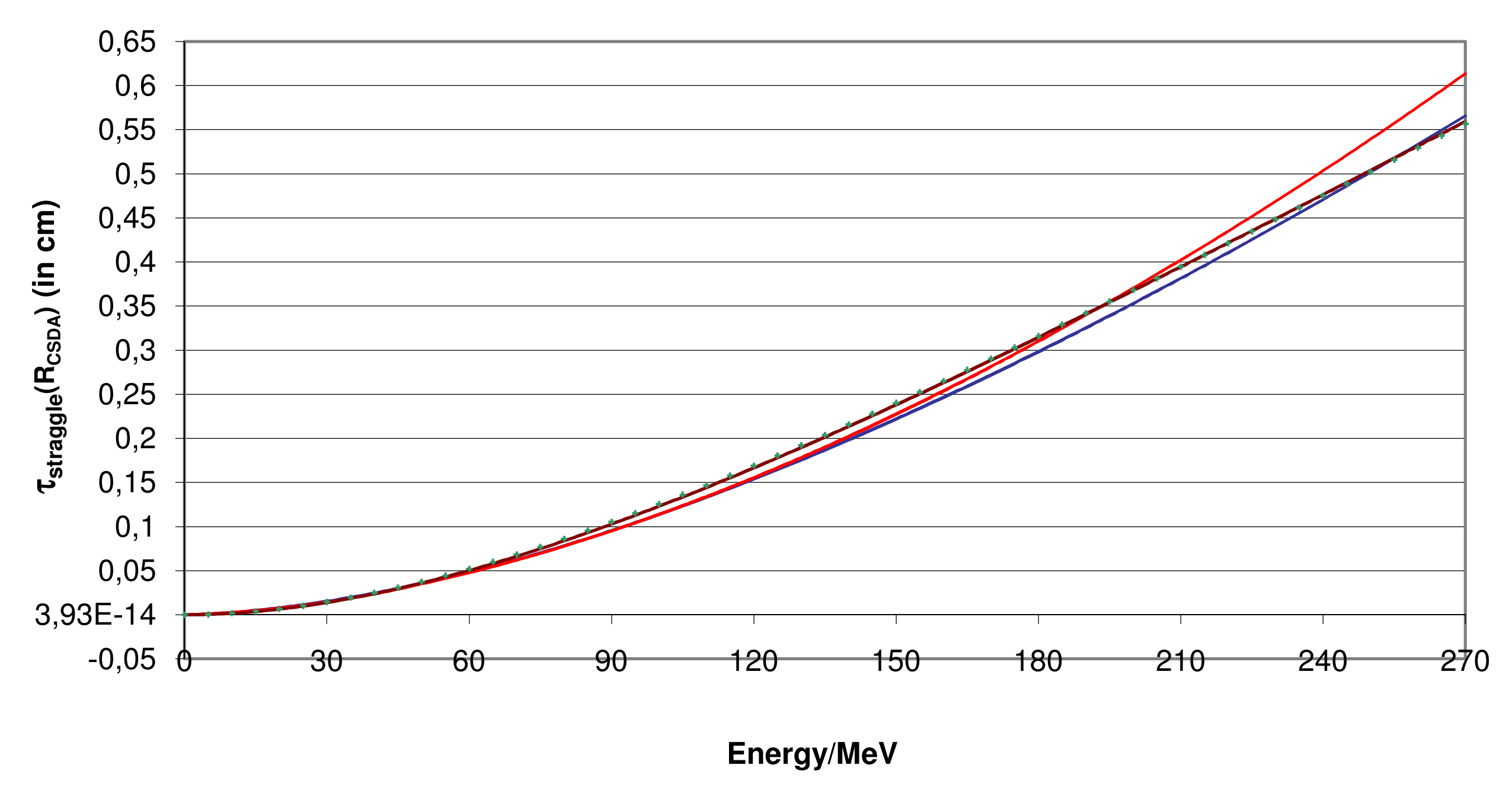

**Fig. 14:** The determination of $\tau_{\text{straggle}}$ according to Bortfeld (1997): solid blue; according to the relativistic Formula (22) and Restriction (109): solid red; according to the exact integration of the BBE (Eq. (40)): solid brown; calculation with the relativistic formula (22) without restriction (109): green crosses.

It has to be noted that that $\tau_{\text{in}}$ represents an additional fitting parameter with importance for all three models. In a subsequent section, we will investigate the dependence of this additional parameter on the machine type.

*1.8.1.2 Determination of the parameters for secondary/recoil protons and heavy recoils*

In the case of secondary (nonreaction) protons, we have to replace Formula (103) by:

$$\tau = \sqrt{\tau_{straggle}^{\;2} + \tau_{in}^{\;2} + 0.5 \cdot \tau_{heavy}^{\;2}} \qquad (111)$$

The rms value $\tau_{heavy}$ is obtained via $Q^{tot}$ as previously defined:

$$\tau_{heavy} = \begin{cases} 0, & \text{if } \; E_0 < E_{Th} \\ 0.55411 \cdot \dfrac{E_0 - E_{Th}}{E_{res} - E_{Th}}, & \text{if } \; E_{\mathrm{Th}} < E_0 \le E_{res} \\ 0.554111 - 0.000585437 \cdot (E_0 - E_{res}), & \text{if } \; \mathrm{E}_{\mathrm{res}} < E_0 \end{cases} \qquad (112)$$

The shift of the z-coordinate by $z_{shift}$, due to the reduced energy of secondary protons, is given by:

$$z_{shift} = \begin{cases} 0 & (if \; E_0 < E_{res}) \\ \displaystyle\sum_{n=1}^{4} a_n E_{res}^{\;n} \cdot [1 - \exp\left( - \frac{(E_0 - E_{res})^2}{5.2^2 \cdot E_{res}^{\;2}} \right)] & (if \; E_0 \ge E_{res}) \end{cases} \qquad (113)$$

If $E_0 \le E_{res}$, we have to put $z_{shift} = 0$, but, in practice, this case does not occur. The application of Eq. (113) implies either Eq. (35) or Eq. (36) and the substitution $E_0 \rightarrow E_{res}$. Finally, we have to give the parameters for the determination of the stopping-power contribution of heavy recoils:

$$C_{heavy} = \begin{cases} 0 & (\text{if } \; E_0 < E_{Th}) \\ 0.00000042643926 \cdot (E_0 - E_{Th})^2 & (\text{if } \; E_0 \ge E_{Th}) \end{cases} \qquad (114)$$

$$S_{heavy}(z, E_0) = \Phi_0 \, C_{heavy} \cdot e^{-\frac{z}{z_{max}}} \cdot \left[ 1 + erf\left( \sqrt{2} \cdot \frac{R_{csda} - z + \frac{E_0}{250} - 1}{\tau_{heavy}} \right) \right] \qquad (115)$$

We emphasize that, with regard to fits necessary to complete Relations (114 – 115), we have partially used results obtained by Monte-Carlo calculations with GEANT4.

*1.8.2 Calculation of the total stopping power of the models M1, M2, M3*

The total stopping power $S_{tot}$ is defined by:

$$\left.\begin{aligned} S_{tot} &= S_{pp} + S_{sp} + S_{rp} + S_{heavy} \\ S_{sp} &= S_{sp,n} + S_{sp,r} \end{aligned}\right\} \quad (116)$$

$S_{pp}$ refers to primary protons, $S_{sp}$ to secondary protons (i.e., $S_{sp,r}$ to reaction and $S_{sp,n}$ to nonreaction protons), $S_{rp}$ to recoil protons, and $S_{heavy}$ to heavy recoil particles given in Eq. (115). Remember that, for $S_{pp}/S_{rp}$, $\tau$ is defined by Eq. (103) and, for $S_{sp,n}$, by Eq. (111). The contributions of $S_{sp,r}$ are defined in the Appendix. Due to the presence of the beamline elements (degrader, modulator wheel, range shifter, etc.), the assumption $\tau_{in} = 0$ is rather meaningless. In this section, we make use of the following abbreviations:

$$\left.\begin{aligned} & z_s = z + z_{shift} \\ & F = R_{CSDA} \cdot [\tfrac{E_0 - E_{Th}}{M \cdot c^2}]^f ; \; f = 1.032 \\ & F_{\upsilon s} = F \cdot \upsilon; \; \upsilon = 0.958 ; \; \upsilon' = \upsilon - 2 \cdot C_{heavy} \\ & F_{\upsilon} = F \cdot (1-\upsilon); F_{\upsilon'} = F \cdot \upsilon' \end{aligned}\right\} \quad (117)$$

The convolution problem of every contribution of primary, secondary, and recoil protons has the general structure:

$$S(z, \tau) = \int_{-\infty}^{R_{CSDA}} S(u) \cdot \frac{1}{\tau \sqrt{\pi}} \exp(-(u - z)^2 / \tau^2) \, du \qquad (118)$$

S(u) solely refers to results obtained by CSDA approach.

*1.8.2.1 Calculation of $S_{pp}$, $S_{sp,n}$ and $S_{rp}$ for model M1*

The basic equation of this subsection is Eq. (22). Using the power expansion (3), Formula (22) can be written

as:

$$dE/dz = -p^{-1} \sum_{n=0}^{\infty} (-1)^n (1\cdot3\cdot5\cdot\cdot(2n-1)) \tfrac{1}{n!} \cdot (R_{CSDA} - z)^{(1+n)/p-1} / [A^{(1+n)/p} \cdot (Mc^2)^n] \Big\} \quad \textit{(119)}$$

Only the nonrelativistic limit case n = 0 produces the difficulty of a singularity at z = $R_{CSDA}$. With respect to a Gaussian convolution, we deal here with the general case, with arbitrary n. The application of Formula (118) to Formula (119) yields the solution function $S_h$, which appears in all three cases:

$$\left.\begin{aligned} & S_h(z,\tau) = \frac{1}{p\sqrt{\pi\tau}} \sum_{n=0}^{N} (-1)^n (1\cdot3\cdot5\cdot(2n-1))(Mc^2)^{-n} A^{-\nu}\Gamma(\nu)(\tau/\sqrt{2})^{\nu}\cdot \\ & \qquad \cdot\exp(-\xi^2/2)\cdot D_{-\nu}(\sqrt{2}\cdot\xi)\cdot\frac{1}{n!} \\ & \xi = (z - R_{CSDA})/\tau; \quad \nu = (n+1)/p \\ & N \to \infty \end{aligned}\right\} \quad \textbf{(120)}$$

The functions $D_{-v}$ and $\Gamma_v$ are the parabolic cylinder function and the Γ function, respectively. Details of these two functions can be found in Abramowitz and Stegun (1970).

*1.8.2.1.1 Primary protons*

With the help of $S_h$, as defined in Eq. (120), we obtain for primary protons:

$$S_{pp}(z,\tau) = (1 - z\cdot F)\cdot S_h(\xi,\tau) \quad (121)$$

*1.8.2.1.2 Secondary nonreaction protons*

$$S_{sp,n} = z\cdot F_{v'}\cdot S_h(\xi,\tau) \quad (122)$$

$$\xi = (z_s - R_{CSDA})/\tau \qquad (122a)$$

*1.8.2.1.3 Recoil protons*

The substitution (122a) is still valid.

$$S_{rp} = z \cdot F_{\upsilon} \cdot S_{h}(\xi, \tau) \quad (123)$$

General remark: In practical applications, a restriction of $S_h$ to the order N = 4 is sufficient; the infinite order is an unwanted complication for therapeutic protons. The evaluation of parabolic cylinder functions is a rather slow process. Thus, the calculation model M1 is slow even in the nonrelativistic limit (N = 0).

*1.8.2.2 Calculation of $S_{pp}$, $S_{sp,n}$ and $S_{rp}$ for model M2*

*1.8.2.2.1 Primary protons $S_{pp}$*

In this subsection, we use $S_{pp} = S_{pp,1} + S_{pp,2}$ and introduce the following abbreviation:

$$\mathrm{Erf}\lambda_k = \tfrac{1}{2} \cdot [1 + \mathrm{erf}((R_{CSDA} - z - 0.5 \cdot \lambda_k \cdot \tau^2)/\tau)] \cdot \exp(-\lambda_k \cdot (R_{CSDA} - z)) \cdot \exp(0.25 \cdot \tau^2 \cdot \lambda_k^2)$$

$S_{pp,1}$:

$$\left.\begin{aligned} S_{pp,1}(z,\tau) = &-\sum_{k=1}^{5} c_k \cdot [1 + \gamma_k \cdot (R_{CSDA} - z) - 0.5 \cdot \lambda_k^2 \cdot \tau^2] \cdot \mathrm{Erf}\lambda_k \\ &-\sum_{k=1}^{5} c_k \cdot \lambda_k \cdot \exp(-\lambda_k \cdot (R_{CSDA} - z)) \cdot \exp(0.25 \cdot \tau^2 \cdot \lambda_k^2) \cdot \exp[-((R_{CSDA} - z - 0.5 \cdot \lambda_k \cdot \tau^2)/\tau)^2] \end{aligned}\right\} (124)$$

$S_{pp,2}$:

$$\left.\begin{aligned} S_{pp,2} = &\mathrm{F} \cdot z \cdot [\sum_{k=1}^{5} c_k \cdot [1 + \lambda_k \cdot (R_{CSDA} - z) - 0.5 \cdot \lambda_k^2 \cdot \tau^2] \cdot \mathrm{Erf}\lambda_k] \\ &+ \mathrm{F} \cdot z \cdot \{\sum_{k=1}^{5} c_k \cdot \lambda_k \cdot \exp(-\lambda_k \cdot (R_{CSDA} - z)) \cdot \exp(0.25 \cdot \tau^2 \cdot \lambda_k^2) \cdot \exp[-((R_{CSDA} - z - 0.5 \cdot \lambda_k \cdot \tau^2)/\tau)^2]\} \end{aligned}\right\} (125)$$

*1.8.2.2.2 Secondary nonreaction protons $S_{sp,n}$*

Abbreviation:

$$\mathrm{Erf}\lambda_{ksp} = \tfrac{1}{2} \cdot [1 + \mathrm{erf}((R_{CSDA} - z_s - 0.5 \cdot \lambda_k \cdot \tau^2)/\tau)] \cdot \exp(-\lambda_k \cdot (R_{CSDA} - z_s)) \cdot \exp(0.25 \cdot \tau^2 \cdot \lambda_k^2)$$

By that, we obtain:

$$\left.\begin{aligned} S_{sp,n} = & -F_{v'} \cdot z \cdot \sum_{k=1}^{5} c_k \cdot [1 + \lambda_k \cdot (R_{CSDA} - z_s) - 0.5 \cdot \lambda_k^{\ 2} \cdot \tau^2] \cdot Erf\ \lambda_{ksp} \\ & - F_{v'} \cdot z \cdot \{ \sum_{k=1}^{5} c_k \cdot \lambda_k \cdot \exp(-\lambda_k \cdot (R_{CSDA} - z_s)) \cdot \exp(0.25 \cdot \tau^2 \cdot \lambda_k^{\ 2}) \cdot \\ & \quad \cdot \exp[-((R_{CSDA} - z_s - 0.5 \cdot \lambda_k \cdot \tau^2)/\tau)^2]\} \end{aligned}\right\} (126)$$

*1.8.2.2.3 Recoil protons $S_{rp}$*

Abbreviation:

$$\mathrm{Erf}\lambda_{krp} = \tfrac{1}{2} \cdot [1 + \mathrm{erf}((R_{CSDA} - z_s - 0.5 \cdot \lambda_k \cdot \tau^2)/\tau)] \cdot \exp(-\lambda_k \cdot (R_{CSDA} - z_s)) \cdot \exp(0.25 \cdot \tau^2 \cdot \lambda_k^{\ 2})$$

By that, we obtain:

$$\left.\begin{aligned} S_{rp} = & -F_v \cdot z \cdot \sum_{k=1}^{5} c_k \cdot [1 + \lambda_k \cdot (R_{CSDA} - z_s) - 0.5 \cdot \lambda_k^{\ 2} \cdot \tau^2] \cdot Erf\lambda_{krp} \\ & - F_v \cdot z \cdot \{\sum_{k=1}^{5} c_k \cdot \lambda_k \cdot \exp(-\lambda_k \cdot (R_{CSDA} - z_s)) \cdot \exp(0.25 \cdot \tau^2 \cdot \lambda_k^{\ 2}) \cdot \exp[-((R_{CSDA} - z_s - 0.5 \cdot \lambda_k \cdot \tau^2)/\tau)^2]\} \end{aligned}\right\} (127)$$

*1.8.2.3 Calculation of $S_{pp}$, $S_{sp,n}$ and $S_{rp}$ for model M3*

It might appear that, compared to model 2, this calculation works with various different expressions. However, this is not true, since these expressions result from different types of functions as previously introduced by Eq. (43). If these expressions are subjected to a Gaussian convolution, according to Eq. (43), we then gain an additional speed-up factor of the order of 2 – 3; the function types, introduced in Eq. (43), are very easy to handle in this type of convolution.

*1.8.2.3.1 Primary protons $S_{pp}$*

$S_{pp}$ is now defined by:

$$S_{pp}(z, E_0) = \Phi_0 \cdot (1 - F \cdot z) \cdot [I_1(z, E_0) + I_2(z, E_0) + I_3(z, E_0) + I_4(z, E_0)] \quad (128).$$

The contributions $I_1$, …, $I_4$ are given by:

$$I_1 = \left( C_1 \cdot \frac{\tau_{straggle} \cdot R_{csda}}{\tau} - C_4 \cdot \frac{\tau \cdot (R_{csda} + z)}{\sqrt{\pi} R_{csda}{}^2} \right) \cdot e^{\left( -\frac{(R_{csda} - z)^2}{\tau^2} \right)} \quad (129a)$$

$$I_2 = \left( C_2 + C_4 \cdot \frac{\tau^2}{2\sqrt{\pi} R_{csda}{}^2} \right) \cdot \left( 1 + erf\left( \frac{R_{csda} - z}{\tau} \right) \right) \quad (129b)$$

$$I_3 = C_3 \cdot \exp\left( \frac{(P_E \cdot \pi \cdot z_{\max} \cdot \tau_{in})^2}{4} - \frac{P_E \cdot \pi \cdot (R_{csda} - z)}{z_{\max}} \right) \cdot \left[ 1 + erf\left( \frac{R_{csda} - z}{\tau} - \frac{P_E \cdot \pi \cdot \tau}{2 \cdot z_{\max}} \right) \right] \quad (129c)$$

$$I_4 = C_4 \cdot \frac{z^2}{R_{csda}{}^2} \cdot \left[ 1 + erf\left( \frac{R_{csda} - z}{\tau} \right) \right] \quad (129d)$$

*1.8.2.2.2 Secondary protons $S_{sp,n}$ and recoil protons $S_{rp}$*

$$S_{sp,n}(z, E_0) = \Phi_0 \cdot (F_{v'} \cdot z) \cdot [I_1(z_s, E_0) + I_2(z_s, E_0) + I_3(z_s, E_0) + I_4(z_s, E_0)] \quad (130)$$

$$S_{rp}(z, E_0) = \Phi_0 \cdot (F_v \cdot z) \cdot [I_1(z_s, E_0) + I_2(z_s, E_0) + I_3(z_s, E_0) + I_4(z_s, E_0)] \quad (131)$$

With regard to the contributions of $I_1$, …, $I_4$, we can use here the Expressions (129a – 129d); in the formulas above, we must only substitute z by $z_s$. The distinctions with respect to τ (i.e., Eq. (111) applies to $S_{sp,n}$, Eq. (103) to $S_{rp}$) are still applicable.

*1.8.3 Inclusion of the Landau-tail corrections and their role in the pristine Bragg curves*

According to Eqs. (101 – 102) it is difficult to decide which order of corrections is sufficient, we will provide, in this section, a summary of the theoretical results with GEANT4. It is an interesting feature that the lowest order (i.e., $S_E(l = 0)$) of Eq. (101) yields Bohr's classical formula of fluctuations and energy straggling. Formula (102) represents the translation in the configuration space. In the following, we keep the terms up to order 2 in Eq. (102), and compare the results with the Monte-Carlo calculations based on GEANT4.

We have not yet accounted for the contributions of the Landau tails in all three models. These tails result from the modification of the energy transfer and the stopping power of protons. The theoretical analysis of the preceding section (previously developed in Ulmer (2007)) concludes that symmetrical – i.e., Gaussian – fluctuations (and the related convolution kernel) of the energy transfer according to the CSDA are only rigorously valid, if the local proton energy and the energy transfer by collisions are nonrelativistic. The maximum energy transfer $E_{max}$ has been shown in Fig. 11 as a function of the local energy for protons in water. $E_{max}$ has a nonlinear term, which becomes more important with increasing energy. In a similar way, the fluctuations of the energy transfer become less symmetrical; collisions occur significantly less frequently. This behavior can be observed more and more for proton energies $E_0$ >> 100 MeV. A consequence of this relativistic effect is that protons in the entrance region (e.g., 250 MeV, $E_{max}$ = 617 keV) undergo fewer collisions with environmental electrons than it would be expected from a symmetrical energy transfer. Thus, less energy is locally stored and a contribution to the buildup effect can be seen as long as the symmetrical fluctuation is not yet reached. However, this effect decreases along the proton track, and when the local energy approaches about 100 MeV, the fluctuations of the energy transfer tend to become symmetrical, i.e., the buildup effect is reduced (and vanishes for $E_0$ << 100 MeV, since Landau tails and reaction protons become negligible). The preceding relativistic treatment of convolutions provides that the inclusion of the Landau tails (Vavilov distribution function) necessitates generalized Gaussian convolutions, i.e., Gaussian convolutions with relativistic correction terms expressed by two-point Hermite polynomials. However, for proton energies below about 300 MeV, the lower-order corrections are sufficient. The results of the generalized Gaussian convolution are the following contributions:

*(a) Primary protons:*

$$S_{Lan,pp} = \Phi_0 \cdot (I_{Lan1}(z) + I_{Lan2}(z)) \cdot (1 - \tfrac{z}{R_{CSDA}}) \quad (132)$$

*(b) Secondary protons:*

$$S_{Lan,sp} = \Phi_0 \cdot (I_{Lan1}(z_s) + I_{Lan2}(z_s)) \cdot \tfrac{z}{R_{CSDA}} \cdot \upsilon' \quad (133)$$

*(c) Recoil protons:*

$$S_{Lan,rp} = \Phi_0 \cdot (I_{Lan1}(z_s) + I_{Lan2}(z_s)) \cdot (1 - \upsilon) \cdot \tfrac{z}{R_{CSDA}} \quad (134)$$

Concerning the calculation formulas for secondary/recoil protons, the substitutions z → $z_s$ has to be performed.

$$I_{Lan1} = C_{Lan1} \cdot \left\{ erf\left(\frac{2 \cdot z}{R_{Lan1}}\right) + erf\left(\frac{2 \cdot (R_{Lan1} - z)}{\tau_{Lan1}}\right) \cdot z / R_{CSDA} \right\} \cdot \left(1 + erf\left(\frac{(R_{CSDA} - z)}{\tau}\right)\right) \quad (135)$$

$$\left.\begin{aligned} C_{Lan1} &= -1.427 \cdot 10^{-6} \cdot R_{csda}^{3} + 1.439 \cdot 10^{-4} \cdot R_{csda}^{2} - 0.002435 \cdot R_{csda} + 0.2545 \\ R_{Lan1} &= R_{csda} \cdot \begin{cases} 0.7 & (\text{if } E_0 < 68\,\text{MeV}) \\ (0.812087912 - 0.001648352 \cdot E_0) & (\text{if } E_0 \geq 68\,\text{MeV}) \end{cases} \\ \tau_{Lan1} &= R_{Lan1} + 0.0492 \cdot \tau_{in} \end{aligned}\right\} (136)$$

$$I_{Lan2} = -\frac{C_{Lan2}}{R_{Lan2}^{2}} \left\{ \left( R_{Lan2}^{2} - z^2 - \frac{\tau_{Lan2}^{2}}{2\sqrt{\pi}} \right) \cdot \left( 1 + \text{erf}\left( \frac{R_{Lan2} - z}{\tau_{Lan2}} \right) \right) + \frac{(z + R_{Lan2}) \cdot \tau_{Lan2}}{\sqrt{\pi}} \cdot e^{\frac{-(R_{Lan2} - z)^2}{\tau_{Lan2}^{2}}} \right\} \quad \textbf{\textit{(137)}}$$

$$C_{Lan2} = \begin{cases} 0.000021791 \cdot E_0, & \text{if } E_0 < 120\,\text{MeV} \\ 0.022 \cdot \left( \frac{E_0 - 118}{168 - 118} \right)^{0.877}, & \text{if } 120\,\text{MeV} \leq E_0 < 168\,\text{MeV} \\ 0.022, & \text{if } 168\,\text{MeV} \leq E_0 \end{cases} \quad (138)$$

$$\left.\begin{aligned} R_{Lan2} &= (3.19 + 0.00161 \cdot E_0) \cdot \left[ 1 - \exp\left( -\frac{E_0^{2}}{165.795268^2} \right) \right] \\ \tau_{Lan2} &= 2.4 \cdot \sqrt{\tau_{straggle}(z)^2 + \tau_{in}^{2}} \end{aligned}\right\} (139)$$

For proton energies larger than about 120 MeV, the correction term $I_{Lan2}$ is of increasing importance (with the energy). Even below 120 MeV, the contribution $I_{Lan1}$ remains noteworthy. The energy spectrum of the protons has a tail due to the beamline elements (range-modulator wheel, etc.). It is certainly not sufficient to take account of all these influences on the basis of a half-width parameter $\tau_{in}$ in a Gaussian convolution. Since the influence of the beamline depends on specific properties of the proton accelerator, an adaptation of the parameters, appearing in $I_{Lan1}$, by a fitting procedure, in addition to the fitting of $\tau_{in}$, is required. As a result, $C_{Lan1}$ (and to a lesser extent $R_{Lan1}$) only represent initial values, which may slightly be modified by fitting procedures, to accommodate machine-specific properties (see Section 2.2).

The correction terms, which have been introduced to account for the Landau tails, have also been subjected to comparisons with the results of Monte-Carlo calculations using the GEANT4 code in the case of monoenergetic beams; in these calculations, the energy was varied between 2 and 250 MeV, with a step of 2 MeV. The result of these comparisons was that the differences never exceeded 2.2 %, the mean standard deviation amounting to 1.3 %. Concerning the polychromaticity of the proton beam, induced by the various beamline elements, a direct comparison with experimental data was done. It should be noted that the already-described contribution to the buildup effect, induced by the Landau tails, could also be verified by the Monte-Carlo calculations, when the Landau tails were taken into account; additionally; the contribution disappeared when the statistical fluctuations were restricted to a Gaussian kernel. The role of reaction protons and Landau tails to buildup is clearly shown in Figs. 15 - 16.

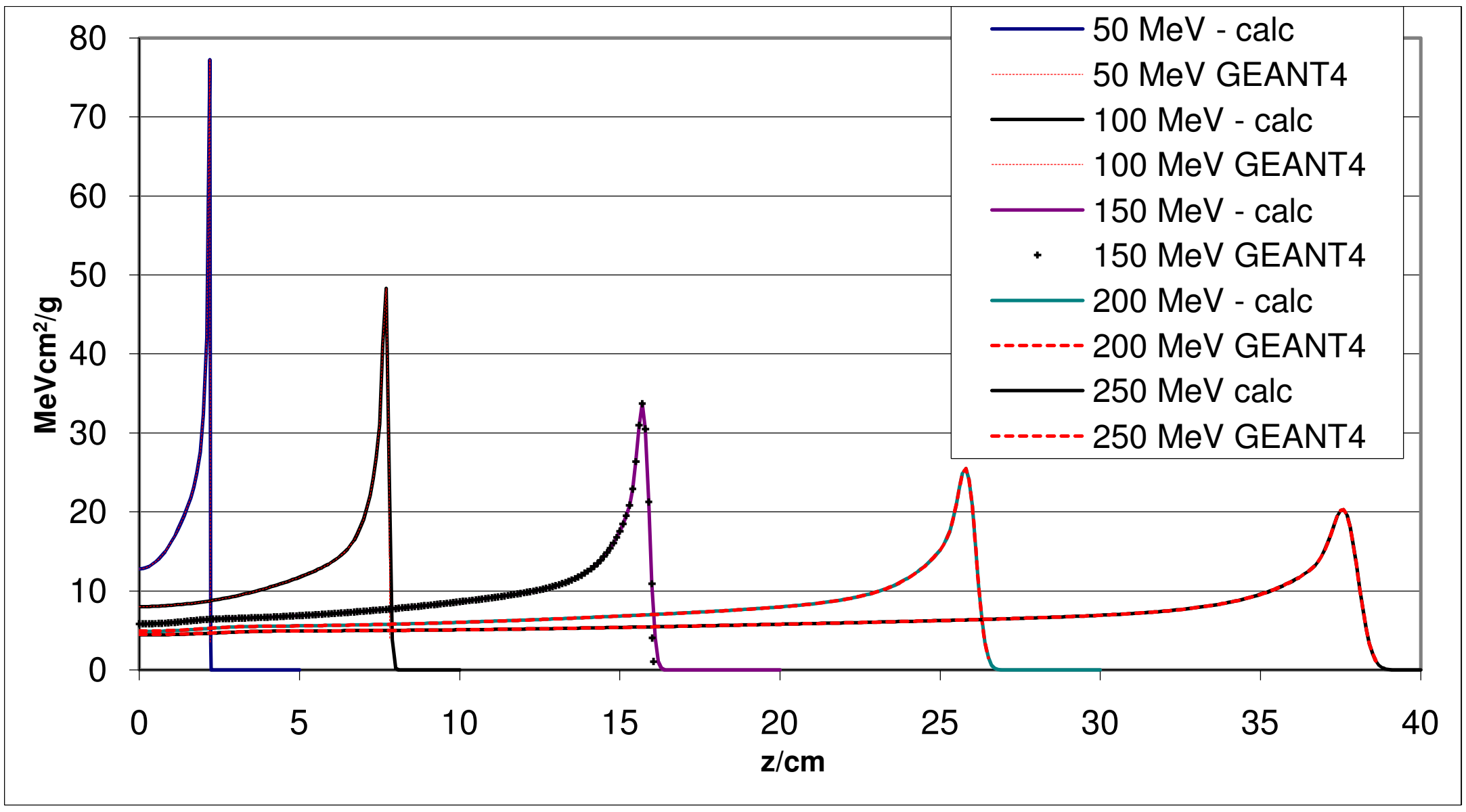

**Fig. 15**: Comparison of the Bragg curves between our calculated buildup with the modeling of the Landau tails for primary/polychromatic protons and GEANT4 (for this purpose, the hadronic generator is switched off). Note that these calculations correspond only to primary protons; $\tau_{in}$ is assumed to be $R_{CSDA} \cdot 0.01$. The buildup in the proton depth doses is only partially a result of nuclear interactions or secondary-proton buildup, as suspected by other authors.

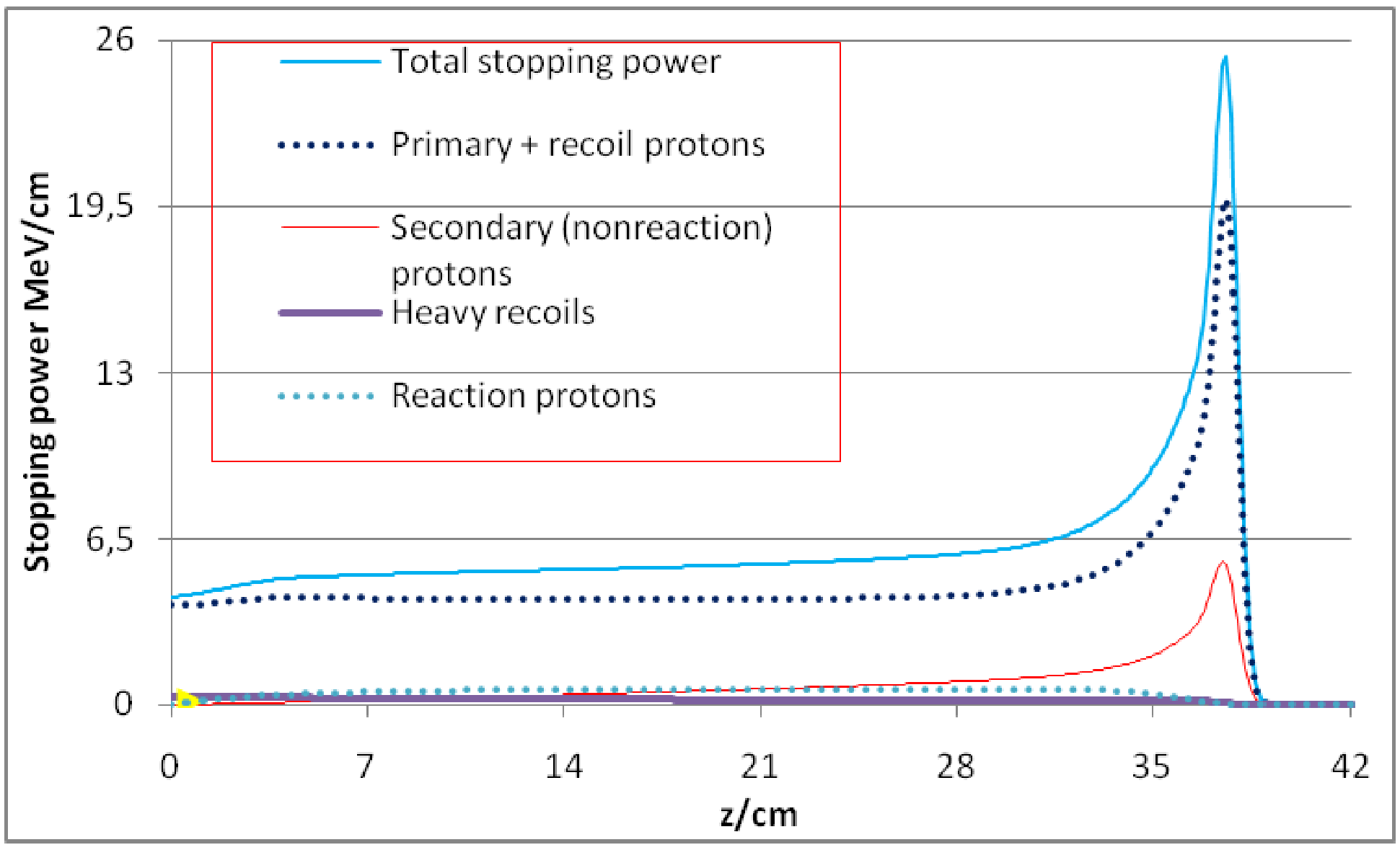

**Fig. 16:** Total stopping power and related partial contributions of a Bragg curve of a monoenergetic 250 MeV proton beam.

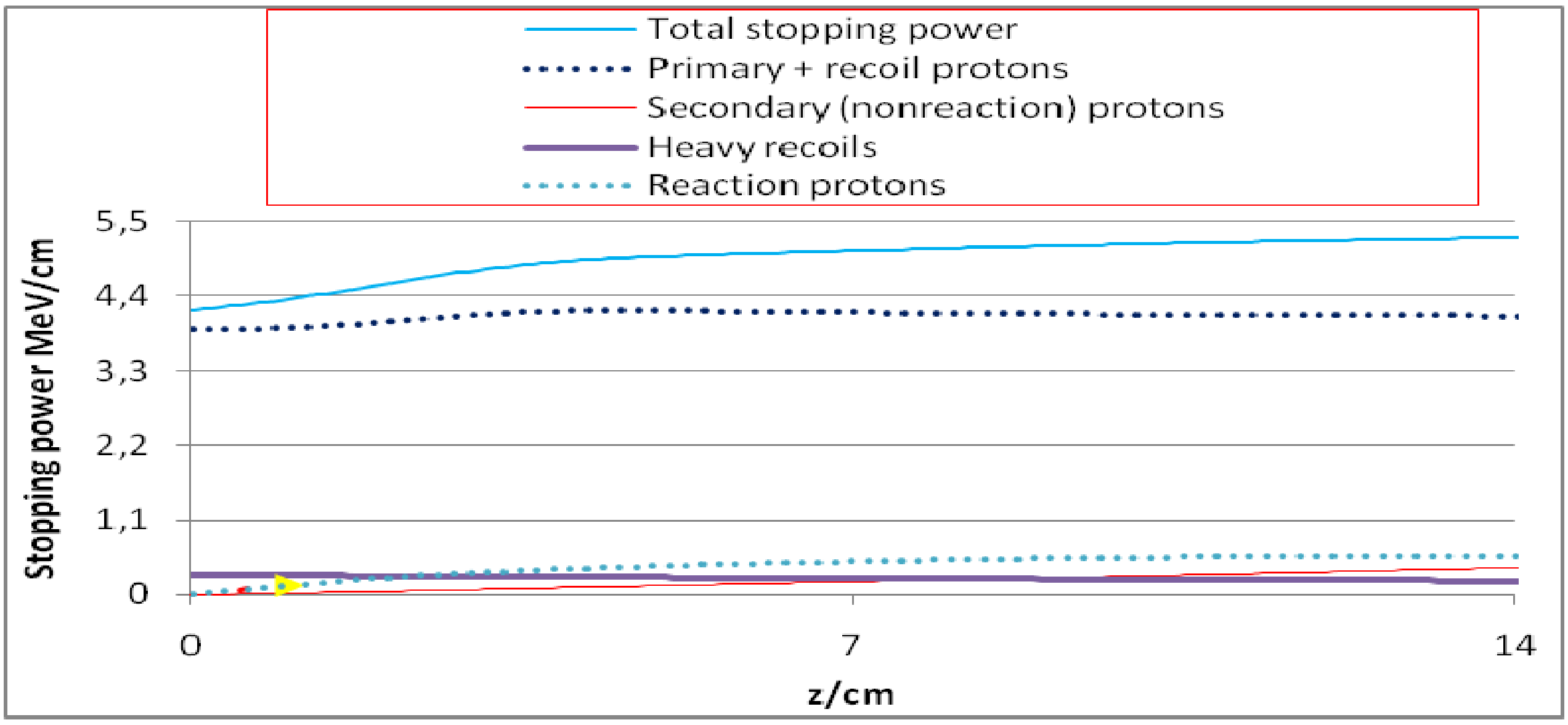

**Fig. 17:** Buildup region of Fig. 16.

Carlsson and Carlsson (1977) have reported a buildup of the Bragg curve of 185-MeV protons (the $E_{max}$ of the incident beam is about 446 keV), and interpreted this effect as an *exclusive* result of secondary protons, since these protons are generated along the proton track rather than at the surface, see also Fippel and Soukup (2004), Medin and Andreo (1997), and Paganetti (2002), where the opinion of Carlsson and Carlsson (1977) has been adopted unquestioningly.

The following remarks should be noted:

1. One cannot label the protons in the measurements, but in Monte-Carlo calculations they are labeled. Thus, one can distinguish them, according to section 1.5. According to Fig. 8, we should expect significant buildup effects of secondary protons between 50 and 150 MeV; thereafter, they should somewhat decrease due to the asymptotic behavior of the nuclear cross section. However, the controversial fact is observed, namely, an increasing buildup with the incident energy. According to Fig. 8, the increase of secondary protons is strongly connected to the decrease of primary protons. Since the fluence of secondary protons is zero at the surface and increases significantly along the proton track, this behavior could be connected with the buildup. However, the fluence decrease of the primaries is concurrent with the behavior of the ‘secondaries’.

2. The question also arises whether the transport of δ-electrons could be responsible for the buildup (similar to the Compton effect of photons). However, $E_{max}$ of 250 MeV protons amounts to 617 keV (Fig. 11) and, with regard to the incident-proton energy of 185 MeV, this energy is lower. The range of these electrons is too small to explain the buildup. In order to produce $E_{max}$ of the order 4 MeV, the proton energy should be of order of 1 GeV.

3. Contributions from γ quanta, resulting either from β+-decays of heavy recoils or from the annihilation of positrons, cannot be significant, as these contributions are expected to be isotropic (i.e., there is no preferred direction of these contributions).

In agreement with GEANT4, the depth-dose curve of primary protons (250 MeV) shows a valley in the middle part of the plateau, resulting from the corresponding fluence decrease (Fig. 9). If the transport of secondary protons is included, the total depth-dose curve does not show this feature (Figs. 15 - 17). A comparison with the results of Medin and Andreo (1997) is noteworthy. If the transport of secondary protons is only partially accounted for or omitted (PTRAN), then this valley can also be observed in the total depth-dose curve. In addition, it is necessary to include the ‘secondaries’ in an accurate manner. According to the aforementioned authors, PTRAN leads (for 200-MeV protons) to a dose contribution of about 10 % at depth z = 20 cm, whereas the early theoretical calculations of Zerby and Kinney (1965) provided 17 % at the same z value. This is in agreement with Fig. 16 and may be calculated on the basis of the formalism developed in Section 1.5.2. In order to be consistent with the total nuclear cross section (Fig. 7), the resulting Fig. 9, and the classification in Section 1.5.2, we consider all protons as secondary protons, originating from nuclear interactions. The so-called ‘reaction protons’, which stand in close heavy recoils, are preferably dominant for E > 150 MeV, else they amount only a small percentage. Their depth-dose curve shows certainly a maximum along their track, but, due to the broad spectral distribution, not a typical Bragg peak (see Appendix, Fig. 46). If most of the secondary protons are nonreaction protons, the total contribution of the secondary protons does yield a Bragg peak, but the whole dose profile is rather different. In various publications on therapeutic protons, this distinction has not been pointed out in a clear manner.

*1.8.3.1 Calculation procedure*

In order to obtain an approximation up to order 2 for the Landau tails according to Eqs. (135 – 138), we have assumed that, in the environment of the impact surface, the first- and second-order terms are the leading ones, whereas the error functions (resulting from first-order corrections) are significant along the entire proton track. All necessary calculation parameters have been determined by fits to the more general theory, as outlined above, and by fits to Monte-Carlo (GEANT4) results, since the option to use a model of Vavilov distributions is available in the code. The terms, which are related to $I_{Lan1}$ given by the usual convolution integrals, are a little bit complicated; therefore we do not go into details here. They contain – besides the pure error functions – also products of error functions with terms proportional to $z/R_{CSDA}$.

The solution procedure of $I_{Lan2}$ is rather a standard task. Resulting from first- and second-order Hermite polynomials, we have to solve:

$$I_{Lan2} = -C_{lan2} \cdot \frac{2}{\sqrt{\pi} \cdot \tau_{Lan2}} \cdot \int (1 - u^2/R_{Lan2}{}^2) \cdot \exp[-(u-z)^2/\tau_{Lan2}{}^2) du \quad (140)$$

In order to carry out the integration, we use the substitution:

$$u' = (u - z) / \tau_{Lan2} \quad (141).$$

Using this substitution, it is easy to solve Eq. (140) via the rule of continued integration to obtain Eq. (137).

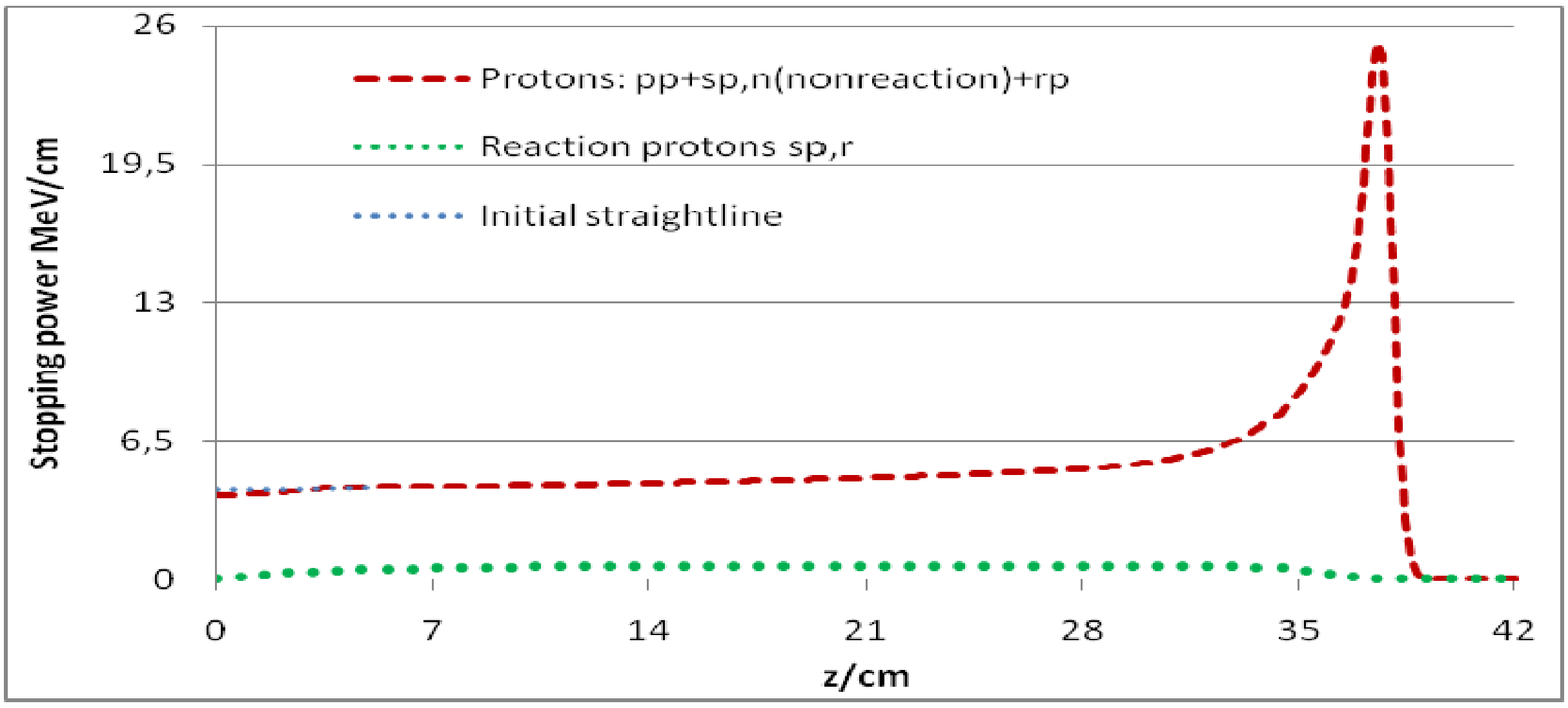

**Fig. 18:** Total stopping power of all protons of Fig. 16, excepted reaction protons, which are shown separately. The buildup is decreased and indicated by a straight line.

It is evident from Fig. 17 that the stopping power of primary protons has a valley in the middle part of the pristine Bragg curve. This valley is only removed by the contributions from secondary/recoil protons. Fig. 18 provides a clear indication that for the buildup two different contributions are essential, namely reaction protons and Landau tails. For polychromatic protons, the role of asymmetric Landau tails may become an increasing importance.

### *1.8.3.2 Energy shifts induced by the Landau tails*

We now introduce the scaling factor, regarding the change from water to a medium M. If $Z_{water}$, $A_{water}$, and $\rho_{water}$ are given for the 'standard medium water', then the scaling factor $s_M$ takes the form:

$$s_M = \rho_M \cdot Z_M \cdot A_{water} /(\rho_{water} \cdot Z_{water} \cdot A_M) \quad (142)$$

For Pb, $s_M$ is equal to 8.078. Fig. 11, referring to the energy dependence of $E_{max}$, can be rescaled by the factor $s_M$ according to $E_{max,M} = E_{max} \cdot s_M$. Some further implications are:

As already pointed out, the existence of the Landau tails implies a significantly higher energy transfer from the projectile protons to the environmental electrons, expressed by $E_{max}$, compared to symmetrical equilibrium conditions (Gaussian fluctuations). A consequence of this energy loss is that the Bragg peak is shifted towards lower energy. If one adapts a calculation model to Bragg curves of homogeneous media, this effect cannot be verified, since one simply associates the whole curve to a lower initial energy, i.e., the range is reduced, as probably expected. However, this shift can be calculated with the help of the area integral over S(z). The difference in the area integral leads to a different total energy $E_0$. We have carried out such calculations for different homogeneous media. For water, we can summarize the resulting range shift $\xi$ related to $R_{CSDA}$ by the relation:

$$\left.\begin{array}{l} \xi(\text{in cm}) = \alpha_0 \cdot E_0 + \alpha_1 \cdot E_0^2 \quad (E_0 \text{ in MeV}) \\ \alpha_0 = 0.00021 \text{ cm/MeV}; \ \alpha_1 = 0.000003521 \text{ cm/MeV} \end{array}\right\} \quad (143)$$

For any other medium M, the scaling factor $s_M$ provides the main contribution, but the average ionization energy $E_I$ introduced by the BBE now enters.

$$\left.\begin{aligned}&\xi_M(\text{in cm}) = \alpha_{0,M}\cdot E_0 + \alpha_{1,M}\cdot E_0^2 \quad (E_0 \text{ in MeV})\\&\alpha_{0,M} = 0.00021\cdot s_M \ \text{cm/MeV};\ \alpha_{1,M} = 0.000003521\cdot(s_M + 0.0021\cdot ln(E_{I,M}/E_{I,water}))\ \text{cm/MeV}\\&E_{I,M} \geq 5.1 \text{ eV}\end{aligned}\right\} \quad (144)$$

The energy shift of a proton beam passing through a thin scatter layer is given in Eq. (145); $E_0$ is the energy before the layer, $E_1$ after it. The shift of a proton beam passing through a medium M with thickness d is then given by

$$\Delta\xi = \xi_0 - \xi_1 = (E_0 - E_1)\cdot[\alpha_{0,M} + \alpha_{1,M}\cdot(E_0 + E_1)] \quad (145)$$

On the other hand, the proton passage through a scatter disc (e.g., 0.5 mm Pb) has to be regarded as an additional energy straggling and obviously implies a Landau distribution. The fitting with the increment of a Gaussian fluctuation approximately holds, see Bethe et al. (1953) and Ulmer (2007):

$$\Delta\tau^2 = K\cdot s_M\cdot[E_0^{-2} - E_1^{-2}]\cdot E_{max,M}\cdot d \quad (146)$$

The calculation of the range shift, according to CSDA with respect to the energy difference between $E_0$ and $E_1$, which we usually would refer to as relative stopping power, does not account for large energy fluctuations in case of a large deviation from the CSDA. This fact can be verified on the basis of the formalism as developed above. Eq. (146) is also applicable in the case of the passage of proton beams through metallic implants, where shifts due to the Landau tails may become important.

## 1.9 Lateral scatter

In the present implementation in Eclipse, the lateral scatter of protons is treated by an approximate version of the multiple-scattering theory (Bethe (1953), Molière (1955), Gottschalk et al. (1993), and Gottschalk et al. (1999)). Highland (1975) used only one Gaussian for the angular distribution of scattered protons:

$$f(\theta)d\theta \approx \exp(-\theta^2/2\theta_0^2)d\theta \quad (147)$$

The angle $\theta_0$ depends on various physical parameters, e.g., on the radiation length L, the incident energy, etc.

In order to describe accurately the lateral tail of the primary and secondary (nonreaction, *sp,n*) protons, we will make use of two Gaussian kernels; Monte-Carlo calculations with GEANT4 indicated that two Gaussian kernels are sufficient in the case of the primary protons. For secondary (reaction, *sp,r*) protons, we restrict the lateral kernel to one modified Gaussian; this is due to their significantly smaller contribution. Our first Gaussian accounts for the inner part of multiple Molière scatter, which is steeper than in the Highland approximation, whereas the second Gaussian has a much larger half-width (to describe the tail). The Highland approximation assumes a slightly broader half-width than necessary for the inner part, in expense of accounting (partially) for the tail. Thus, our calculation model for the lateral scatter in water is given by:

$$K_{lat,prim}(r, z) = \frac{C_0}{\pi\tau_{lat}(z)^2} \cdot e^{-\frac{r^2}{\tau_{lat}(z)^2}} + \frac{(1-C_0)}{\pi\tau_{lat,LA}(z)^2} \cdot e^{-\frac{r^2}{\tau_{lat,LA}(z)^2}} \qquad (148)$$

For the contribution of the main Gaussian, we use a weight coefficient $C_0$ of 0.96. The calculation of $\tau_{lat}(z)$ (inner-part) and $\tau_{lat,LA}(z)$ (large-angle) is carried out as follows:

$$\left.\begin{array}{l} \left.\begin{array}{l} \tau_{lat}(z) = f \cdot 0.626 \cdot \tau_{max} \cdot Q(z) \\ f = 0.9236 \\ \tau_{max} = (E_0 / 176.576)^p \end{array}\right\} \\ p = 1.5 + \begin{cases} 0.00150 \cdot (176.576 - E_0), & \text{if } E_0 \le 176.576 \\ 0.03104 \cdot \sqrt{E_0 - 176.576}, & \text{if } E_0 > 176.576 \end{cases} \\ Q(z) = \dfrac{e^{\frac{1.61418 \cdot z}{R_{CSDA}}} - 1}{e^{1.61418} - 1} \cdot 0.5 \cdot \left[ erf\left( \dfrac{R_{CSDA} - z}{\tau} \right) + 1 \right] \end{array}\right\} \qquad (149)$$

A good approximation for a model with a single Gaussian for the primary protons can be obtained from the equations above by substituting $C_0$ and f by 1. In agreement with the results of Gottschalk et al. (1993), the change from water to other materials is obtained by the scaling of Q(z) on the basis of $R_{CSDA}$. The error-function term models the Gaussian distribution of the stopping distribution due to range straggling. The protons, which have undergone only small-angle scattering, are closer to the central axis of the beamlet and travel further.

A fit to Monte-Carlo results shows that $\tau_{lat,LA}(z)$ can be determined by the kernel:

$$\tau_{lat,LA}(z) = \frac{0.90563}{1 - e^{\frac{-1}{0.252}}} \cdot \tau_{max} \cdot \left[ e^{-\frac{(1 - z/R_{CSDA})^2}{0.252}} - e^{-\frac{1}{0.252}} \right] \cdot 0.5 \cdot \left[ \text{erf}\left( \frac{R_{CSDA} - z}{\tau} \right) + 1 \right] \qquad (150)$$

The scaling properties of Q(z) still hold, since only the ratio $z/R_{CSDA}$ enters Eq. (150). For the secondary protons, one should use different Gaussian kernels for those particles which did or did not undergo nuclear reactions (see $S_{sp}$ and $S_{heavy}$):

$$\left.\begin{aligned}\tau_{sp}(z) &= \tau_{max} \cdot Q(z) \\ \tau_{heavy}(z) &= \tau_{heavy}(E(z))\end{aligned}\right\} \quad (151)$$

$\tau_{max}$ and Q(z) are the same as defined in Eq. (149); $\tau_{heavy}(E(z))$ is given by a similar equation as above, except that it depends on the local energy E(z) instead of $E_0$.

It might appear that Eqs. (148 - 151) are just obtained by fitting methods. However, the spatial behavior of the multiple-scatter theory involves a Gaussian and Hermite polynomials $H_{2n}$:

$$\left.\begin{aligned}&K(r,z,lat) = N \cdot exp(-r^2/lat(z)^2) \cdot [a_0 + a_2 \cdot H_2(r/lat(z)) + a_4 \cdot H_4(r/lat(z)) + \\ &\qquad + a_6 \cdot H_6(r/lat(z)) + 0(\text{higher order}) \\ &N = 1/(\sqrt{\pi} \cdot lat(z)); \quad r^2 = x^2 + y^2\end{aligned}\right\} \quad (152)$$

The parameters of the Hermite-polynomial expansion of multiple-scatter theory are: $a_0 = 0.932$; $a_2 = 0.041$; $a_4 = 0.019$; $a_6 = 0.008$. The Gaussian half-width is the same as assumed for the inner part. The task now is to determine a linear combination of two Gaussians, according to Eq. (148), with different half-widths, on the basis of an optimization problem. This way, we are able to go beyond the Highland approximation.

The calculation of $\tau_{max}$ is carried out in the following way. The differential cross section is given by:

$$\left.\begin{aligned}\frac{dlat(z)^2}{dz} &= \frac{dlat_E^{\,2}}{dE} \cdot \frac{dE}{dz} \\ \frac{dlat_E^{\,2}}{dE} &= \frac{\alpha_{Medium}}{E^2}\end{aligned}\right\} \quad (153)$$

In our calculations, we have only used $\alpha_{Water}$; $\alpha_{Medium}$ is proportional to $Z_M \cdot \rho_M / A_M$, where $Z_M$, $\rho_M$, and $A_M$ are respectively the nuclear charge, the mass density, and the mass number of the medium. Values for E(z) and dE/dz in Eq. (153) may be obtained by using the Bragg-Kleeman rule:

$$\left.\begin{aligned}R_{CSDA} &= 0.00259 \cdot E_0^{\,p} \\ R_{CSDA} - z &= 0.00259 \cdot E(z)^p\end{aligned}\right\} \quad (154)$$

The inversion of Eq. (154) leads to:

$$E(z) = \left( \frac{R_{CSDA} - z}{0.00259} \right)^{1/p} \quad (155)$$

An accurate application of this rule requires consideration of the dependence of p on $E_0$. The quantity dE(z)/dz can be computed from Eq. (155). Finally, Eq. (153) yields:

$$lat_{\max} = \sqrt{\int_0^{R_{CSDA}} \left[ \frac{\alpha_{Water}}{E(z)^2} \right] \cdot \frac{dE(z)}{dz} \cdot dz} \quad (156)$$

One might expect that the lateral-scatter functions (Eqs. (149 – 150)) continuously increase up to z = $R_{csda}$. Actually, this assumption would be valid, if the energy spectrum for the scattered protons would be identical at depth z, independent of the scatter angle and the fluctuations due to the energy/range straggling $\tau_{straggle}$ and $\tau_{in}$. In reality, there are small fluctuations of the lateral-scatter functions along the proton tracks. In particular, from the Bragg peak down to the distal end, there is a significant difference between the protons, which have only undergone small-angle scatter in this domain, and the ones with larger scatter angle. The latter protons have deposited their energy in an oblique path; therefore, they stop earlier and cannot reach the distal end of the Bragg curve. It is clear that the scatter functions for primary and secondary protons ($\tau_{lat}$, $\tau_{lat,LA}$ and $\tau_{sp}$) depend on $\tau_{straggle}$ and $\tau_{in}$, which induce these fluctuations and cause significant changes in the energy spectrum at the end of the proton tracks. In order to describe this behavior by a mathematical model, we prefer to use a Gaussian convolution, which is certainly justified in the domain of Bragg peak (low-energy region of the proton tracks). As an example, we use the function Q(z) in Eq. (148), which fixes both $\tau_{lat}$ and $\tau_{sp}$. We denote the fluctuation parameter by $\tau$; the connection to straggle parameters will be considered thereafter.

The crude model assumes:

$$\left. \begin{aligned} Q_0(z) &= \frac{\exp(1.61418 \cdot z / R_{csda}) - 1}{\exp(1.161418) - 1} \quad (\text{if } z \le R_{csda}) \\ &= 1 \quad (\text{if } z > R_{csda}) \end{aligned} \right\} \quad (157)$$

A more realistic model, taking all the arguments with regard to fluctuations into account, is obtained by:

$$Q(z) = \frac{1}{\tau \cdot \sqrt{\pi}} \cdot \int_{-\infty}^{R_{csda}} Q_0(\xi) \exp[-(z - \xi)^2 / \tau^2] d\xi \quad (158)$$

Using standard methods, we obtain the modified Eq. (148) for Q(z), instead of $Q_0(z)$ according to Eq. (158):

$$Q(z) = \frac{exp(1.61418 \cdot z / R_{csda}) - 1}{exp(1.161418) - 1} \cdot \frac{1}{2} \cdot [1 + \mathrm{erf}((R_{csda} - z)/\tau)] \quad (159)$$

This result implies that Q(z) increases exponentially along the proton track, as long as the error function is 1 (or nearly 1). Only in the environment of z = $R_{csda}$ Q(z), does it decrease rapidly. However, this behavior actually depends on $\tau$. The connection between $\tau$ and the aforementioned convolution parameters for energy/range straggling is a valid question. One might assume that, for proton pencil beams with energy/range straggling, we can set:

$$\tau^2 = \tau_{straggle}^{\;2} + \tau_{in}^{\;2} \quad (160)$$

This assumption might be reasonable, since the proton history, resulting from the beamline and expressed by $\tau_{in}$, should be accounted for. As already pointed out, we have made use of the GEANT4 results for the adaptation of the scatter functions with regard to monoenergetic protons. It turned out that the best adaptation in the domain from the Bragg peak to the distal end can be obtained, if we put (monoenergetic protons):

$$\tau^2 = \tau_{straggle}^{\;2} + \tau_{Range}^{\;2} \quad (161)$$

The mean standard deviation for protons with $E_0$ = 50 MeV up to 250 MeV in intervals of 25 MeV amounts to 2.6 %. $\tau_{Range}$ has been already defined (Eq. (47)). Therefore, it may be justified to modify Eq. (160) by writing:

$$\tau^2 = \tau_{straggle}^{\;2} + \tau_{in}^{\;2} + \tau_{Range}^{\;2} \quad (162).$$

## 1.10 Beamlines for protons

At several places in this study, we have mentioned that we will give information on the description of the available beamlines, which are used in proton treatment. Prior to entering this subject, we now summarize the essential aspects, which are relevant and have been given in the previous sections. With regard to the pristine Bragg peak, the parameter $\tau_{in}$ takes account of the Gaussian convolution (Eqs. (103, 111), as well as of the Landau tail (Eqs. (135, 137)). However, a sole Bragg curve does not fulfill the requirement of creating a homogeneous dose distribution in the target. Therefore, we have to consider the creation of an SOBP, via the superposition of different Bragg peaks with additional modulation of the proton energy.

Besides the energy modulation, the lateral distribution of the proton beam is a very important issue. In particular, a sufficient description of the tail, resulting from multiple scattering, goes beyond the single-Gaussian approximation (Eqs. (147 – 159) and Fig. (32)). There are three different methods commonly used to solve this task:

1. Active-scanning technique

2. Passive-scanning technique

3. Broad-beam technique (double scattering, uniform scattering (formerly known as wobbling)).

*Active-scanning technique*

The specific proton accelerator is a synchrotron, which provides accurately the required energy of each beam to form an SOBP. Since no further range shifters are necessary, the energy spectrum of the impinging proton beam approaches best the monoenergetic case (see Fig. 33).

*Passive-scanning technique*

In this case, a cyclotron provides only some fixed energies (e.g., 100 MeV, 175 MeV, 250 MeV). The desired energy to form an SOBP is obtained by a range modulator (e.g., the modulator wheel). The energy spectrum of the impinging beam is broadened (Fig. 33).

*Broad-beam technique (double scattering, wobbling)*

In double scattering, a (small) beamlet is transformed into a broad beam by (at least one) scatter disc. An additional range shifter can reduce the whole proton beam to form an SOBP, and the resulting broad beam is usually shaped by the primary collimator (see section 2.3). The use of a compensator is an important feature in this technique. We decompose the whole broad beam in pixel areas (beamlets), and each pixel is the origin of a proton pencil beam. The compensator then controls the range of each beamlet in the given region of interest (Fig. 19).

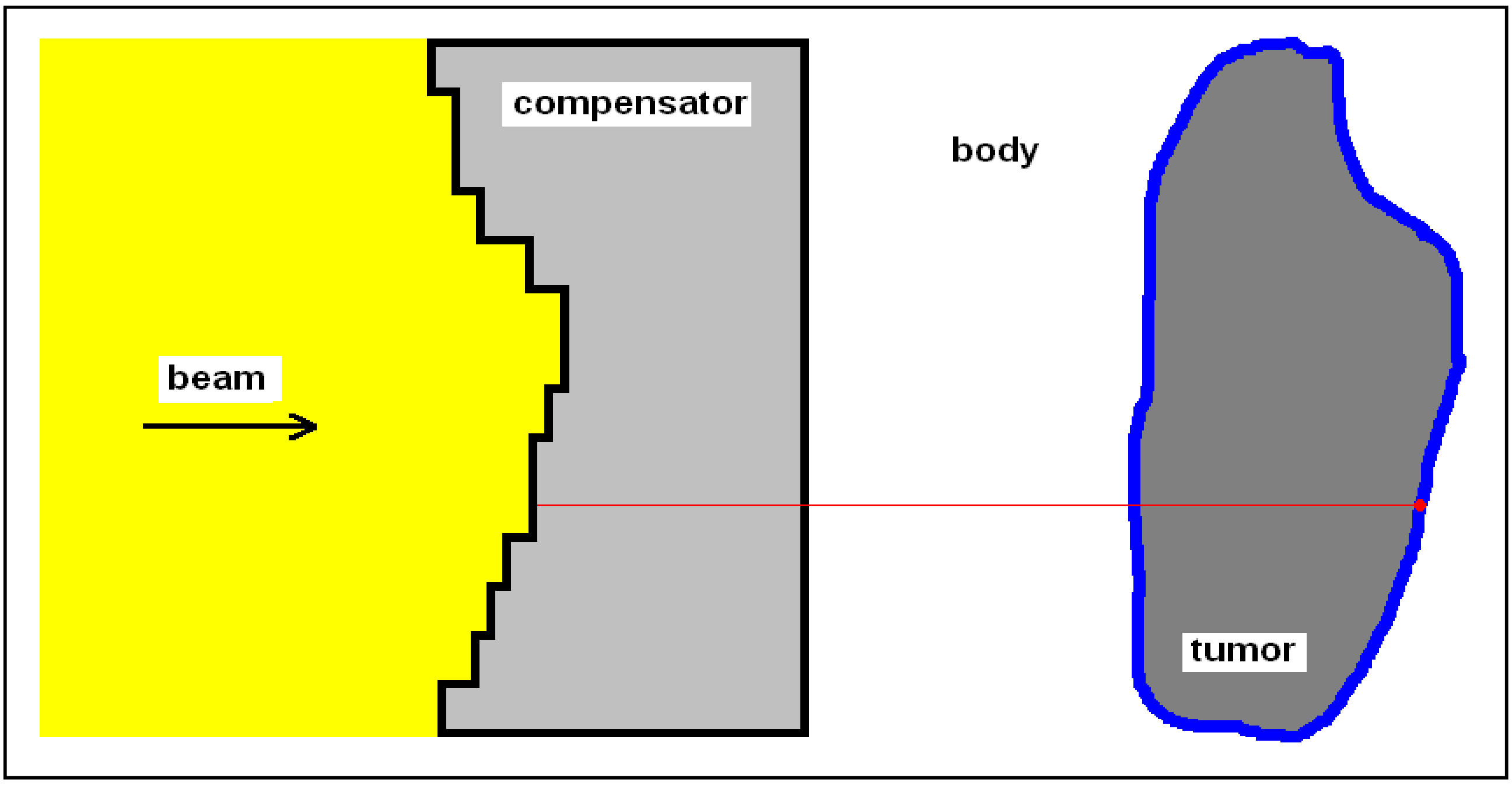

**Fig. 19:** The variability in the water-equivalent thickness of the target is counter-balanced by the profile of the compensator. As a result, the use of a compensator enables the positioning of all Bragg peaks (corresponding to the beamlets) on the distal surface of the target.

The calculation procedure according to Eqs. (163 – 164) is a superposition/convolution method to account for the boundary conditions of the pencil beam in a given voxel of the target. In wobbling, the lateral spreading of the original beam is achieved on the basis of magnetic deflection in two (x/y) directions.

## 1.11 Calculations of beamlets and 3D dose distributions by superposition/convolution

Our dose model is a superposition/convolution model, i.e., a superposition of individual 3D proton beamlets convolved with the fluence at the position of the beamlet. The term ‘fluence’ refers here to the undisturbed fluence in air. The calculation and configuration of the fluence in air is described in detail in Schaffner (2008). The fluence in air incorporates all effects which contribute to the lateral distribution of the protons after they exit the beamline. The main effects modeled in the fluence are the initial lateral penumbra – following the effective-source size concept introduced by Hong et al. (1996), the scatter in the compensator, and the phase space and weight of scanning pencil beams. The concept in the dose calculation has been published, see Schaffner et al. (1999) and Ulmer et al. (2005). In short, it is given by:

$$D_{total}(x,y,z) = \sum_{i,j} \Phi_{air}(x_i, y_j, z) \cdot D_{beamlet\ ij}(r(x,y),z) \quad (163)$$

For the practical computation, we substitute the convolution by a sum over the beamlets at each point of the calculation grid. The fluence $\Phi_{air}$ is always taken at the position corresponding to the central axis of each beamlet, i.e., $x_i$ and $y_j$ for the beamlet ij.

In order to save computation time, we do not use separate models for the lateral distribution of recoil protons, heavy recoils, and reaction protons. We describe the lateral extension of recoil protons by the same kernel as for primary protons, since it can be assumed that the production of recoil protons follows the distribution of primary protons and the energy of recoil protons is deposited locally. Heavy recoils deposit most of their dose through the $\beta^+$-decay and annihilation, or by neutron emission, see also listing (52). This means that their lateral distribution is very broad. Due to the very small overall contribution of heavy recoils, we simply add the dose of heavy recoils to the dose deposited by secondary protons, which have, in general, the broadest distribution. The 3D beamlet is calculated from the results of the previous sections and with the simplifications discussed above as follows:

$$\left.\begin{array}{l} D_{beamlet}(r,z)=\dfrac{1}{\rho_{H_2O}}\left[S_{pp}(d(z))+S_{sp,n}(d(z))+S_{rp}(d(z))\right]\cdot K_{lat,prim}(r,d(z))+ \\ \left[S_{sp,r}(d(z))+S_{heavy}(d(z))\right]\cdot K_{lat,sec}(r,d(z)) \end{array}\right\} \quad (164)$$

The distance from the central axis of the beamlet is denoted by r, z is the position along the central axis of the beamlet, and d(z) is the water-equivalent distance up to the position z. Fig. 20 represents plots of a building block of the beamlet calculation, the lateral distribution due to scattering in water.

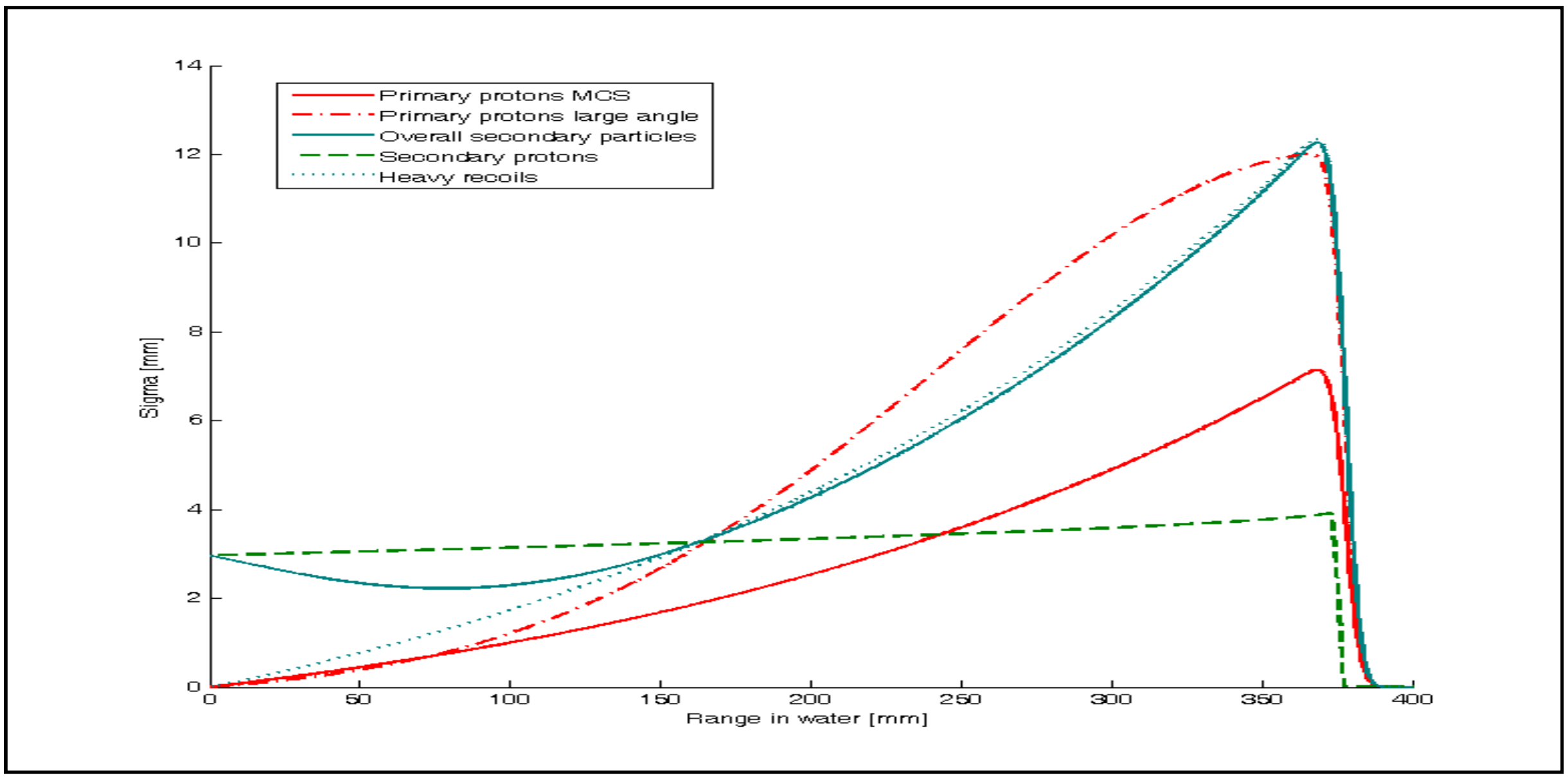

**Fig. 20:** Example of the lateral dose distribution in a beamlet of a 250-MeV proton beam. The secondary protons exhibit a nonzero width at entrance; this is due to the contribution of $\tau_{heavy}$, see Eq. (112). The emission of heavy recoil particles from a nuclear reaction is assumed to be isotropic. The quantity shown in this figure is defined as $\tau_{lat}/\sqrt{2}$.

Since the calculated dose refers to water, not to the medium, the density of water is used in Eq. (164), following the convention used in photon dose-calculation algorithms. The path-length correction or scaling of the beamlet in depth for water media other than water is taken into account in the beamlet calculation by using d(z) instead of the distance from the surface. The path-length correction is applied in the same way for the scaling in depth and in the lateral distribution. An improved model for the scaling of the lateral distribution in inhomogeneous media has been put forth by Szymanowski and Oelfke (2002).

## *2. Applications and comparisons with measurements*

### 2.1 Bragg curves of protons

In Ulmer (2007), we have pointed out that the measured Bragg curves can be adapted optimally, if a certain approach is followed. The starting point is the energy $E_0$ of the incident proton beam and the assumption of an incident spectral distribution described by $\tau_{in}$. The rough estimate of $C_{Lan1}$, according to Formula (136), can also be considered as a starting value. Small variations of $E_0$ and $\tau_{in}$ lead to minimum standard deviation. If the Landau tails are included, the mean standard deviation did not exceed 0.1 - 0.3 %. The source-surface-distance (SSD) may also be subjected to a small variation; in the scanning technique, SSD $\rightarrow \infty$. We will next show that the aforementioned parameters are not arbitrary and can be estimated on the basis of the beamline characteristics of each proton-treatment machine.

#### *2.1.1 Model M1*

The case of the 141-MeV beam (Figs. 21 and 22) has been taken from a previous publication, see Ulmer (2007). Both figures demonstrate the importance of the Landau tail (which is omitted in Fig. 22).

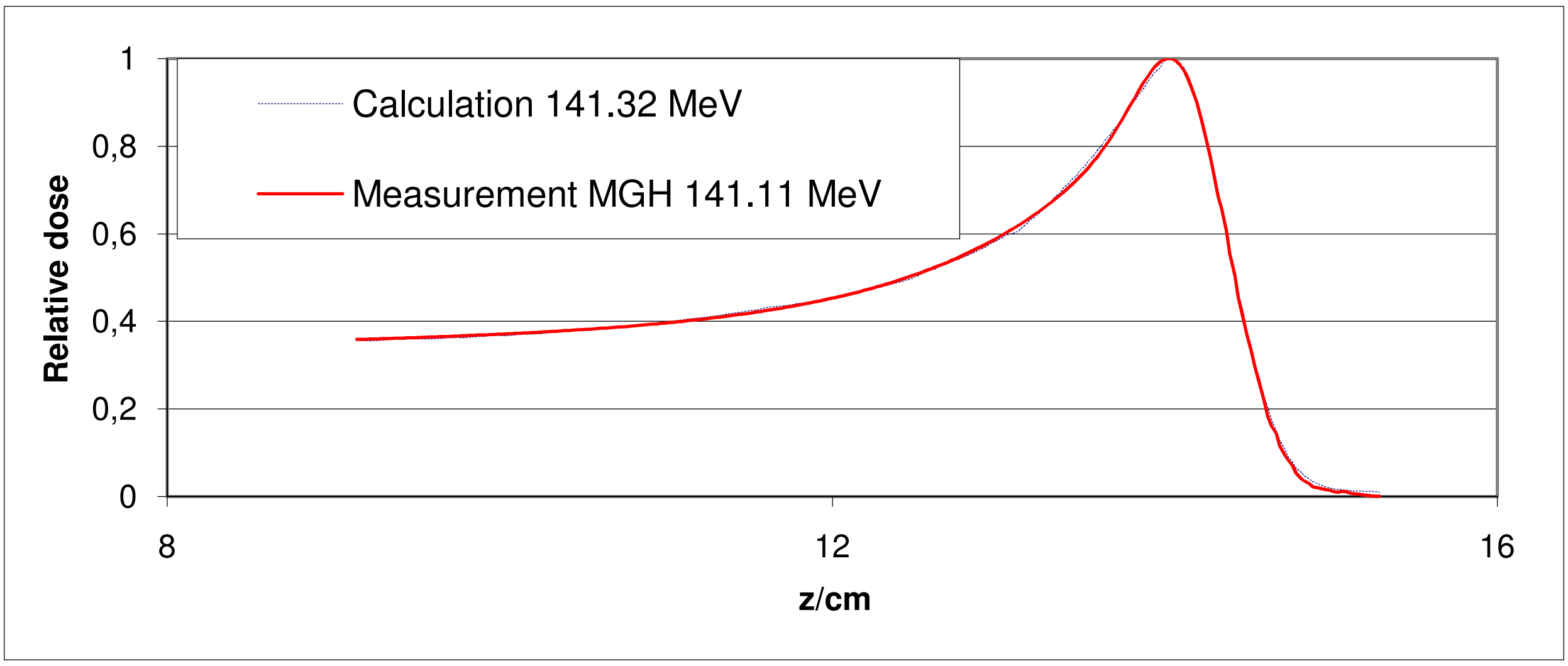

**Fig. 21:** Best adaption with the model M1. The Landau tails have been included.

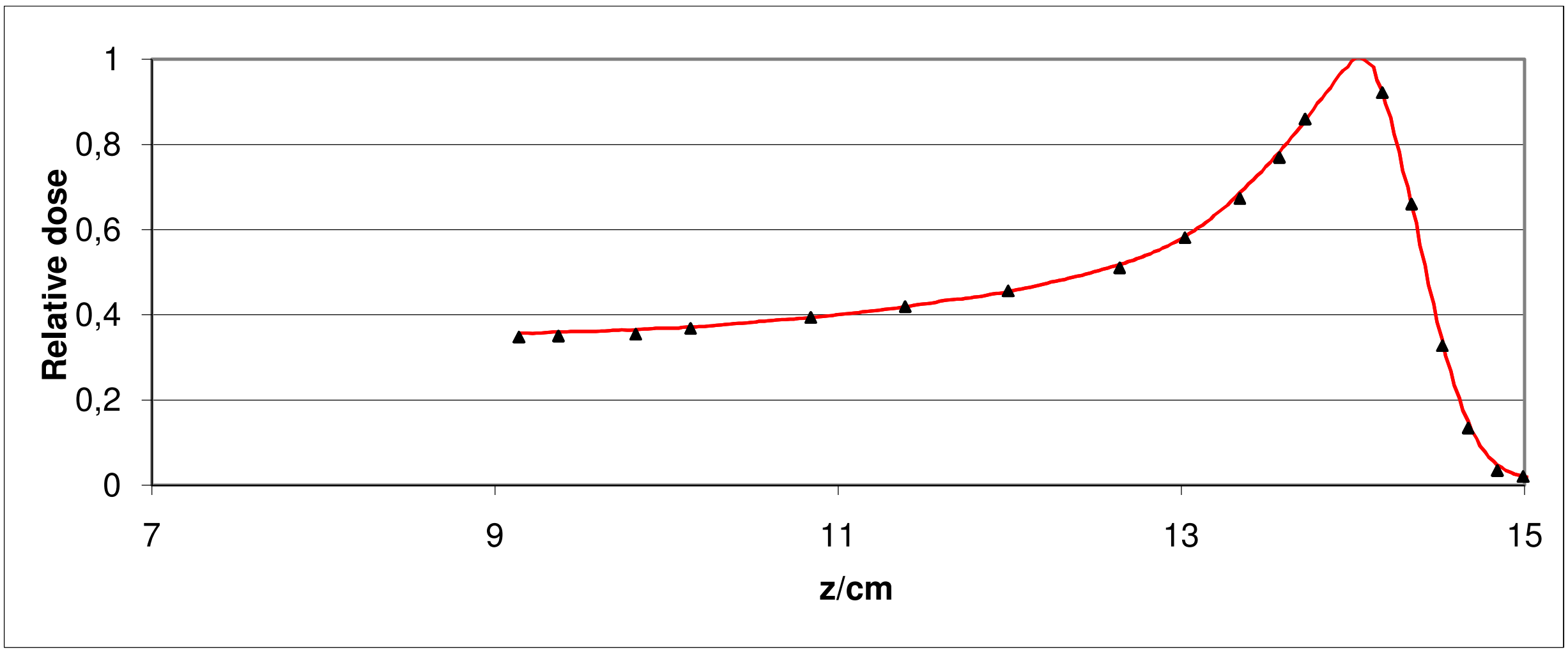

**Fig. 22:** Best adaptation of the measured data of Fig. 20 with the model M1 without the Landau tails (samples: calculated Bragg curve).

Unfortunately, the measured data start with the nozzle as a reference point. Therefore, the behavior at the surface is unknown; there is only an indication of the existence of a buildup. The contribution $C_{Lan1}$, used in Fig. 21, has been based on Formula (136), without any further optimization. Additional parameters are: $\tau_{in}$ = 0.263 cm (Fig. 21), $\tau_{in}$ = 0.289 cm (Fig. 22) and SSD = 214.5 cm.

*2.1.2 Model M2*

As already mentioned, we have accounted for all the terms relating to the stopping power of the BBE according to the recommendations of ICRU49. Since the numerical-simulation procedure in GEANT4 has to account for all these terms separately, a compensation of the logarithmic term by the other terms in the low-energy region cannot occur, and a cutoff in the proton energy at 1 MeV has been imposed. Fig. 23 shows relative depth-dose curves for monoenergetic protons in water. The agreement between GEANT4 and the analytical calculation is excellent up to the distal end. In this domain, we could only verify very small deviations of the order of 0.01 %; these are probably due to numerical inaccuracy. In the distal end, the calculated values exceed the GEANT4 results by about 1 - 8 %. We assume that this difference originates from the energy cutoff in GEANT4.

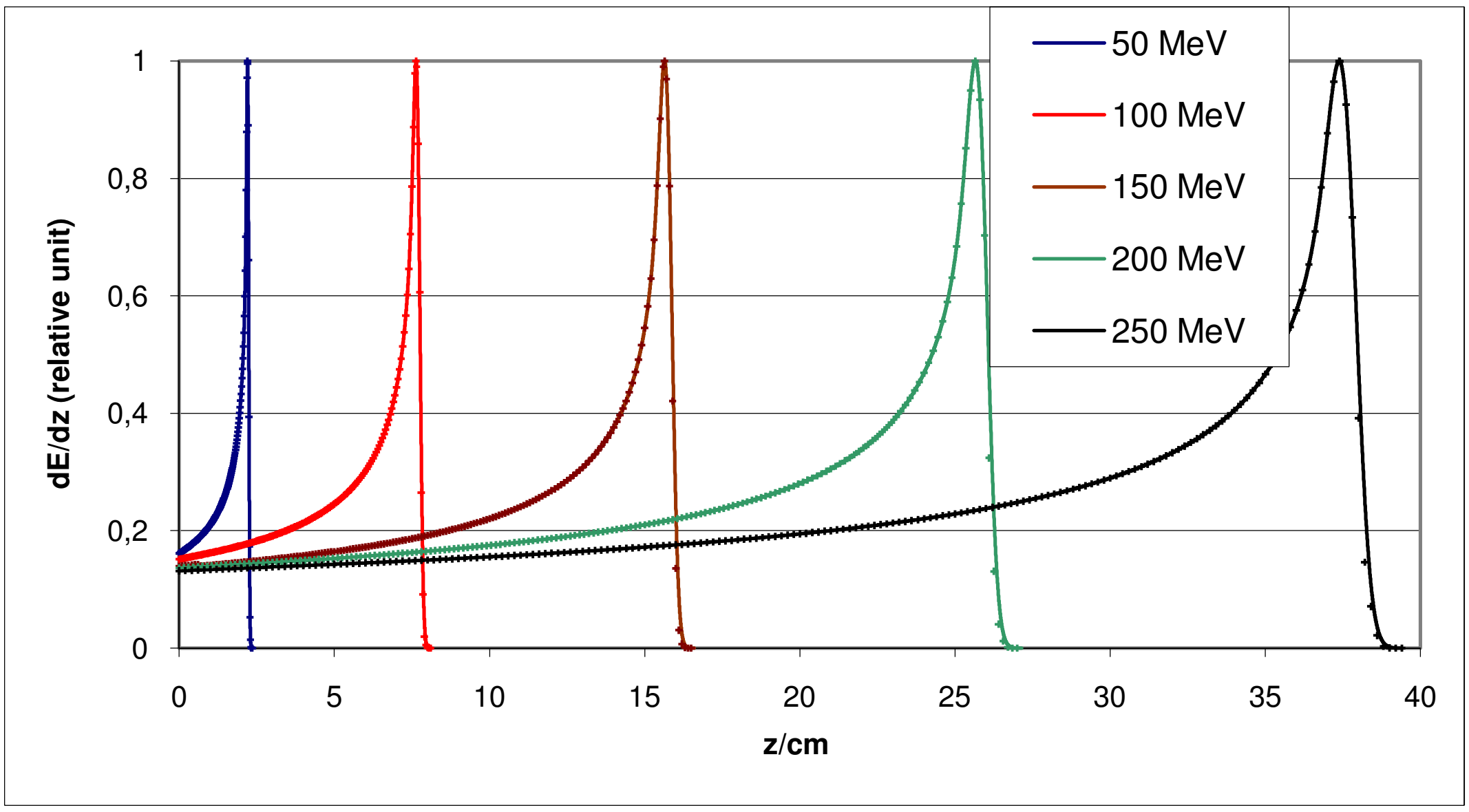

**Fig. 23:** Comparison of the stopping powers for monoenergetic proton beams between elaborated analytical integration and numerical handling with GEANT4 (discrete points, obtained with a cutoff of 1 MeV).

The role of the energy/range straggling in LET calculations is an important feature, which can be best studied with model M2. Energy/range straggling transforms a monoenergetic proton beam to a polychromatic one. Fig. 24 shows that the LET is significantly affected by polychromatic proton spectra. We have subjected the stopping power dE/dz of protons in water to convolutions of monoenergetic protons (10 MeV, 50 MeV, and 270 MeV) and a polychromatic spectrum with a composite convolution (the restriction to the CSDA approach has no practical importance; however, at $E \rightarrow 0$ dE/dz amounts to about 654 MeV/cm). This figure shows that for energies exceeding 30 MeV, the energy/range straggling has no significance; only for energies below 30 MeV, is the LET significantly reduced in the distal end. Figs. 25, 26, and 27 present pristine Bragg curves, obtained at PSI by degrading a 600 MeV proton beam. Unfortunately, a buildup effect could not be verified in this beamline, and it should be mentioned that only the contribution $I_{Lan1}$ plays a certain role with regard to the calculation procedure, since it slightly modifies the energy spectrum in the middle part of a Bragg curve. The comparably high values for $\tau_{in}$ result from the degrading.

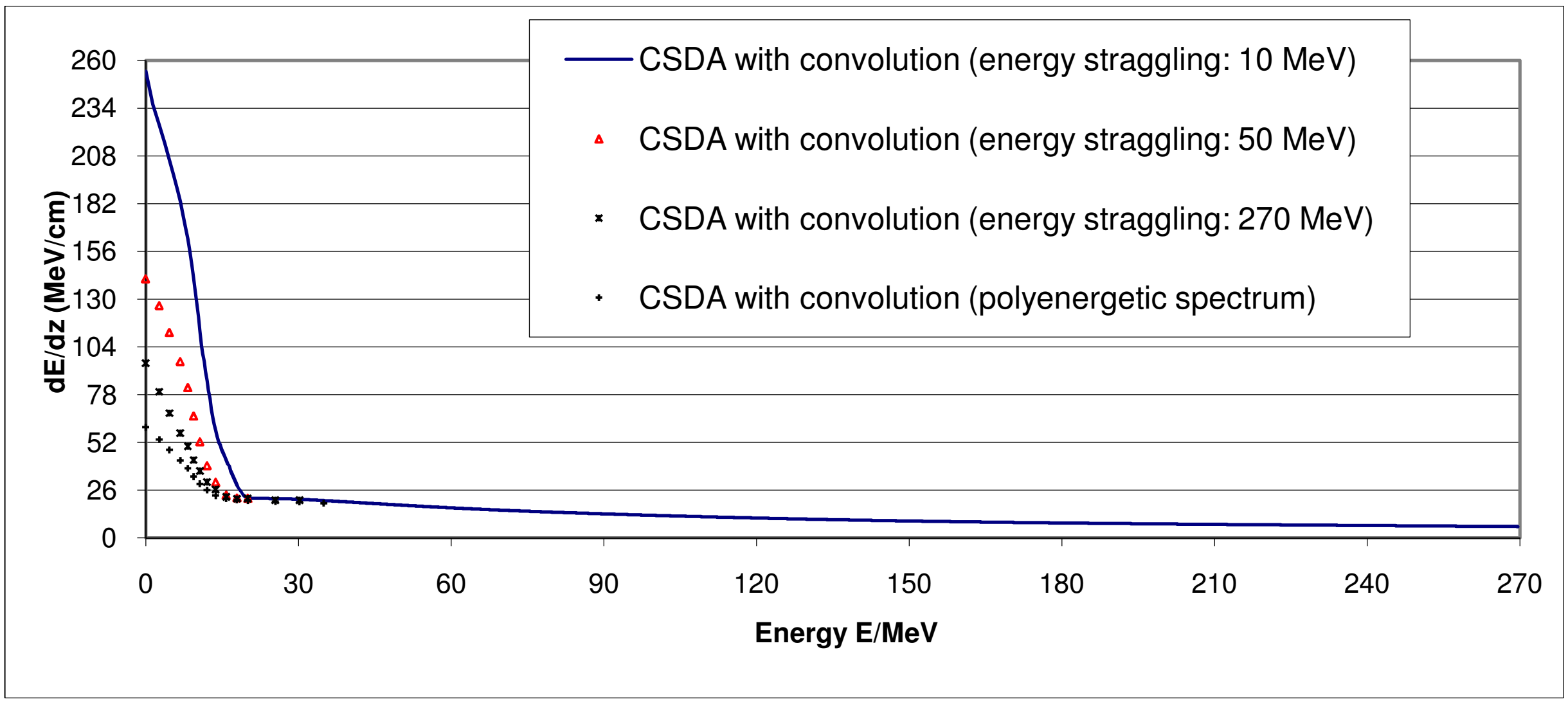

**Fig. 24:** Calculation of the LET (monoenergetic proton beams with energy straggling of 10 MeV, 50 MeV, and 270 MeV and polychromatic proton beam with energy/range straggling corresponding to 270 MeV; $\tau^2 = \tau_{straggle}{}^2 + \tau_{in}{}^2$).

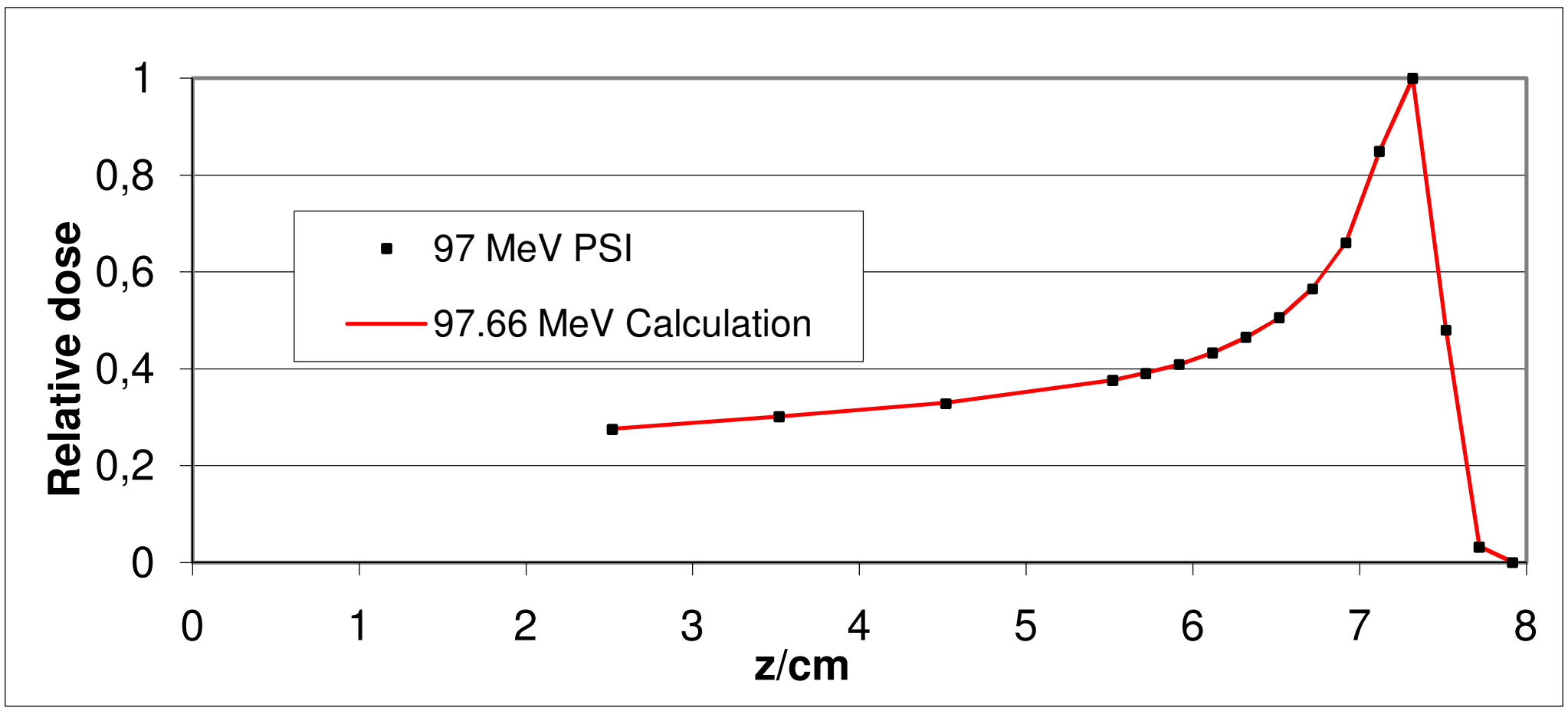

**Fig. 25:** 97 MeV – PSI (discrete points: measurements): dev=0.14 %, $\tau_{in}$ = 0.159 cm.

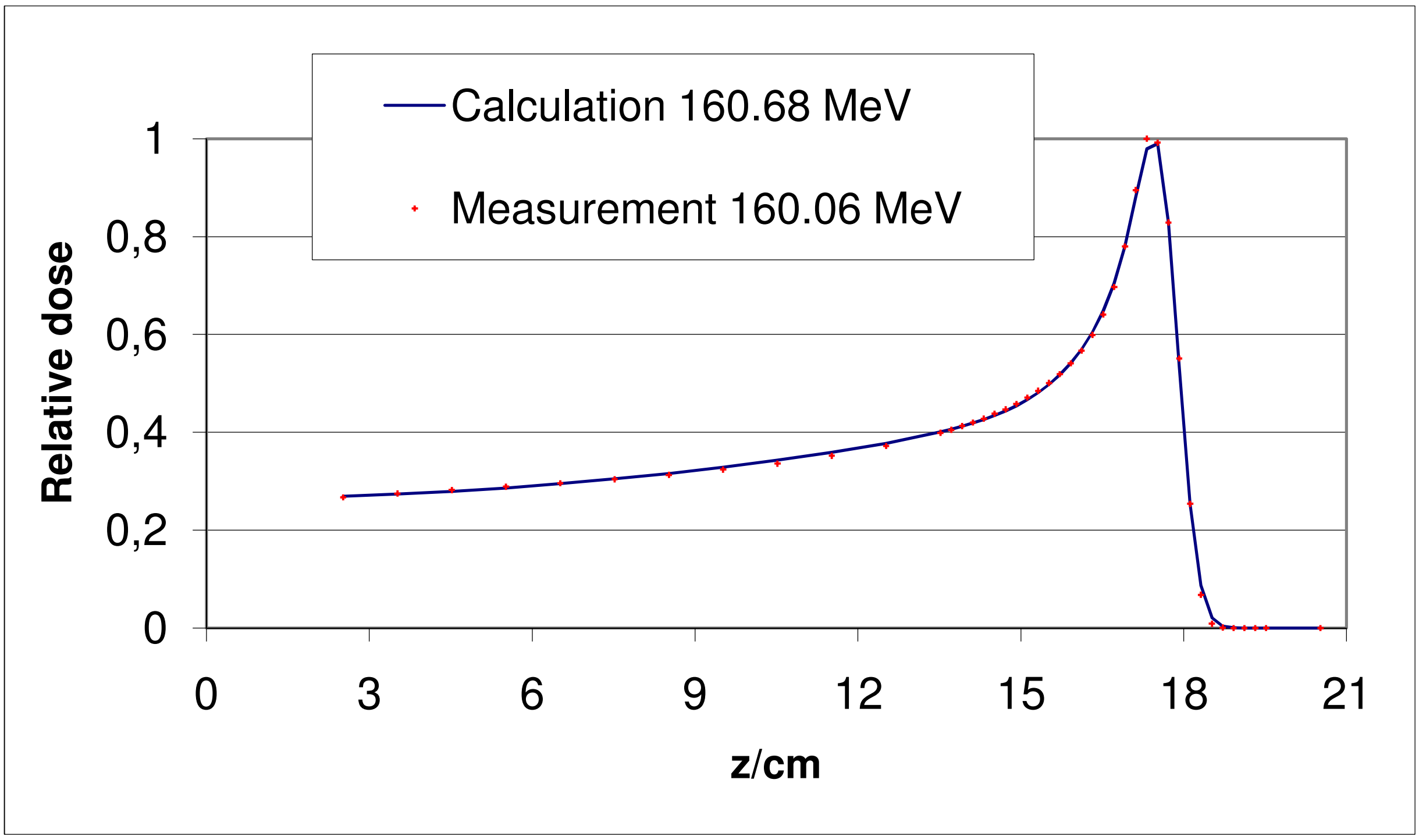

**Fig. 26:** 160 MeV – PSI (discrete points: measurements): dev=0.16 %, $\tau_{in}$ = 0.231 cm.

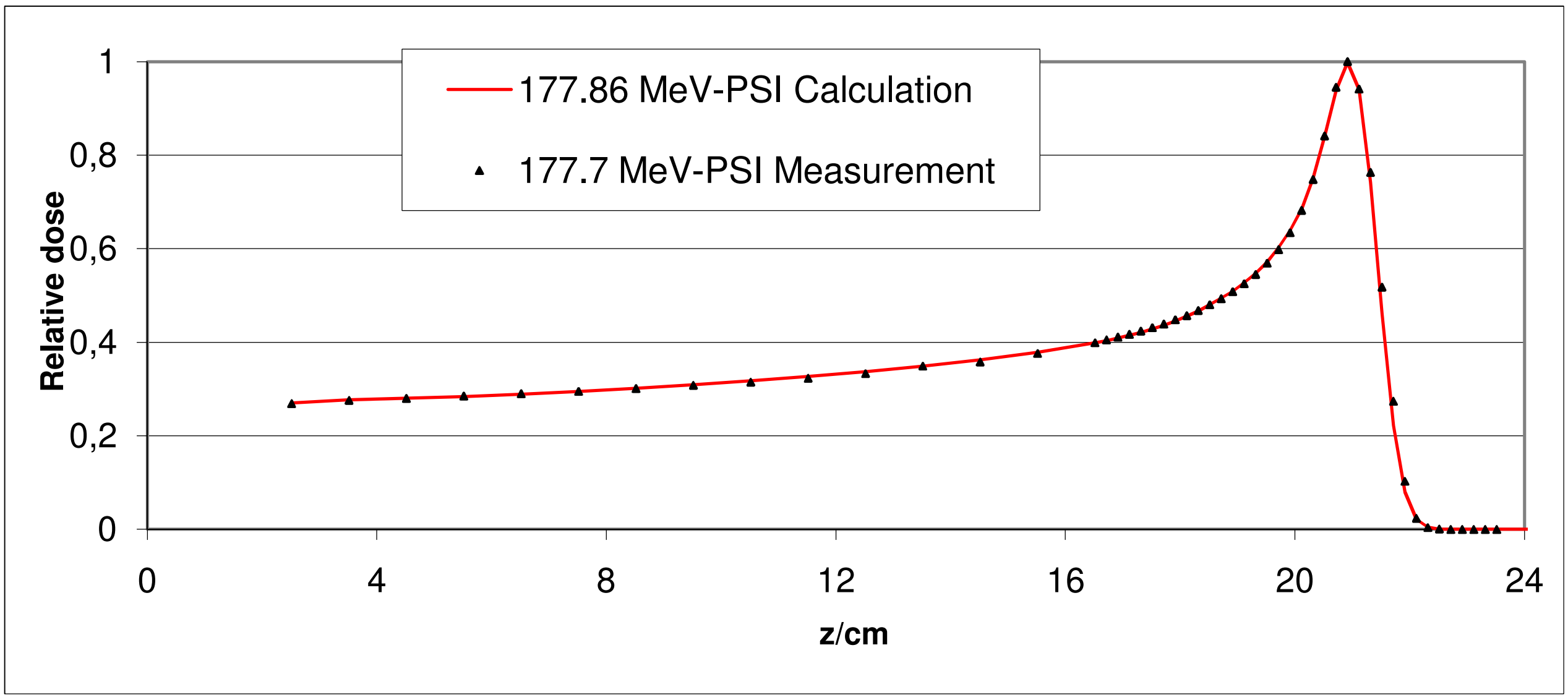

**Fig. 27:** 177 MeV – PSI (discrete points: measurements): dev=0.16 %, $\tau_{in}$ = 0.347 cm.

### *2.1.3 Model M3*

Since the model M3 represents an accelerated algorithm of the version M2, it has been implemented in Eclipse. The fitting of the model to measured pristine Bragg peaks is one central aspect, which we now consider.

We have used measured pristine Bragg curves to test our model. The depth-dose measurements were performed differently, depending on the delivery technique of the treatment machine. Double-scattering pristine Bragg curves are measured with a thimble ionization chamber. In order to compare those curves to our model, we correct the measured pristine Bragg curve for SAD effects by using the $1/r^2$–law and shift the measurement depth by the water-equivalent thickness of the nozzle (NeT). The available NeT values are provided by the machine manufacturer and always lead to a good agreement (better than 1 MeV) between the energy obtained from our fits and the nominal energy provided by the manufacturers. In the case of uniform and modulated scanning beams, the pristine Bragg peak is typically measured with a large parallel plate chamber. Only the shift in depth with the water-equivalent thickness of the nozzle applies here.

When fitting the beamlet model to a measured depth-dose curve, we only allow a variation of the following parameters:

- The spectrum of the initial beam – the main impact of the beamline properties on the shape of the depth-dose curve (see also Fig. 33).
- A normalization factor, which allows the conversion of the calculated depth dose from MeV/cm to the measured unit of Gy/MU.
- The energy of the initial beam. The energy is fitted, and usually agrees to better than 1 MeV with the energy claimed by the machine manufacturer.

These three parameters have the largest impact on the shape and on the absolute scaling of the depth-dose curve. Minor corrections are possible by allowing the following parameters to vary:

- The fraction of secondary protons reaching the water phantom. We assume, that a certain percentage of the secondary protons are lost along the beamline, due to the fact that they are scattered broader than the primary component and, therefore, hit the primary collimator. The fraction of the secondary protons is typically fixed to 1 for scanning beamlines (as in Figs. 28 and 29).
- The Landau parameter ($C_{Lan1}$): There is some dependence of the amount of the Landau correction on the beamline. However, its impact is small (see also Figs. 28 and 29).

The depth-dose model is fitted in a two-step approach to the measured pristine Bragg curves after their processing as explained above. In a first round of fitting, the free parameters are the energy at nozzle entrance ($E_0$), the initial range spectrum $\tau_{in}$ and a normalization factor. Optionally, we allow fitting of the contribution of secondary protons. The use of this option makes sense whenever a considerable fraction of secondary protons has not been detected, as they might have been stopped along the beamline. This condition typically applies to scattering beamlines. In the second round of fitting, we can fit $C_{Lan1}$ freely while allowing only a 2 % variation in the energy, the range spectrum, and the normalization, as well as 50

% variation of the contribution of secondary protons. Some results of the fitting of our model to measured pristine Bragg curves are shown in Fig. 28 (modulated scanning, high-energy beam from an Accel machine), Fig. 29 (modulated scanning, low-energy beam from an Accel machine), and Fig. 30 (double scattering, high-energy beam from an IBA machine). The match between calculation and model is excellent in all cases.

It has to be pointed out that the term $I_{Lan1}$ depends on the residual range of the beam when entering the medium. In scattering techniques, there is always a considerable amount of absorption through the elements of the nozzle. This means that $R_{CSDA}$ must be replaced by $R_{CSDA}(E_0)$ – NeT. This shift in the position of the Landau correction can be seen in Fig. 28, where the zero position of the depth axis corresponds to the entry position of the beam into the nozzle. In order to compare measurements and calculations, one must shift the calculated values taking the NeT into account. A comparison between measurements and the calculated curves may be performed only after this range shift is taken into account.

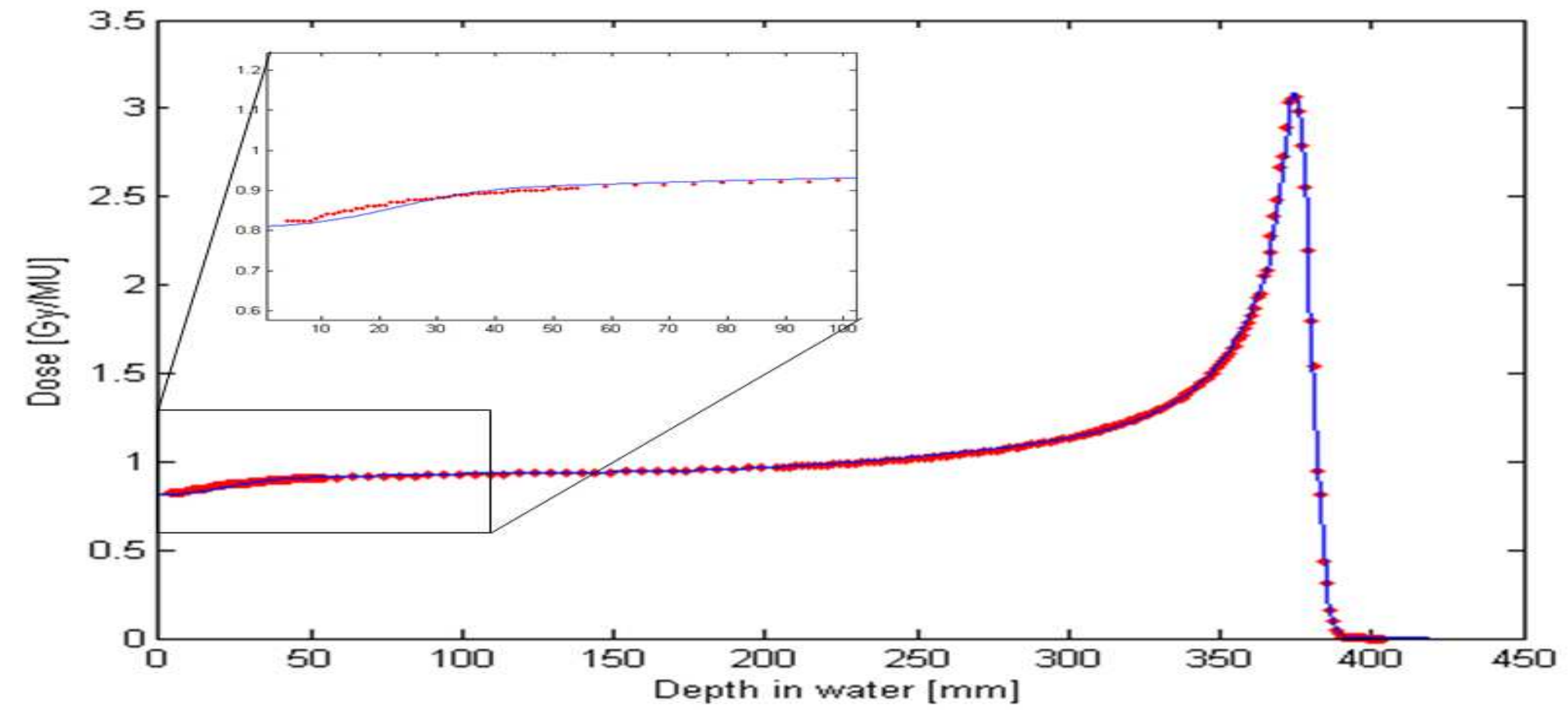

**Fig. 28:** Comparison between a measured and a calculated pristine Bragg peak. The dose is measured with a large parallel plate chamber across a single 250-MeV pencil beam delivered by the Accel machine. The buildup is most visible for high-energy beams without absorbing material in the beamline; it is well described by the model.

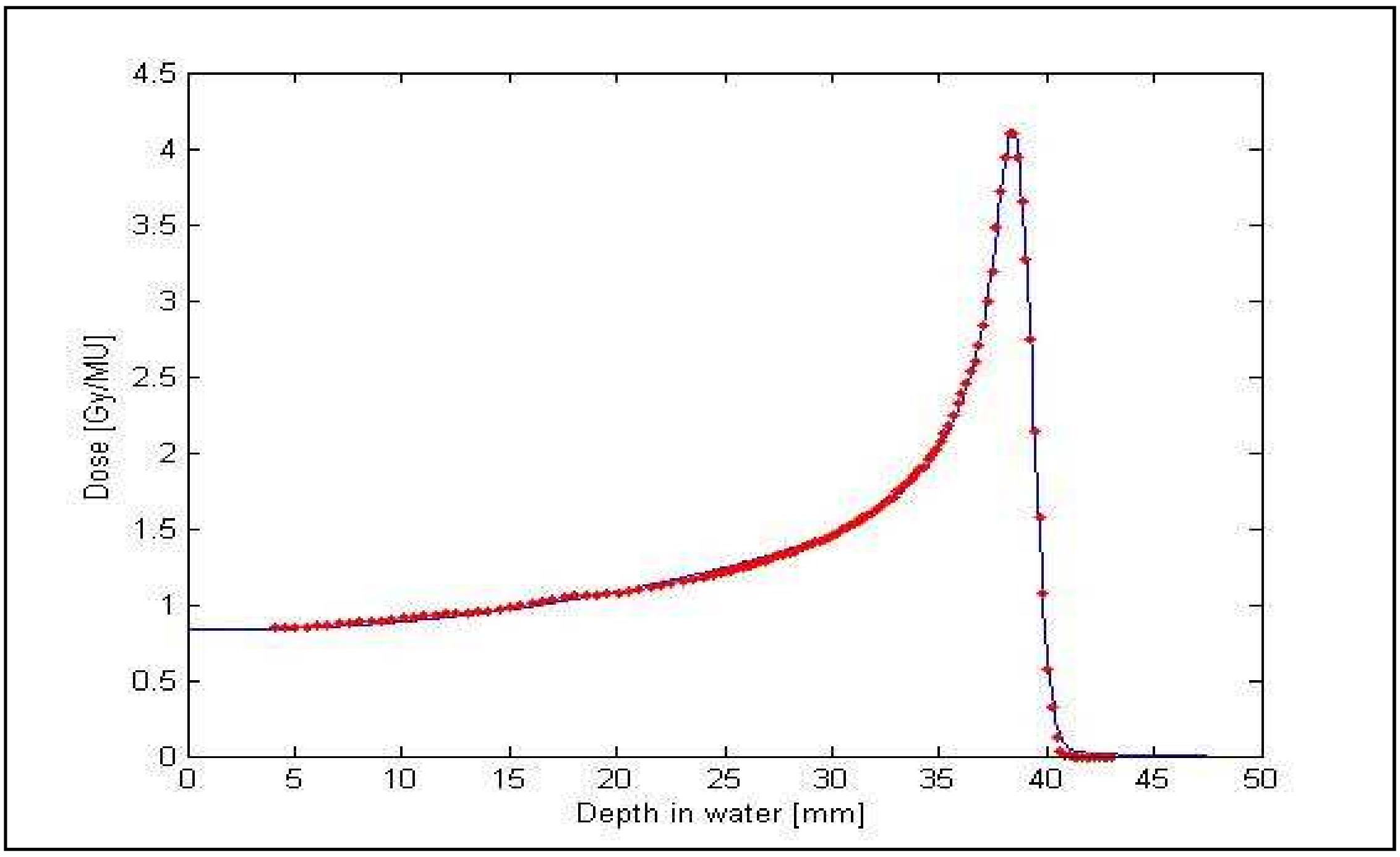

**Fig. 29:** Comparison between a measured and a calculated pristine Bragg peak. The dose is measured with a large parallel plate chamber across a single 68-MeV pencil beam delivered by the Accel machine.

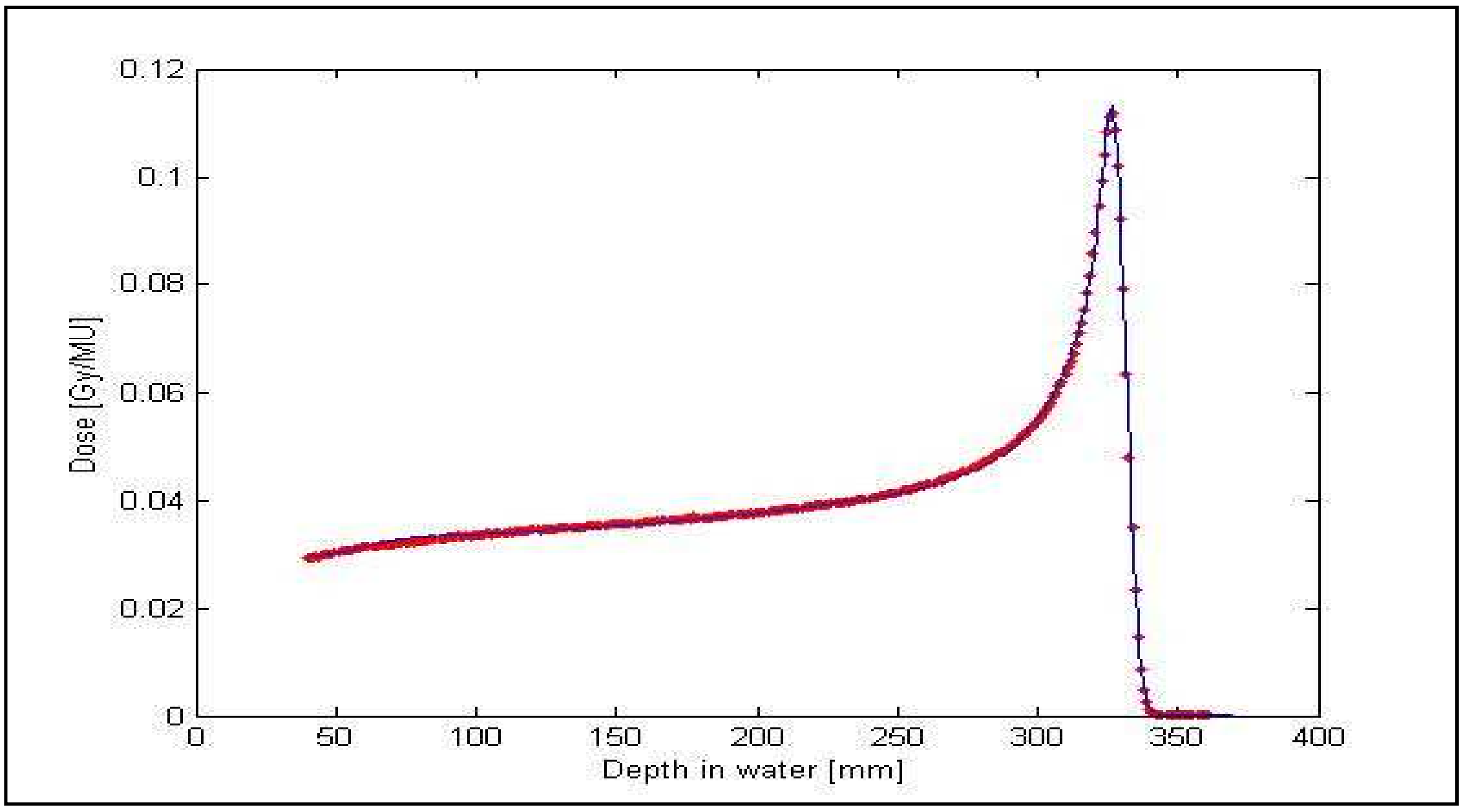

**Fig. 30:** Comparison between a measured and a calculated pristine Bragg peak for the IBA double-scattering beamline at the Proton Therapy Center of the National Cancer Center (NCC), South Korea. The initial beam energy is 230 MeV. The measured data points are corrected for SAD effects and are shifted in depth by the nozzle-equivalent thickness of 40.7 mm. Note that the buildup effect is also clearly visible in this scattering beamline and should not be neglected at higher energies.

## 2.2 Some common features (lateral scattering and parameter determination) applicable to each of the three models

Since lateral scattering and the determination of the parameters $\tau_{in}$ and $C_{Lan1}$ (the proposed calculation procedure might not provide sufficiently accurate results) are issues for all three models (M1, M2, and M3), it is necessary to introduce a per-case (for each machine, separately) procedure to obtain their values.

The experimental verification of the lateral-scattering calculation is very challenging due to the small contributions of the large-angle scattered primary protons and of the secondary protons. Furthermore, there are other contributions to the lateral distribution of protons (initial phase space of a scanned beam or effective-source size and block-scattering effects in a scattered beam) which affect the measurement. The following plots show comparisons between calculated and fitted (i.e., by using our models) beam width.

Fig. 31 shows how the beam width is reduced when passing from a single-Gaussian model (dashed line, x) to a model with higher-order contributions (thick solid line and o for the main Gaussian).

Fig. 32 shows the magnitude of the contributions from higher-order scattering. Despite the fact that these contributions may be negligible in the situation treated here, they are nevertheless expected to play an important role at higher energies, as well as when a range shifter is present in the planning. Especially in the latter case, an impact is also expected on the estimation of the MU factor, see Pedroni et al. (2005).

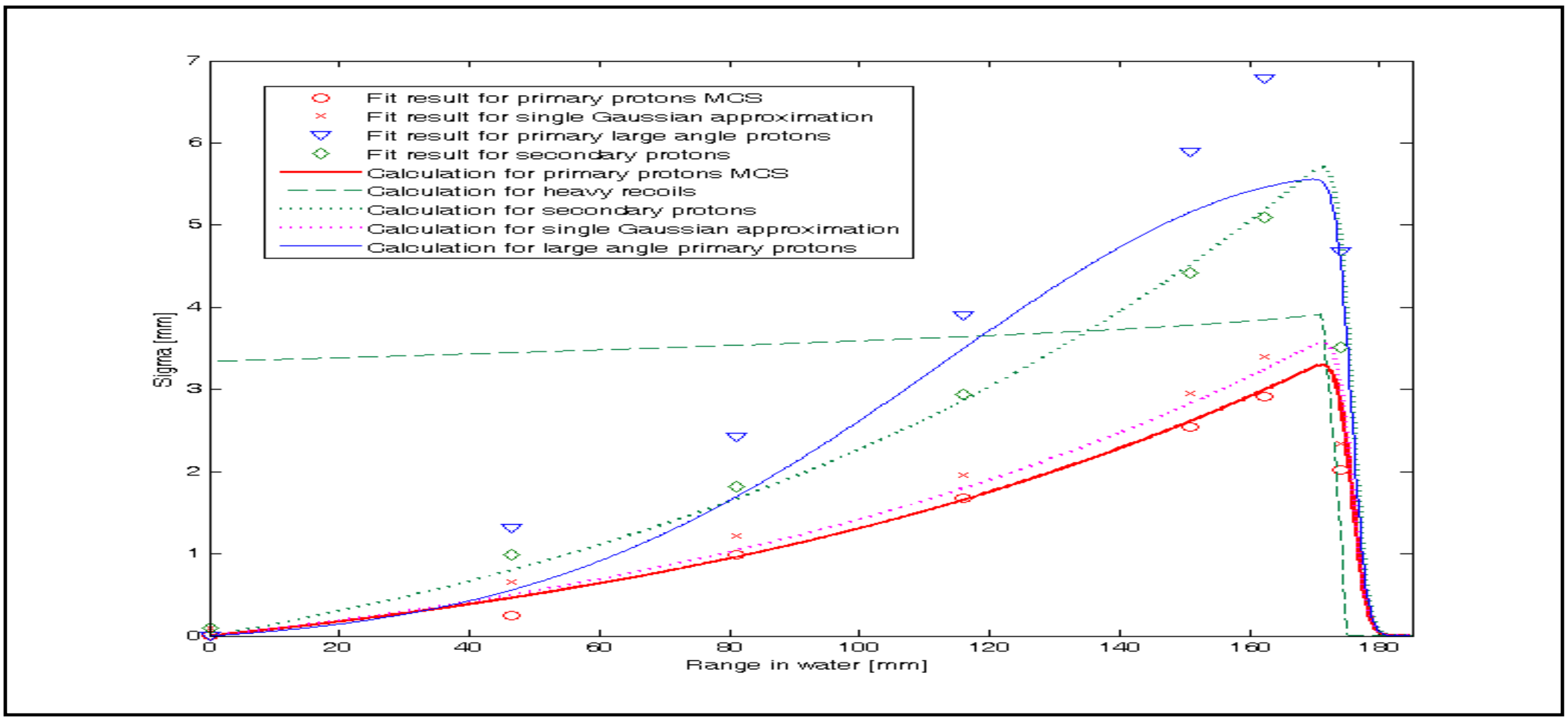

**Fig. 31:** Comparison of the theoretical beam width for a 160-MeV scanning beam (Accel) and the results obtained by fitting a single Gaussian (x) or our theory (open symbols) to the measured profiles. Due to the smallness of the majority of the contributions, the fit results are extremely sensitive to the starting values. Furthermore, limits had to be defined in order to keep the secondary-proton and the large-angle contributions to reasonable values. The calculated width for a single Gaussian was obtained from Eq. (149) by setting $C_0$ to 1. The quantity shown in this figure is defined as $\tau_{lat}/\sqrt{2}$.

Inspection of the results for the fitted parameter leads to the conclusion that the difference between our fitted-energy values and those claimed by the machine manufacturer is kept below 1 MeV.

Interesting observations can be made by plotting the initial range spectrum $\square_{in}$ as a function of the nominal energy for a number of different machines and techniques (Fig. 33). The Hitachi machine is a synchrotron, which produces naturally a narrow Bragg peak. The other machines are cyclotrons; the lower energies are obtained by degrading the initial beam. This process creates a broad energy spectrum, which needs to be narrowed through energy-selection slits. The setting of the energy-selection slits is a compromise between the width of the Bragg peak and the beam intensity. It seems that this compromise leads to similar width values for the three machines by Accel, PSI, and IBA. The width of the PSI beam is somewhat broader, probably due to the fact that the data from the PSI beamline originates from the 600-MeV physics-research accelerator and, therefore, has to be degraded more than the beams of the other machines.

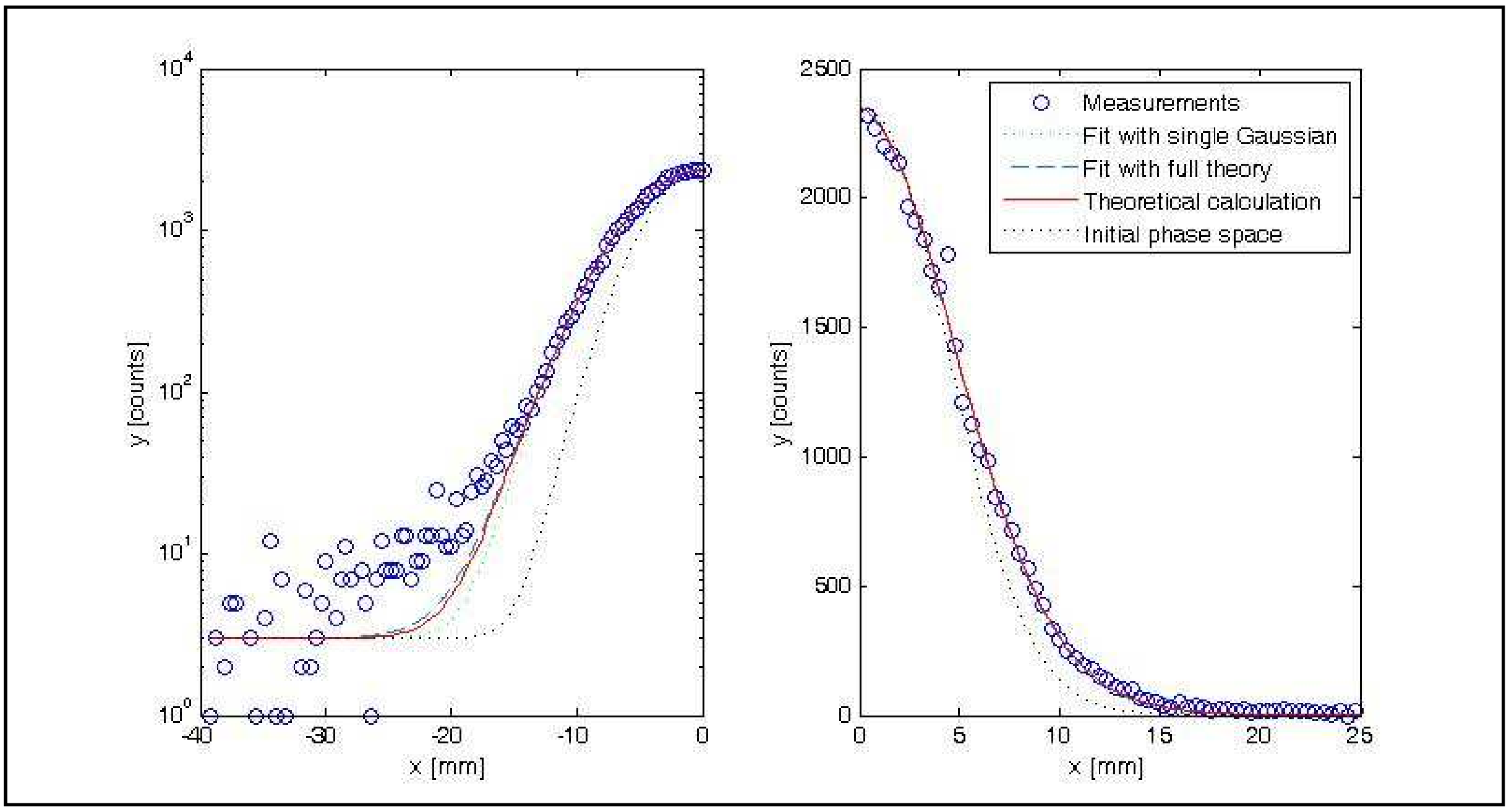

**Fig. 32**: Logarithmic and linear plots of a spot-size measurement of a 160-MeV beam at the water-equivalent depth of 151mm for an Accel machine. The black dotted line shows the initial phase space, which is subtracted quadratically from the other contributions.

The data points from the Hitachi machine show that the range spectrum seems to increase with the amount of high-density material in the beamline (i.e., the large-field configuration (□) has more Pb than the medium field size (◊); the scanning-beam line (∇) has no extra material). This demonstrates convincingly that the range straggling in higher-density materials is larger than in the same (water-equivalent) amount of low-density materials. Since our model does not distinguish between different compositions of the beamline (i.e., $\tau_{straggle}$ is not changed), the additional straggling component is accounted for by the fitting procedure in beam configuration through an increase in $\tau_{in}$. We found that the resulting range spectra, depending on the nominal energy, can be fitted well by a third-degree polynomial for both machine types, i.e., synchrotron and cyclotron; this enables us to model the pristine Bragg curves for intermediate (not configured) energies.

The fitting of the Landau parameter $C_{Lan1}$ for double scattering (Fig. 34) shows quite a bit of scatter between different machines and also for different hardware configurations of the beamline (often called 'options') within the same machine. However, the magnitude of $C_{Lan1}$ is the same for all machines and there is a trend towards a minimum at residual-range values of 100-150 mm. The effect of these variations of $C_{Lan1}$ on the total depth-dose calculation is very small and we normally use the nominal $C_{Lan1}$ according to Eq. (136) for scattering- and uniform-scanning dose calculations.

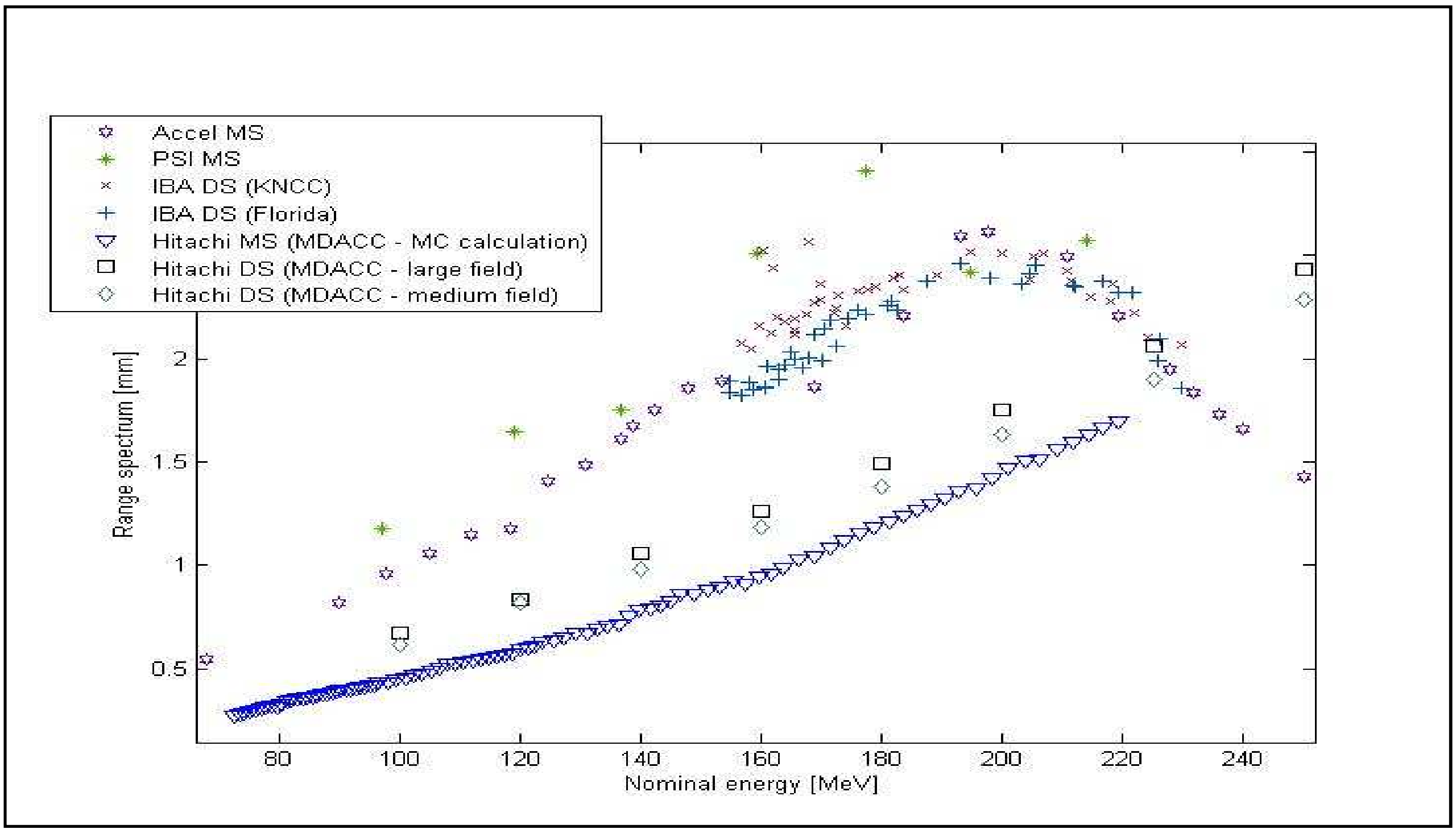

**Fig. 33:** The range spectra ($\tau_{in}$) obtained from our depth-dose model for a variety of treatment machines and techniques (MS: modulated scanning, DS: double scattering). It is well known that a synchrotron (Hitachi) usually creates a smaller initial range spectrum than a cyclotron (all other machines). It is interesting to observe that all cyclotrons produce similar range spectra. It has to be commented that, as measured data for MS in the Hitachi machine has not yet been available to us, the data, shown in this figure, has been obtained via a Monte-Carlo simulation of the pristine Bragg curves.

The results for the pristine Bragg measurements from the modulated-scanning beamlines show a much clearer trend for $C_{Lan1}$ as a function of the residual range (Fig. 34). We usually substitute $C_{Lan1}$ from Eq. (136) by another third-degree polynomial with parameters obtained from fits to the data points of each machine – as indicated by the straight lines in Fig. 35.

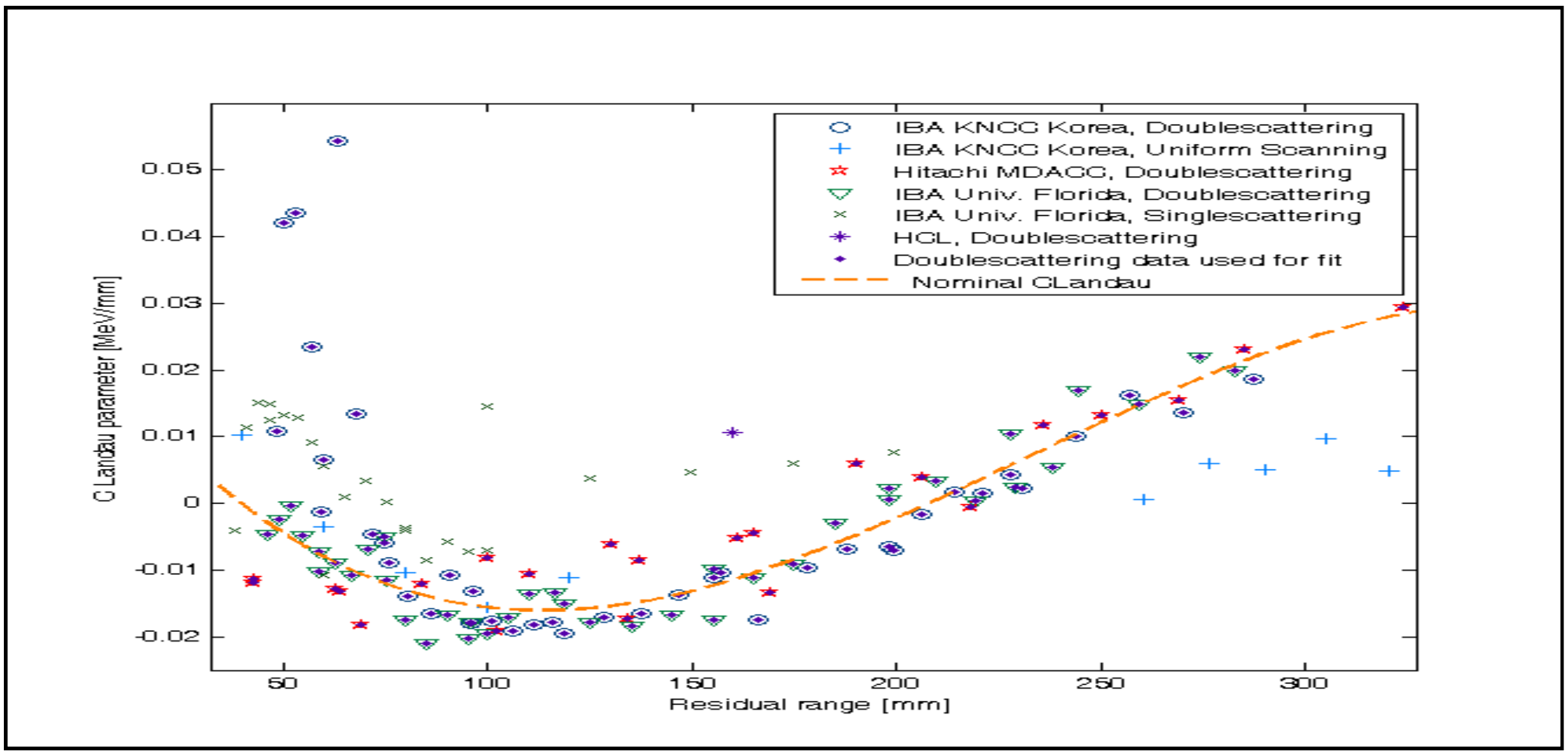

**Fig. 34**: Fit results for $C_{Lan1}$ for a number of different double-scattering beamlines, as well as for one uniform-scanning beamline. The line corresponds to $C_{Lan1}$ according to Eq. (136); it has been obtained from a polynomial fit to the data sets of Hitachi MDACC, IBA NCC (double scattering) and IBA Florida. The other data sets are plotted only for comparison. Due to a restriction to horizontal beam geometry on the other double-scattering beamlines, the pristine Bragg measurements could not be measured up to the surface of the water phantom; this implies that the most-relevant data points for the determination of $C_{Lan1}$ are missing.

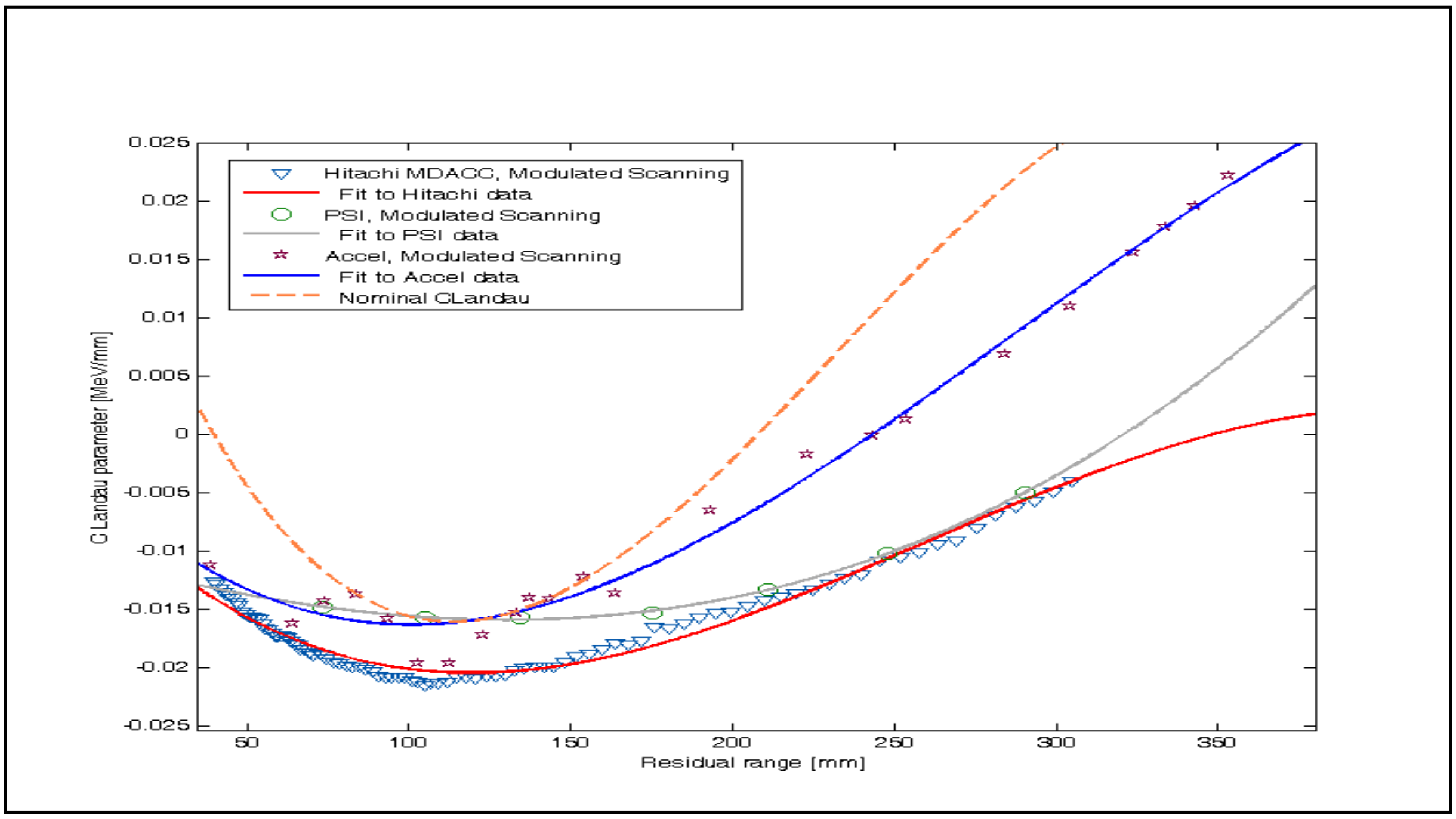

**Fig. 35**: The value of $C_{Lan1}$ for different modulated-scanning beamlines; the Hitachi MDACC values originate from Monte-Carlo calculated pristine Bragg peak. The fitted lines are third-degree polynomials; they replace the nominal $C_{Lan1}$ from Eq. (136) in the dose calculation for the respective machines.

In a recent paper, Hollmark et al. (2004) point out that, in the domain of the Bragg peak, the Gaussian solution (one Gaussian) is sufficient for both longitudinal energy straggling and lateral scatter. The corresponding arguments are based on the transition of the more general Boltzmann transport theory to the Fermi-Eyges theory in the low-energy limit, see Eyges (1948). However, it appears that the conclusion is only partially true, since the history of the proton track has an influence on the behavior in the low-energy domain (see above results referring to the Landau tails of the energy transfer), and, according to the results of Saitoh (2001) and Matsuura and Saitoh (2006), all types of transport equations also have more general solutions than given by one Gaussian in the diffusion limit. Yet the linear combination of two Gaussians with different half-widths, as used in the present study, is not a solution of the Fermi-Eyges theory, but a corresponding one of a nonlocal Boltzmann equation. This is an integro-differential equation with different transition probabilities for the local and nonlocal part (long-range interaction). In the diffusion limit, the nonlocal part provides, at least, one additional Gaussian (e.g., see the adaptation of multiple-scatter theory by two Gaussians).

We have developed analytical models for the depth-dose distribution of a proton beam – the pristine Bragg peak. The models depend on a few beamline-specific parameters (nominal energy, energy/range spread, Landau parameter, contribution of secondary protons), which need to be obtained by fitting the model to the measured pristine Bragg curves. We have shown that the models can reproduce the pristine Bragg curves for different accelerator and beamline designs. An interpolation of the key parameters permits the accurate calculation of any intermediate pristine Bragg peaks; this is particularly important for delivery machines which feature an analog energy tuning. The lateral distribution of the protons is modeled separately for primary and secondary protons; in order to describe better the large-angle scattering, the lateral distribution of the primary protons is modeled by a sum of two Gaussians. However, it has been shown by Pedroni et al. (2005) and Kusano et al. (2007) that a correct modeling of the large-angle scattered primary protons and the scattering of the secondary protons has an impact on the determination of the MU factor.

## 2.3 Collimator effects

To shape the beam so that it matches the characteristics of the specific treatment in radiation therapy (thus achieving the delivery of the prescribed dose to the target (tumor) and maximal protection of the surrounding healthy tissue and vital organs), beam-limiting and beam-shaping devices (BL/BSDs) are routinely used. Generally speaking, the beam is first restricted (in size) by the primary collimator, a beam-limiting device giving it a rectangular shape. The beam may subsequently encounter the multi-leaf collimator (MLC), which may be static or dynamic (i.e., undergoing software-controlled motion during the treatment session). More frequently than not, the desirable beam shaping is achieved by inserting a metallic piece (with the appropriate aperture and thickness) into the beamline, directly in front of the patient; this last beam-shaping device is called a patient collimator or simply a block. Being positioned close to the patient, the block achieves efficient fall-off of the dose (sharp penumbra) outside the target area. The simultaneous use of MLC and block is not common.

The presence of BL/BSDs in treatment plans induces three types of physical effects:

a) Confinement of the beam to the area corresponding to full transmission (i.e., the aperture of the device).

b) Effects associated with the nonzero thickness of the device (geometrical effects).

c) Effects relating to the scattering of the beam off the material of the device.

Type-(a) effects (direct blocking of the beam) are dominant and have always been taken into account. The standard way to do this is by reducing the BL/BSD into a 2D object (i.e., by disregarding its thickness) and assuming no transmission of the beam outside its aperture. Type-(b) and type-(c) effects induce corrections which, albeit at a few-percent level of the prescribed dose, may represent a sizable fraction of the *local* dose; due to their complexity and to time restrictions during the planning phase, these corrections had (so far) been omitted in clinical applications.

The detailed description of a method, which may be used for the determination and the application of the type-(b) and type-(c) corrections to the fluence of the pristine beam (protons which do not hit the BL/BSD), may be found in Matsinos (2008). It has been shown that the application of these corrections is greatly facilitated by decomposing the effects of two-dimensional objects into one-dimensional, easily-calculable contributions (via the concept of *miniblocks*).

Given the time restrictions during the planning, the derivation of the scattering corrections necessitated the introduction of a two-step approach. The first step occurs at the beam-configuration phase. At first, the value of the only parameter of the model ($\lambda$), employed in the description of the beamline characteristics, is extracted from half-block fluence measurements. A number of Monte-Carlo runs follow, the output of which consists of the parameters pertaining to convenient parameterizations of the fluence contributions of the scattered protons. These runs take account of the variability in the block material and thickness, incident energy, and NeT in all the options (combinations of the hardware components of the beamline, leading to ranges of available energies and of NeTs, as well as imposing restrictions on the field size) for which a proton-treatment machine is configured. To enable the easy use of the Monte-Carlo results, the output is put in the form of expansion parameters in two quantities which are involved in the description of the scattering effects. The scattering corrections for all the blocks in a particular plan are determined from the results, obtained at beam-configuration phase, via simple interpolations.

The verification of the method should involve the reproduction of dedicated dose measurements. At present, given the lack of such measurements, the only possibility for verification rested on re-using the half-block fluence measurements, formerly analyzed to extract the $\lambda$ value; this is a valid option because parts of the input data had been removed from the database to suppress the (present in the measurements) block-scattering contributions. The goodness of the reproduction of the measurements was investigated on the basis of the $\chi^2$ function; we concluded that the inclusion of the scattering effects leads to substantial improvement (e.g., see Figs. 36 and 37; the measurements shown have been obtained at the NCC).

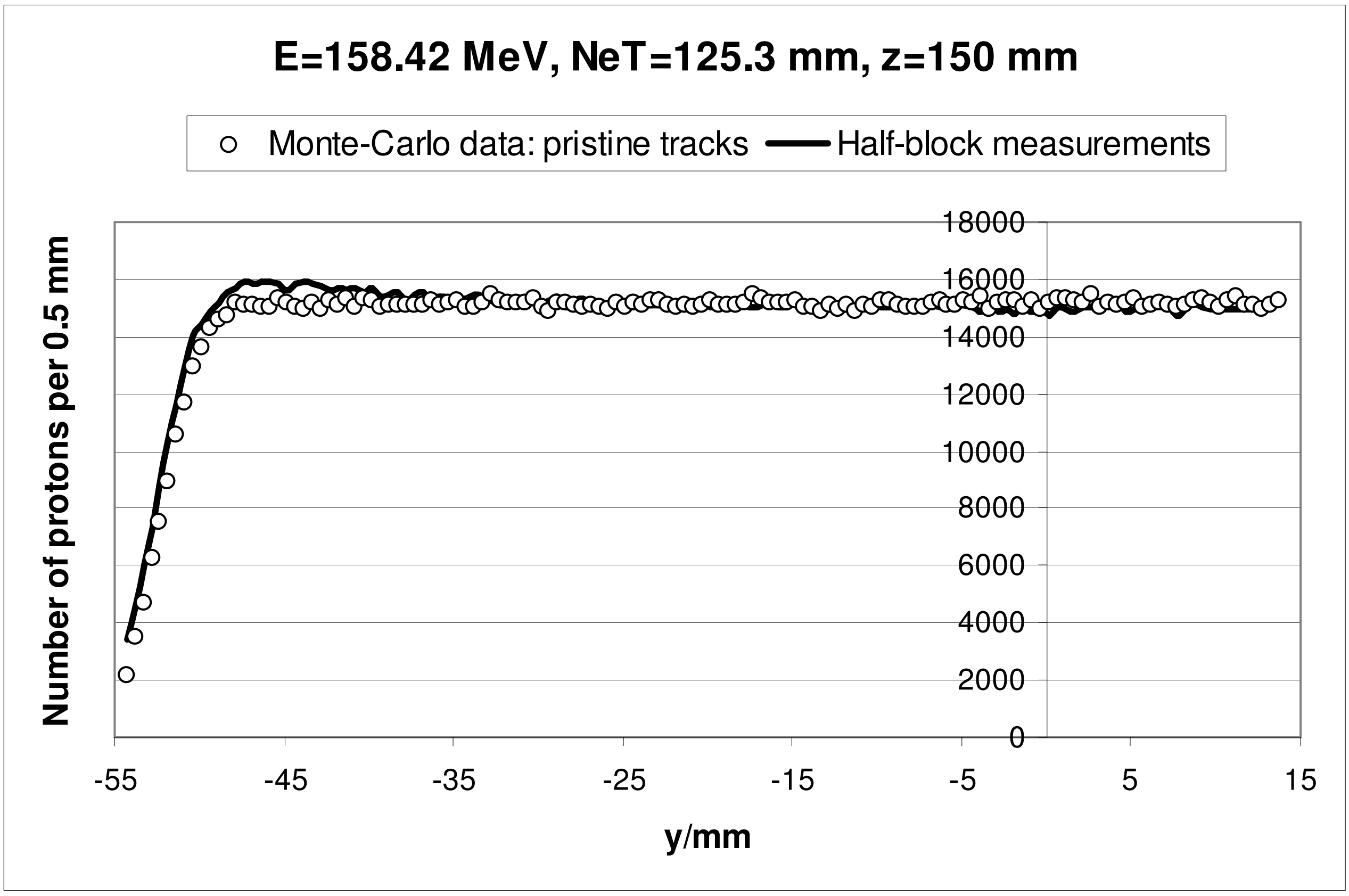

**Fig. 36:** The lateral fluence measurements (continuous line) corresponding to one energy-NeT combination of one option of the NCC machine, taken 100 mm away from the downstream face of the block. The Monte-Carlo data shown correspond only to the pristine-beam fluence obtained at the same incident-energy, NeT, and depth values; the measurements have been scaled up by a factor which is equal to the ratio of the median values (of the two distributions), estimated over the fluence plateau.

The current version of Eclipse was extensively modified to include the derivation (in beam configuration) and the application (in planning) of both block-relating corrections. The method above was then applied to one plan involving a simple water phantom; the different contributions from the type-(b) and type-(c) effects have been separately presented and compared. It was found that these effects amount to a few percent of the prescribed dose and are significant in the area neighboring the border of the block.

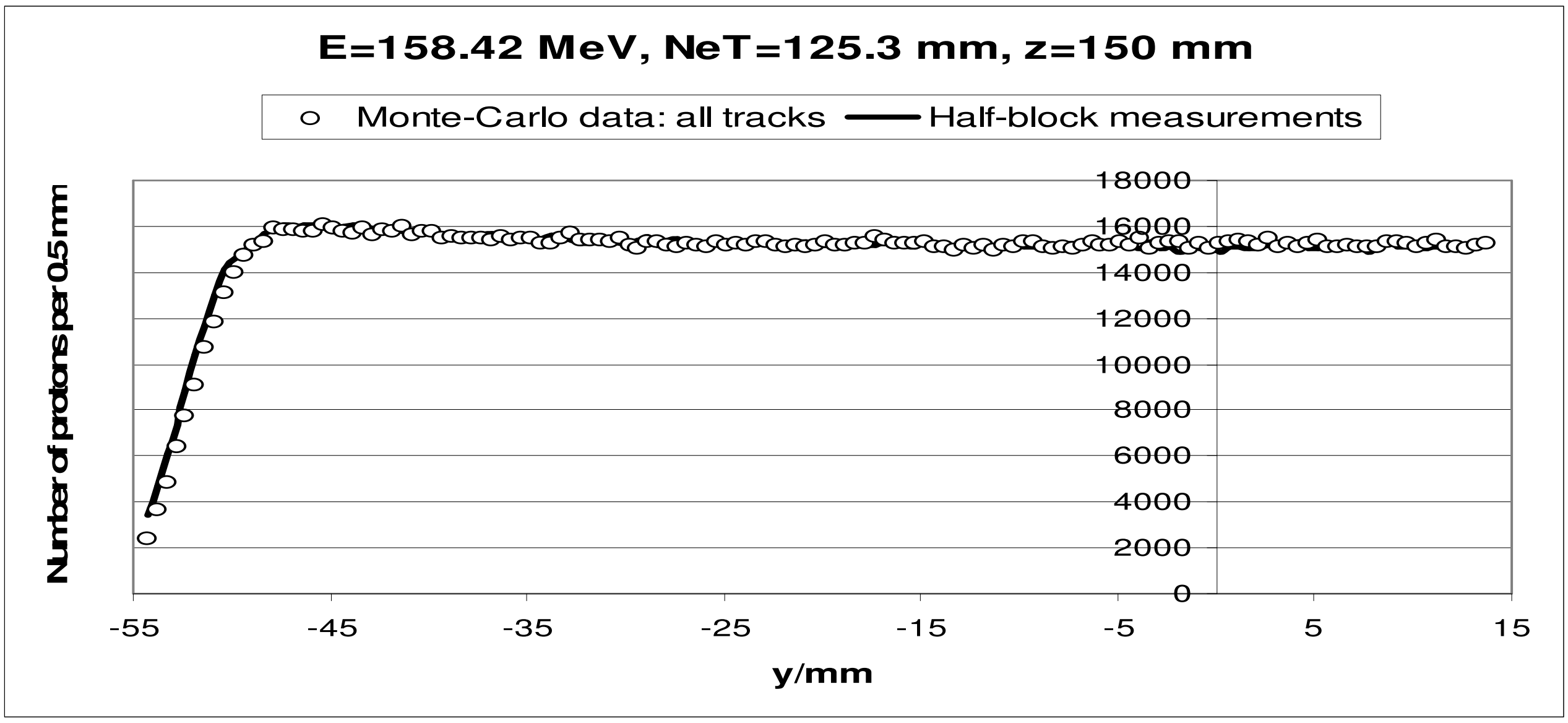

**Fig. 37:** The lateral fluence measurements (continuous line) corresponding to one energy-NeT combination of one option of the NCC machine, taken 100 mm away from the downstream face of the block. The Monte-Carlo data shown correspond to the total (pristine-beam plus scattered-protons) fluence obtained at the same incident-energy, NeT, and depth values; the measurements have been scaled up by a factor which is equal to the ratio of the median values (of the two distributions), estimated over the fluence plateau.

## *3. Summary*

This review work provides broad information on various aspects of therapeutic protons:

1. Energy-range relations (nonrelativistic and relativistic extension).

2. Integration of the Bethe-Bloch equation and calculation procedures based on this integration.

3. Theoretical aspects of the energy/range straggling and its role on the beamline.

4. Lateral scatter based on the Molière theory of multiple scatter.

5. Irradiation techniques for protons, the connection to the beamline, and the role of the collimator scatter in experimentally verified Bragg curves.

6. Some aspects on the theoretical tools of importance in this work (theory of convolutions and deconvolutions and nuclear physics).

*Point 1*

With the help of a generalized nonrelativistic Langevin equation, we have derived the famous Bragg-Kleeman rule. An extension to a relativistic generalization provides deviations for initial proton energies $E_0 >$ 50 MeV. By inversion of the relation, the residual energy E(z) can be calculated as a function of $R_{CSDA} - z$.

The long debate in literature with respect to the power p, adequate to reflect most accurate the range $R_{CSDA}$, has been answered by the following verification: The power p is in both cases (relativistic and nonrelativistic) dependent on the initial energy $E_0$. The results are the foundation of the calculation model M1 of the stopping power.

*Point 2*

The integration of the BBE provided two ways of calculating the range $R_{CSDA}$; one of these could be inverted to yield the residual energy E(z) as a function of $R_{CSDA} - z$. A modification of the formula for the stopping-power calculation gave an accelerated algorithm with only slight reduction of the precision. Together with the energy/range straggling, the fluence decrease of primary protons, and the increase of secondary protons, the calculation models M2 and M3 have been developed. The calculation speed of M2 and M3 is faster than that of the model M1.

*Point 3*

A unique theory of energy/range straggling has been developed, based on nonrelativistic and relativistic considerations of the convolution problem. The latter case leads to Landau tails and has the implication of buildup effects in the region of the initial plateau.

*Point 4*

The presented approach goes beyond the Highland approximation by assuming a linear combination of two Gaussian kernels in the description of the lateral scatter of protons. The lateral scatter of secondary protons is modeled by one Gaussian, which is sufficient due to their minor importance. The whole procedure does not yet completely fulfill the requirements of Molière's multiple-scatter theory with regard to long-range tails. However, this is mainly a question of saving calculation time.

*Point 5*

Currently, as far as proton therapy is concerned, four dose-delivery techniques are in use: single-scattering, double-scattering, uniform-scanning (formerly known as wobbling), and modulated-scanning (formerly simply known as scanning). In the modulated-scanning technique, magnets deflect a narrow beam onto a sequence of pre-established points (spots) on the patient (for pre-determined optimal times), thus 'scanning' the (cross section of the) region of interest. Uniform scanning involves the spread-out of the beam using fast magnetic switching. The broadening of the beam in the single-scattering technique is achieved by one scatterer, made of a high-Z material and placed close to the entrance of the nozzle. Currently, the most 'popular' technique involves a double-scattering system.

In a double-scattering system, a second scatterer is placed downstream of the first scatterer in order to achieve efficient broadening of the beam; studies of the effects of the second scatterer may be found in the literature. The second scatterer is usually made of two materials: a high-Z (such as Pb) material at the center (i.e., close to the central beam axis), surrounded by a low-Z (such as Al, lexan, etc.) material (which is frequently, but not necessarily, shaped as a concentric ring). The arrangement produces more scattering at the center than the periphery, leading (after sophisticated fine-tuning) to the creation of a broad flat field at isocenter.

The method to derive and apply the collimator effects in proton planning (see Matsinos (2008)) was originally developed for the double-scattering technique; however, it is also applicable in single scattering and uniform scanning. Given that the scattering effects involve broad fields, they have no bearing on modulated scanning.

*Point 6*

Appendix provides information about the calculation of the total nuclear cross section of some nuclei which are of interest in radiotherapy. An analysis of the extended nuclear shell theory has been carried out to elucidate the different contributions of the nuclear cross section: potential scatter, resonance scatter, and nuclear reaction channels. The results gave a clear indication that theoretical models represent a useful contribution in the interpretation of experimental data and implementations of Monte-Carlo codes.

## Acknowledgments

We would like to express our gratitude to many researchers, who have made their experimental data available to us. Above all, we would like to thank Barbara Schaffner for having performed measurements at the Harvard cyclotron and worked out many contributions with regard to the Eclipse implementation of the model M3. We would also like to thank Jürgen Heese (Varian-Accel), Tony Lomax (PSI), Daniel Morf, (Varian/Baden), Nicolas Denef, Michel Closset, Victor Breev and Gilles Mathot (IBA), Se Byeong Lee, Dongho Shin, Jungwon Kwak and Dongwook Kim (NCC, Seoul), Daniel Yeung and Roelf Slopsema (University of Florida), Martin Bues and George Ciangaru (MD Anderson), and Wen Chien (Hsi). Without their diligent work, the verification of the models M1, M2, M3, and of the lateral scatter would have been an impossible task.

## *Appendix:* **Aspects** *on nuclear physics (theory and approximations)*

### *A1. Fitting procedure for the determination of $E_{Th}$ and $Q^{tot}$*

The calculation of the total nuclear cross section requires some information acquired by fitting the Los-Alamos data, results of extended nuclear shell theory, and empirical rules. Let us at first consider Eqs. (53 – 54) to compute $E_{Th}$ and k. We assume an isoscalar nucleus, i.e., one in which the numbers of protons and neutrons are equal ($A_N = 2 \cdot Z$). This assumption holds for almost all light nuclei. For nuclei with spherical symmetry, the nuclear radius $R_{strong}$ is given by:

$$R_{strong} = 1.2 \cdot 10^{-13} \cdot \sqrt[3]{A_N} \quad (in \quad cm) \quad (165)$$

Thus for $r > R_{strong}$, the contribution of strong interaction is negligible. The nuclear shell theory with oscillator potential gives for $r = R_{strong}$:

$$(-U_0 + \tfrac{1}{2} \cdot M \cdot A_N \cdot \omega_0^2 \cdot R_{strong}^2) \cdot C_F =$$
$$(Z \cdot e_0^2 / R_{strong} + Z \cdot (Z-1) \cdot e_0^2 / R_{strong}) / F(A_N, Z) \quad (166)$$

The first term on the right-hand side of this equation represents the Coulomb repulsion of an incoming proton; the second one is the mutual Coulomb repulsion of Z protons in the nucleus. $U_0$ is the depth of the potential and is put equal to $A_N \cdot E_B \cdot R_{strong}^2$; $C_F$ is a proportionality factor and $E_B$ the binding energy per nucleon. $E_B$ is equal to 8 MeV, if $A_N \geq 12$, smaller for $A_N < 12$. This fact is the reason that we have to calculate $E_{Th}$ with Eq. (54), instead of Eq. (53). We are able to rescale $\omega_0$ (if $E_B$ is constant) in such a way that $U_0$ vanishes. A least-squares fit of all available Los-Alamos data yielded k = 1.659, instead of k = 1 + 2/3. This might result from crude assumptions in the creation of our fitting model: we have assumed $M_{proton} = M_{neutron}$ and, furthermore, neglected the spin-orbit coupling. The equation above may also provide the means for the calculation of the rescaled $\omega_0$. If the number of neutrons is slightly different from Z (i.e., $A_N = 2 \cdot Z + \varepsilon_N$ and $\varepsilon_N << Z$), then a correction term is required, which is already taken into account in the form factor function $F(Z, A_N)$; the rotational symmetry still has to hold:

$$\left.\begin{aligned} & E_{Th} = C_F \cdot Z^{\kappa} \cdot F(Z, A_N) \\ & F = (A_N / 2Z) \cdot \\ & \cdot (a_0 + a_1 / A_N + a_2 / A_N^{p1} + a_3 / A_N^{p2} + a_4 / A_N^{p3} \end{aligned}\right\} (167)$$

**Table 9**: Parameters of the form factor function $F(Z, A_N)$

| $a_0$ | $a_1$ | $a_2$ | $a_3$ | $a_4$ | p1 | p2 | p3 |
|---|---|---|---|---|---|---|---|
| 2.1726 | -335.0440 | 479.5400 | -194.9400 | 11.7125 | 0.76965 | 0.5575 | 0.3405 |

Function $F(Z, A_N)$ represents a form factor function of the proton – nucleus interaction, which becomes important for $Z \geq 6$. We have used the oxygen nucleus as the reference system, since $F(A_N = 16, Z = 8) = 1$. We point out that the model above is not applicable without strong modification to heavy nuclei (e.g., W, Pb, U), where the proton number Z is much smaller than the number of neutrons. These nuclei cannot be characterized by an (approximate) spherical symmetry.

With regard to the determination of the total nuclear cross section $Q^{tot}$, we have borrowed some elements from the collective model of nuclear interactions. First of all, we have to know $Q^{tot}{}_{max}$ required for the calculation of other quantities. We start with the following 'Ansatz', and thereafter we give some explanations:

$$\left.\begin{array}{l} Q^{tot}{}_{\max} = a \cdot A_N + b \cdot A_N{}^{2/3} + c \cdot A_N{}^{1/3} + d \cdot Z^{\kappa} / A_N{}^{1/3} \\ (Q^{tot}{}_{\max} \;\; in \;\; mb) \end{array}\right\} \quad \textbf{(168)}$$

What is the physical interpretation? With the aid of Formula (168), we obtain the following properties:

Term a: Connection of $Q^{tot}{}_{max}$ to the complete volume of the nucleus. It is important in the resonance domain; it includes resonance scatter via nuclear deformations (vibrations) of the whole nucleus, resonance excitations by changing the spin multiplicity (all effects are inelastic), and transformation of a nuclear neutron according the listing 52 (case 1, inelastic).

Term b: Proportional to the area of the geometric cross section. It contains potential scatter (major part, elastic), rotations induced by Coulomb repulsion/strong-interaction attraction (elastic and inelastic), and nuclear reactions by changing the isospin multiplicity (inelastic).

Term c: Proportional to the nuclear radius $R_{strong}$. Excitations by spin-orbit coupling, when the whole nucleus changes its angular momentum, inelastic resonance effect, and elastic spin-spin scatter.

Term d: Proportional to $Z^k/R_{strong}$. Excitation of nuclear vibrations by Coulomb repulsion (resonance effect, inelastic) and elastic scatter.

The asymptotic behavior $Q^{tot}{}_{as}$ of $Q^{tot}$ is given by the relation:

$$Q^{tot}{}_{as} \cong C_{as} \cdot A_N{}^{2/3} \;\; (C_{as} = 85.203266 \; mb) \;\; \textbf{(169)}$$

This connection mainly contains the term b above. We have verified the validity of this property for nuclei up to Zn. The order of magnitude of the term b is a clear indication that elastic potential scatter of the nucleus via the strong interaction is the main contribution of the total nuclear cross section. However, Eq. (219) can also be used for the determination of $Q^{tot}{}_{c}$ and $Q^{tot}{}_{as}$, and only the four coefficients are different. Therefore we write Eq. (174) in a modified form (the parameters are given in Table 10):

$$Q^{tot}{}_{type} = a \cdot A_N + b \cdot A_N{}^{2/3} + c \cdot A_N{}^{1/3} + d \cdot Z^{\kappa} / A_N{}^{1/3} \quad (170)$$

**Table 10:** Parameters a, b, c and d for some different types of $Q^{tot}$.

| $Q^{tot}{}_{(type)}$ | a | b | c | d |
|---|---|---|---|---|
| $Q^{tot}{}_{max}$ | 2.61696075942438 | 81.2923967886543 | 2.94220517608668 | - 1.95238820051575 |
| $Q^{tot}{}_{c}$ | 2.61323819764975 | 76.4164500007471 | 2.40550058121611 | - 1.26209790271275 |
| $Q^{tot}{}_{as}$ | 0.26244059384442 | 46.6811789688200 | 0.37714379933853 | - 0.14166405273391 |

$E_{res}$ is given by $E_{res} = E_m + E_{Th}$. Results obtained by using the extended nuclear theory and Los-Alamos data indicate the following connection:

$$\left.\begin{aligned} E_m &= 11.94 + 0.29 \cdot (A_N - 12) \quad (if\ A_N \geq 12) \\ E_m &= [(A_N - 1/11]^{2/3} \cdot 11.94 \quad (if\ A_N < 12) \end{aligned}\right\} \quad (171)$$

The parameter $\sigma_{res}$, according to Eq. (56), results from a fitting procedure of calculated and measured data. $I_c$ is defined by the continuity condition for the Gaussian (Eq. (57)) and the hyperbolic-tangent function (Eq. (58)).

It is known that the nuclear cross section can be described by a series of exponential functions, before the asymptotic behavior is reached. We have verified that the hyperbolic-tangent function, which can be expanded in terms of exponential functions, provides optimal results, and it easily accommodates the continuity conditions at $E = E_c$ and $Q^{tot}{}_c = Q^{tot}{}_{max} \cdot I_c$. The Gaussian behavior in the resonance domain is due to the numerous resonance excitations occurring at $E \approx E_{res}$ according the generalized Breit-Wigner formula, see Flügge (1948).

## *A2 Results of the generalized nuclear shell theory*

### *A2.1 Harmonic oscillator models*

This section is appropriate only for interested readers. At first, we consider the 3D harmonic oscillator, described by the Hamiltonian $H_{osc}$:

$$H_{osc} = \frac{1}{2M} \cdot \sum_{k=1}^{3} p_k^{\,2} + \frac{M}{2} \cdot \omega_0^{\,2} \cdot \sum_{k=1}^{3} q_k^{\,2} \quad (172)$$

In this equation, isospin symmetry is assumed to hold, i.e., $M_p = M_{neutron} = M$. There are three ways to obtain the general solution of this equation, well-known from standard textbooks of quantum mechanics and nuclear physics; herein, we only present the results.

1. Use of creation and annihilation operators (algebraic method) based on the commutation relation:

$$p_k \cdot q_l - q_l \cdot p_k = \frac{\hbar}{i} \cdot \delta_{kl} \quad (173)$$

2. Replacement of $p_k$ by $-i \cdot \hbar \cdot \partial/\partial q_k$ in the Hamiltonian and solving the resulting Schrödinger equation by a Gaussian function multiplied with Hermite polynomials.

3. Solving the Schrödinger equation in terms of spherical harmonics and Laguerre polynomials.

*First method*

We rewrite the Hamiltonian of the 3D oscillator as:

$$\left.\begin{aligned} b_k &= (M \cdot \omega_0 / 2\hbar)^{1/2} \cdot q_k + (i/(2 \cdot M \cdot \omega_0 \cdot \hbar) \cdot p_k \\ b^+{}_k &= (M \cdot \omega_0 / 2\hbar)^{1/2} \cdot q_k - (i/(2 \cdot M \cdot \omega_0 \cdot \hbar) \cdot p_k \end{aligned}\right\} (k = 1,..,3) \quad (174)$$

These operators obey the commutation relations for bosons:

$$\left.\begin{aligned} &[b_k \,, b^+{}_l] = \delta_{kl} \quad (k, l = 1,..3) \\ &[b^+{}_k \,, b^+{}_l] = [b_k \,, b_l] = 0 \end{aligned}\right\} \quad (175)$$

With the help of these operators, the Hamiltonian assumes the shape:

$$H = \frac{1}{2} \cdot \hbar \cdot \omega_0 \cdot \sum_{k=1}^{3} (b^+{}_k \cdot b_k + b_k \cdot b^+{}_k) = \hbar \cdot \omega_0 \cdot (\sum_{k=1}^{3} b^+{}_k \cdot b_k + \frac{3}{2}) \quad (176)$$

The operator of the angular momentum is given by:

$$\mathfrak{I}_{kl} = q_k \cdot p_l - p_k \cdot q_l = i \cdot (b_k b^+{}_l - b_l b^+{}_k) \quad (177)$$

The angular-momentum operator commutes with the Hamiltonian and, therefore, it only connects degenerate states of the Hamiltonian H, by transforming a quantum state k to the state l and vice versa. The operator $b^+_k$ (absorption operator) and $b_k$ (emission operator) modify (increase and decrease, respectively) the energy $\hbar\cdot\omega_0$. There are nine independent types of bilinear products $b^+_k\cdot b_l$ (i.e., k = 1, …, 3 and l = 1, ..., 3), which implies that they can be the generators of $SU_3$ in the configuration space. This means that there is a correspondence between $SO_3$ (rotational symmetry in the configuration space) and $SU_3$, in analogy to the one between the group $SO_2$ (x/y – plane) and $SU_2$ for the two-dimensional harmonic oscillator. In nuclear physics, the group $SU_2$ is connected to the isospin, referring to both nucleons obeying anticommutation rules. Although bilinear products of fermion operators satisfy the above commutation rules, physical differences exist. Applied to a whole nucleus, the oscillator model is rather a collective description of physical properties as oscillations/vibrations via deformation or creation of rotational bands (quanta of the angular momentum of the whole nucleus) due to interactions with comparably low-energy protons (e.g., see resonance scatter of $Q^{tot}$, in particular the first term of Eq. (170)). A further critical aspect is that the oscillator potential has a minimum for E = 0, not for E << 0; bound states exist for arbitrarily high energies. Nevertheless, with the help of some modifications this model will become suitable for practical problems.

*Second method*

This method is the configuration-space representation of the first method and implies the corresponding solution of the Schrödinger equation in Cartesian coordinates:

$$-(\hbar^2/2M)\cdot\Delta\psi+\frac{M}{2}\cdot\omega_0^{\,2}\cdot(q_1^{\,2}+q_2^{\,2}+q_3^{\,2})\cdot\psi=E\cdot\psi \quad (178)$$

We make use of the substitutions

$$\rho_k=\sqrt{\omega_0\cdot M/\hbar}\cdot q_k \quad \textit{and} \quad \rho^2=\rho_1^{\,2}+\rho_2^{\,2}+\rho_3^{\,2} \quad (179)$$

The complete solution is then given by:

$$\left.\begin{array}{l} \psi_{j,k,l}=N\cdot H_j(\rho_1)\cdot H_k(\rho_2)\cdot H_l(\rho_3)\cdot\exp(-\rho^2/2) \\ (j,k,\,l=0,1,.....,\infty) \\ E_{j,k,l}=\hbar\cdot\omega_0\cdot(3/2+j+k+l) \end{array}\right\} \quad (180)$$

N is a normalization factor; $H_j$, $H_k$, and $H_l$ are Hermite polynomials as already used in previous sections. The advantage of this representation is the possible connection to other problems, involving Gaussian functions.

*Third method*

This method uses the framework of the H atom, i.e., the separation of the wavefunction ψ in a radial function and spherical harmonics. The only difference is that the potential $Z \cdot e_0^2/r$ is replaced by the oscillator potential; this implies that the Laguerre polynomials have an argument different to that of the H atom:

$$\left.\begin{aligned} &\psi_{n,l,m} = L_{n+l-1/2}^{l+1/2}(\rho^2) \cdot Y_m^l(\cos\vartheta, \varphi) \\ &E_\mu = (\mu + \tfrac{3}{2}) \cdot \hbar \cdot \omega_0 \\ &\mu = 0, 1, 2, .... \\ &l = \mu, \mu - 2, \mu - 4, ... \\ &n = \tfrac{1}{2} \cdot (\mu - l + 2) \end{aligned}\right\} \quad (181)$$

The energy eigenvalues are independent of the quantum number m (m = -l, -l+1, …, 0, l-1, l). The correspondence between Hermite polynomials and spherical harmonics may be found in Abramowitz and Stegun (1970).

The concept of spin can be introduced to the 3D harmonic oscillator by the commutation rule, see Ulmer and Hartmann (1978):

$$\sigma \otimes p \cdot q_l - q_l \cdot \sigma \otimes p = \tfrac{\hbar}{i} \cdot \sigma_l \ \ (l = 1, 2, 3) \quad (182)$$

In this case, σ represents the three Pauli spin matrices and the unit matrix. The result is the Pauli equation of a 3D harmonic oscillator. Similarly, we introduce the isospin τ by the substitution:

$$\sigma \Rightarrow \sigma \otimes \tau \quad (183)$$

By that, we obtain the Pauli equation for a 3D oscillator for a proton and neutron:

$$\left.\begin{aligned} &\tfrac{1}{2M} \cdot ((\tfrac{\hbar}{i}\nabla - \tfrac{e0}{c}A)^2 \psi_p + +(V_{osc} - \mu_p \cdot B) \cdot \psi_P = -i\hbar \tfrac{\partial}{\partial t}\psi_p \\ &-\tfrac{\hbar\hbar^2}{2M} \cdot \Delta\psi_n + \ (V_{osc} - \mu_n \cdot B) \cdot \psi_n = -i\hbar \tfrac{\partial}{\partial t}\psi_n \\ &\mu_p = g_p \cdot \mu_0 \cdot s; \qquad \mu_n = g_n \cdot \mu_0 \cdot s \\ &B = \nabla \times A \end{aligned}\right\} \quad (184)$$

In nuclear many-particle theory, the Pauli principle is generalized, i.e., the total wavefunction has to be antisymmetric with regard to spin and isospin. The property $g_p/g_n = -3/2$ does not follow from the scope of the present theory and further principles have to be introduced (see Feynman (1972)), which we do not consider here. Even by extending Eq. (184) to a many-particle equation (including the spin-orbit coupling) and to a Slater determinant (Hartree-Fock: ground state), the problem of nuclear reactions, due to the properties of the harmonic oscillator potential $V_{osc}$, cannot be solved. The problem of simple oscillator models can be verified in the following Fig. 37, which shows the effective nuclear potential energy for O. The abscissa is expressed in units of $r = 1.2 \cdot 10^{-13} \cdot A_N^{1/3}$ cm ($A_N = 16$).

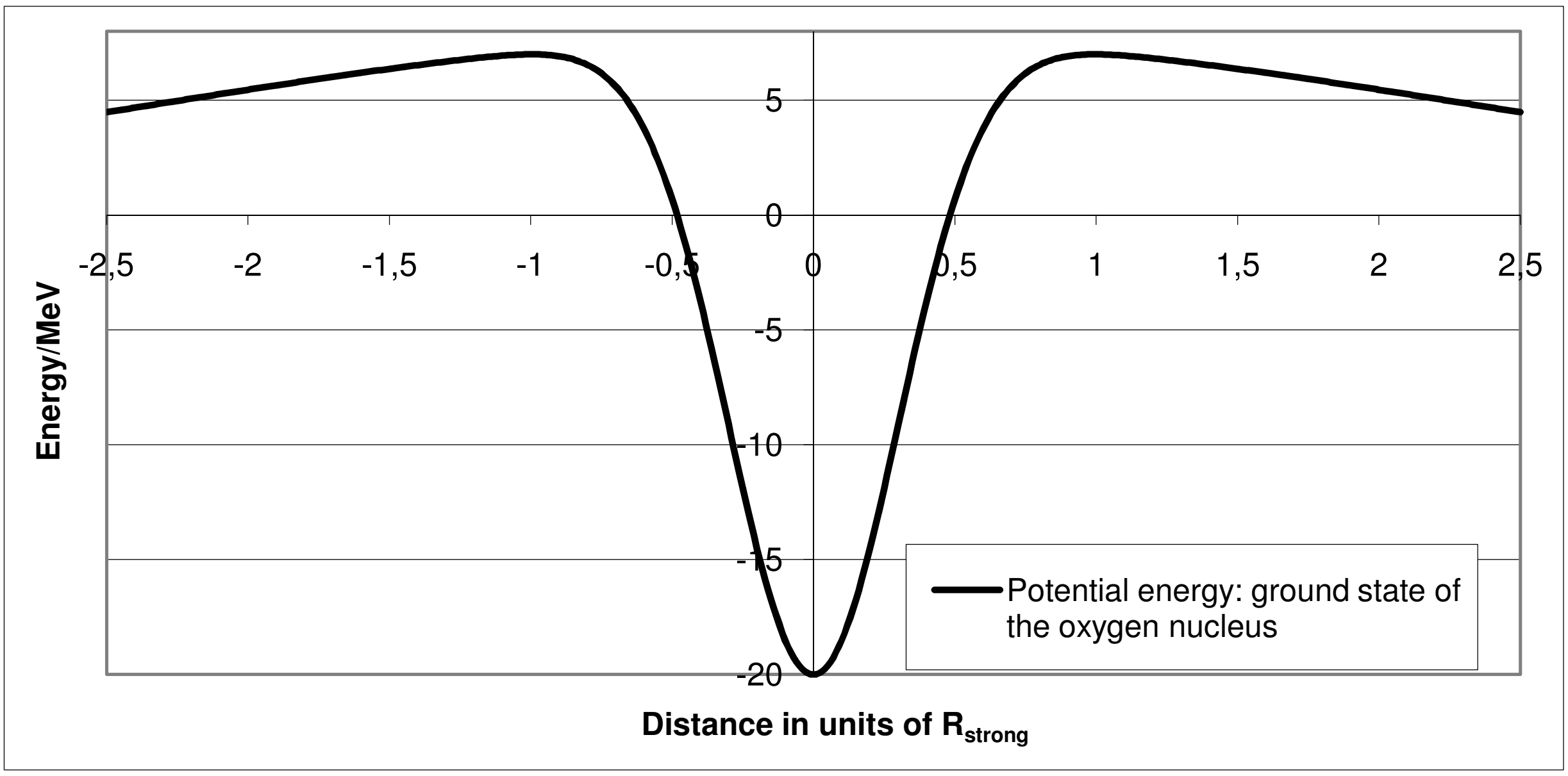

**Fig. 38:** Total (effective nuclear potential plus Coulomb repulsion) for O.

We approximate the potential according to:

$$\left.\begin{aligned} \varphi(r) &= V_0(1 - r^2/\sigma_0^2) \\ \sigma_0 &\approx 0.423901 \end{aligned}\right\} \quad (185)$$

The complete potential function can be expressed as a linear combination of two Gaussians:

$$\left.\begin{aligned} \varphi(r) &= V_0 \cdot \exp(-r^2/\sigma_0^2) + V_1 \cdot \exp(-r^2/\sigma_1^2) \\ \sigma_1 &= 3.37402\ ;\ V_0 = -27.75925\ ;\ V_1 = 7.75935 \end{aligned}\right\} \quad (186)$$

A property of the Gaussian function is that its curvature changes sign at $r = r_c$. For a single Gaussian, as the first one in Eq. (186), $r_c$ is given by:

$$r_c = \sigma_0 / \sqrt{2} = 0.29974 \qquad (187)$$

Only for $r \leq r_c$, is a harmonic-oscillator approach useful, and the deviation to a Gaussian in this domain small. This is, however, not true for $r > r_c$. In the case of the linear combination of two Gaussians, $r_c$ is broader:

$$\left.\begin{aligned} & r_c \approx \frac{1}{\sqrt{2}} \cdot \sigma_0 \cdot \sigma_1 \cdot \sqrt{(-\sigma_1{}^2 \cdot V_0 - \sigma_0{}^2 \cdot V_1 \cdot A_\sigma)/(-\sigma_1{}^4 \cdot V_0 - \sigma_0{}^4 \cdot V_1 \cdot A_\sigma)} \\ & r_c \approx \mathbf{0.4582}; \qquad A_\sigma = \exp[(\sigma_1{}^2 - \sigma_0{}^2)/2 \cdot \sigma_1{}^2)] \end{aligned}\right\} \quad (188)$$

Fig. 38 shows that, for r < 0.4982, strong interactions are dominant (all other interactions are negligible). If $0.4682 \leq r < 1$, strong interactions are still present with decreasing tendency, whereas Coulomb repulsion is increasing; finally, for r = 1, strong interactions are negligible. We will come back to these results in the next section.

*A2.2 Nonlinear/nonlocal Schrödinger equation, anharmonic oscillators with self-interaction and Hartree-Fock method (inclusive configuration interaction)*

Let us first consider the usual Schrödinger equation for a bound system:

$$\left.\begin{aligned} & E \cdot \psi + \frac{\hbar^2}{2 \cdot M} \cdot \Delta \psi = \varphi(x, y, z) \cdot \psi \\ & \varphi \leq 0 \ \ (x, y, z \in R_3); \ \ \Delta: \ 3D - Laplace \ \ operator \end{aligned}\right\} \quad (189)$$

During the past decades, this type has been encountered in many fields of physics, such as superconductivity, nuclear and plasma physics, e.g., see Ulmer and Hartmann (1978), Milner (1990), and various other references. A nonlinear Schrödinger equation is obtained by introducing the potential φ, proportional to the density of solutions:

$$\varphi = \lambda \cdot |\psi(x,y,z)|^2 = \lambda \cdot \int \delta(x - x') \cdot \delta(y - y') \cdot \delta(x - x') \cdot |\psi(x', y', z')|^2 dx'dy'dz' \quad (190)$$

The coupling constant λ is negative (in which case, the solutions are bound states with E < 0); Eq. (190) can be interpreted as a contact interaction. It is known from many-particle problems (e.g., quantum electrodynamics, Hartree/Hartree-Fock method, etc.) that the mutual interactions between the particles in configuration space lead to nonlinear equations in quantum mechanics. However, in these cases, there are not at all contact interactions; the nonlinear Schrödinger equation above is an idealistic case. By taking ε → 0, the Gaussian kernel is transformed into a δ kernel:

$$\varphi=\lambda\cdot\int(1/(\sqrt{\pi}^3\cdot\varepsilon^3)\cdot\exp[-((x-x')^2+(y-y')^2+(z-z')^2)/\varepsilon^2]|\psi(x',y',z')|^2dx'dy'dz' \quad (191)$$

The nonlinear/nonlocal Schrödinger equation can be interpreted as a self-interaction of a many-particle system with internal structure, and it is possible to generalize this type by incorporation of additional internal symmetries (e.g., the introduction of the spin to obtain spin-orbit coupling, $SU_2$, $SU_3$, and also discrete-point groups).

According to the principles developed in this work, we are able to write Eq. (191) in the form of an operator equation (the Gaussian kernel is Green's function):

$$\varphi=\lambda\cdot\exp(\tfrac{1}{4}\cdot\varepsilon^2\cdot\Delta)\cdot|\psi(x,y,z)|^2 \quad (192)$$

Expanding this operator in terms of a Lie series and keeping only the terms up to Δ, Eq. (192) becomes a stationary Klein-Gordon equation, which describes the interaction between the particles obeying the Ψ-field:

$$\left.\begin{array}{l}\exp(\tfrac{1}{4}\cdot\varepsilon^2\cdot\Delta)=1+\tfrac{1}{4}\cdot\varepsilon^2\cdot\Delta+0(higher-order\ \ terms)\\(1+\tfrac{1}{4}\cdot\varepsilon^2\cdot\Delta)\cdot\varphi=\lambda\cdot|\psi|^2\end{array}\right\} \quad (193)$$

By rescaling the Klein-Gordon equation, we obtain the more familiar form: $1 + 0.25\ \varepsilon^2\ \Delta \rightarrow k^2 + \Delta$; Green's function is of the form:

$$\left.\begin{array}{l}G(r,r')=N\cdot\exp(-k\cdot|r-r'|)\cdot\frac{1}{|r-r'|}\\N: normalization\ \ factor\end{array}\right\} \quad (194)$$

By setting $k \rightarrow 0$, the Poisson equation of electrostatics is obtained, if $|\psi|^2$ is interpreted as a charge density. The Gaussian kernel K also represents the exchange of virtual particles between the nucleons. In view of this fact, we point out that we have incorporated a many-particle system from the beginning. Which information now does this nonlinear/nonlocal Schrödinger equation provide? In order to combine Eqs. (189 - 192) with the oscillator model of nuclear shell theory, we analyze the kernel K in detail. In the Feynman-propagator method (see Feynman and Hibbs (1965)), the expansion of K in terms of generating functions is an important tool:

$$\left.\begin{aligned} K(\varepsilon,x'-x,y'-y,z'-z) &= (1/\sqrt{\pi}^3\cdot\varepsilon^3)\sum_{n1=0}^{\infty}\exp(-x'^2/\varepsilon^2)\cdot H_{n1}(x'/\varepsilon)\cdot x^{n1}/(\varepsilon^{n1}\cdot n1!)\cdot\\ &\cdot\sum_{n2=0}^{\infty}\exp(-y'^2/\varepsilon^2)\cdot H_{n2}(y'/\varepsilon)\cdot y^{n2}/(\varepsilon^{n2}\cdot n2!)\cdot\\ &\cdot\sum_{n3=0}^{\infty}\exp(-z'^2/\varepsilon^2)\cdot H_{n3}(z'/\varepsilon)\cdot z^{n3}/(\varepsilon^{n3}\cdot n3!)\end{aligned}\right\}\quad(195)$$

Inserting this expression into the nonlinear/nonlocal Schrödinger equation, we obtain:

$$\left.\begin{aligned} &E\cdot\psi+\frac{\hbar^2}{2\cdot M}\cdot\Delta\psi=\varphi(x,y,z)\cdot\psi=\lambda\cdot(1/\sqrt{\pi}^3\cdot\varepsilon^3)\sum_{n1=0}^{\infty}\sum_{n2=0}^{\infty}\sum_{n3=0}^{\infty}\Phi_{n1,n2,n3}\cdot x^{n1}\cdot y^{n2}\cdot z^{n3}\cdot\psi\\ &\Phi_{n1,n2,n3}=\frac{1}{n1!}\cdot\frac{1}{n2!}\cdot\frac{1}{n3!}\cdot\frac{1}{\varepsilon^{n1+n2+n3}}\cdot\int\left|\psi(x',y',z')\right|^2\cdot\exp(-(x'^2+y'^2+z'^2)/\varepsilon^2)\cdot\\ &\quad\cdot H_{n1}(x'/\varepsilon)\cdot H_{n2}(y'/\varepsilon)\cdot H_{n3}(z'/\varepsilon)dx'dy'dz'\end{aligned}\right\}\quad(196)$$

The equation above represents a highly-anharmonic oscillator equation of a self-interacting field. Since the square of the wavefunction is always positive definite, all terms with odd numbers of n1, n2, and n3 vanish due to the antisymmetric properties of those Hermite polynomials. For $r_c \le \varepsilon/\sqrt{2}$ (domain with positive curvature), the whole equation is reduced to a harmonic oscillator with self-interaction; the higher-order terms are small perturbations. We summarize the results and refer to previous publications, see Ulmer (1979), Ulmer (1980), and Milner (1990):

$$E\cdot\psi+\frac{\hbar^2}{2\cdot M}\cdot\Delta\psi=\varphi(x,y,z)\cdot\psi=\lambda\cdot(1/\sqrt{\pi}^3\cdot\varepsilon^3)\cdot[\Phi_{0,0,2}(x^2+y^2+z^2)+\Phi_{0,0,0}]\cdot\psi\quad(197)$$

The solutions of this equation are those of a 3D harmonic oscillator; the classification of the states by $SU_3$ and all previously developed statements with regard to the angular momentum are still valid. The only difference is that the energy levels are not equidistant; this property can easily be verified in one dimension. The usual ground state energy is $\hbar\omega_0/2$. This energy level is lowered by the term ~$\lambda\cdot\Phi_{0,0,0}$, depending on the ground-state wavefunction. The energy difference between the ground and the first excited state amounts to $\hbar\omega_0$; this is not true in the case above, since the energy level of the excited states depends on the corresponding eigenfunctions (these are still the oscillator eigenfunctions!). Next, we will include the terms of the next order, which are of the form ~ $\lambda\cdot(\Phi_{0,2,2}, \Phi_{2,2,0}, \Phi_{2,0,2})$:

$$\left.\begin{aligned}&E\cdot\psi+\frac{\hbar^2}{2\cdot M}\cdot\Delta\psi=\varphi(x,y,z)\cdot\psi=\lambda\cdot(1/\sqrt{\pi}^3\cdot\varepsilon^3)\cdot[\Phi_{0,0,0}+\Phi_{0,0,2}(x^2+y^2+z^2)+T]\cdot\psi\\&T=\Phi_{2,2,0}\cdot x^2\cdot y^2+\Phi_{2,0,2}\cdot x^2\cdot z^2+\Phi_{0,2,2}\cdot y^2\cdot z^2\end{aligned}\right\}\quad(198)$$

The additional term T represents tensor forces. The whole problem is still exact soluble. In further extensions of the nonlinear/nonlocal Schrödinger equation, we are able to account for spin, isospin, and spin-orbit coupling.

The spin-orbit coupling, as an effect of an internal field with nonlocal self-interaction, is plausible, since the extended nucleonic particle has internal structure; consequently, we have to add $H_{so}$ to the nonlinear term:

$$H_{so}\cdot\psi=g_\tau\cdot\frac{\hbar\cdot\sigma}{4\cdot M\cdot c^2}\cdot\nabla\varphi\cdot x\,p\cdot\psi \quad (199)$$

$\Psi$ is now (at least) a Pauli spinor (i.e., a two-component wavefunction), and together with $H_{so}$ the $SU_3$ symmetry is broken. We should like to point out that the operation $\nabla\varphi$ acts on the Gaussian kernel K:

$$\nabla\varphi=-\frac{2}{\varepsilon^2}\cdot[H_1((x-x')/\varepsilon),H_1((y-y')/\varepsilon),H_1((z-z')/\varepsilon)]\cdot\varphi \quad (200)$$

The expression in the bracket of the previous equation represents a vector, and p ($p\rightarrow -i\hbar\nabla$) acts on the wavefunction. Since the neutron is not a charged particle, the spin-orbit coupling of a neutron can only involve the angular momentum of a proton. In nuclear physics, these nonlinear fields are adequate for the analysis of clusters (deuteron, He, etc.). Milner (1990) has extended the theory to describe nuclei with odd spin.

The complete wavefunction $\Psi_c$ is now given by the product of a function in configuration space $\Psi$ multiplied with the total spin and isospin functions. We should like to add that an extended harmonic oscillator model with tensor forces has been regarded in Elliott (1963). The application of oscillator models in nuclear physics goes back to Heisenberg (1935); Feynman and Schwinger, see Feynman (1962), have verified that the use of Gaussians in the description of meson fields provides many advantages compared to the Yukawa potential (Green's function according to Eq. (194)).

In a final step, we consider the generalized Hartree-Fock method to solve the many-particle problem. In order to derive all required formulas, it is convenient to use second quantization. The method of second quantization is only suitable to derive the calculation procedure: extension of the Pauli principle to isospin besides spin, inclusion of spin-orbit coupling, and exchange interactions. This is the consequence of dealing with identical particles, in which case every state can only occupy one quantum number. In order to get numerical results (i.e., the minimum of the total energy of an ensemble of nucleons, the extraction of the excited states, the scatter amplitudes, etc.), we have to use representations of the wavefunction by at least one determinant in the configuration space. In the 'language' of second quantization of fermions, we would have to regard expressions like:

$$\left.\begin{array}{l} a^{+}{}_{k,\sigma,\tau} \cdot a_{l,\sigma',\tau'} + a_{l,\sigma',\tau'} \cdot a^{+}{}_{k,\sigma,\tau} = \delta_{kl} \cdot \delta_{\sigma\sigma'} \cdot \delta_{\tau\tau'} \\ a_{k,\sigma,\tau} \cdot a_{l,\sigma',\tau'} + a_{l,\sigma',\tau'} \cdot a_{k,\sigma,\tau} = 0 \\ a^{+}{}_{k,\sigma,\tau} \cdot a^{+}{}_{l,\sigma',\tau'} + a^{+}{}_{l,\sigma',\tau'} \cdot a^{+}{}_{k,\sigma,\tau} = 0 \end{array}\right\} \quad (201)$$

The operators of the form $a_k^+$ and $a_k$ (k being a set of quantum numbers) are creation and destruction operators in the state space. The nonlinear/nonlocal Schrödinger equation with Gaussian kernel for the description of the strong interaction, including the spin-orbit coupling, can be written by these operators, leading from an extended particle with internal structure to a many-particle theory. Before we start to explain the calculations by including one or more configurations, we recall that, according to Fig. 38, we have an increasing contribution of the Coulomb repulsion for $r > r_c$,, though in the domain $r < r_c$, the contributions of the Coulomb interactions are negligible. Since all basis elements of the calculation procedures, i.e., the calculation of eigenfunctions in the configuration space, two-point kernels of strong interactions between nucleons, and the spin-orbit coupling can be expressed in terms of Gaussians and Hermite polynomials, we want to proceed in the same fashion with regard to the Coulomb part. According to results of elementary-particle models (e.g., see Feynman (1972)), the charge of the proton is located in an extremely small sphere with radius $r_p = 10^{-14}$ cm, not at one 'point'. Therefore, we write the decrease of the proton Coulomb potential by $1/(r+r_p)$; for r = 0, we then obtain $10^{14}$ $cm^{-1}$, not infinite. In a sufficiently small distance of $r = 2.4 \cdot 10^{-13}$ cm, we can approximate the Coulomb potential with high precision by:

$$\frac{1}{r+r_p} = c_0 \cdot \exp(-r^2/r_0^2) + c_1 \cdot \exp(-r^2/r_1^2) + c_2 \cdot \exp(-r^2/r_2^2) \quad (202)$$

The mean standard deviation amounts to $10^{-5}$, if the parameters of Formula (202) are chosen as:

$$\left.\begin{array}{l} c_0 = 0.5146\cdot 10^{14};\; c_1 = 0.3910\cdot 10^{14};\; c_2 = 0.0944\cdot 10^{14} \\ r_0 = 0.392\cdot 10^{-13}\,cm;\; r_1 = 0.478\cdot 10^{-13}\,cm;\; r_2 = 2.5901\cdot 10^{-13}\,cm \end{array}\right\} \quad (203)$$

If necessary, it is possible to rescale $r_0$, $r_1$, and $r_2$ by dividing by $(A_N)^{1/3}$. The contribution with $c_2$ incorporates a long-range correction. In the absence of an external electromagnetic field, the Hamiltonian reads as:

$$\left.\begin{array}{l} H = \sum_j -\frac{\hbar^2}{2M}\Delta_j + H_{so} + H_{Coul} + H_{strong} \\ H_{Coul} = e_0^{\,2}\cdot \sum_{j,proton}\ \sum_{l,proton\neq j}\ \sum_{k=0}^{2} c_k \cdot \exp(-(r_j - r_l)^2 / r_k^{\,2}) \\ H_{strong} = -g_s \cdot \sum_j \sum_{l\neq j} \exp(-(r_j - r_l)^2 / \sigma_s^{\,2}) \end{array}\right\} \quad (204)$$

Note that it is possible to distinguish between the proton and the neutron masses by indexing M; the ε, previously used in Eq. (204), has been replaced by $\sigma_s$. The coupling constant of $g_s$ is 1, if the Coulomb interaction is scaled to

$$\left.\begin{array}{l} g_s = 1 \\ e_0^{\,2} / (\hbar\cdot c) = 1/137 \end{array}\right\} \quad (205)$$

Thus, in theoretical units with $e_0 = c = h/2\pi = 1$, the coupling constant $g_s$ assumes 137. This relation can be best seen in the Dirac equation containing a Coulomb repulsion potential ~ $e_0^2$ and a strong interaction term ~ $-g_s$. The aforementioned relation is obtained by dividing the kinetic-energy operator $c\cdot\alpha\cdot p \rightarrow -c\cdot\alpha\cdot\hbar\cdot\nabla$ and $\beta\cdot mc^2$ by $(c\cdot\hbar)$. In the calculations for deuteron, $He^3$, and He, we have assumed the range length $\sigma_s$:

$$\left.\begin{array}{l} \sigma_s = \sigma_{sp} = \frac{\hbar}{m_p\cdot c} \approx 10^{-13}\ cm \\ m_p\ :\ mass\ \ of\ \ \pi - meson \end{array}\right\} \quad (206)$$

This assumption turned out to be not sufficient; a replacement of $\sigma_s$ was justified to distinguish between the range length $\sigma_{sp}$ ($\pi$-mesons) and $\sigma_{sk}$ (K-mesons):

$$\left.\begin{aligned} &-g_S \cdot \exp(-(r_j - r_l)^2 / \sigma_s^2) \Rightarrow -g_S \cdot [c_{sp} \cdot \exp(-(r_j - r_l)^2 / \sigma_{sp}^2) + c_{sk} \cdot \exp(-(r_j - r_l)^2 / \sigma_{sk}^2)] \\ &c_{sp} = 1 - (\sigma_{sk} / \sigma_{sp})^2; \; c_{sk} = (\sigma_{sk} / \sigma_{sp})^2 \\ &\sigma_{sp} = 1.02 \cdot 10^{-13} cm; \; \sigma_{sk} = 0.29 \cdot 10^{-13} \, cm \end{aligned}\right\} \quad (207)$$

The range length $\sigma_{sk}$ is proportional to $1/m_k$ ($m_k$: mass of the K-meson).

The Hartree-Fock method provides the best one-particle approximation of the closed-shell case.

$$\Phi = \frac{1}{\sqrt{N!}} \begin{vmatrix} \varphi_{k1}(1) \ldots\ldots\ldots\ldots \varphi_{k1}(N) \\ \varphi_{k2}(1) \ldots\ldots\ldots\ldots \varphi_{k2}(N) \\ \ldots\ldots\ldots\ldots\ldots\ldots\ldots\ldots\ldots\ldots \\ \varphi_{kN}(1) \ldots\ldots\ldots\ldots \varphi_{kN}(N) \end{vmatrix} \quad (208)$$

The one-particle functions $\varphi_{k1}(1)$, …, $\varphi_{kN}(N)$ contain all variables (configuration space of position coordinates, spin, and isospin). By using a complete system of trial functions, e.g., a Gaussian multiplied with Hermite polynomials, the Hartree-Fock limit is obtained. In view of our question to calculate the S-matrix and the cross section of the proton-nucleon interactions (elastic, inelastic, resonance scatter, and nuclear reactions), this restriction is insufficient. In particular, we have to add excited configurations and virtually-excited configurations. The role of excited states is clear. As an example, we regard the O nucleus, where the total spin is 0. If a proton or neutron of the highest-occupied shell is excited, then the spin may change, and both, highest-occupied and lowest-unoccupied shell, are occupied by one nucleon. The emitted nucleon may be regarded as a ‘hole’. This procedure can be repeated to higher-unoccupied states and to linear combinations of configurations with different nucleon numbers. A virtually-excited state is produced, if the configuration of the excited state only formally exists for the calculation procedure, but cannot be reached physically. An example of this case is already the deuteron with isospin 0 and spin 1. An excited state with spin 1 or 0, where proton and neutron occupy different energy levels (shells), does not exist. In spite of this situation, the Hartree-Fock method does not provide the correct ground state, and linear combinations of determinants with different spin states (S = 1, -1, 0) and ‘holes’ have to be included. These virtual states also enter the calculation of the S-matrix and of the cross section.

We have performed Hartree-Fock-configuration-interaction calculations (HF - CI) for the nuclei: deuteron, $He^3$, He, Be, C, Si, O, Al, Cu, and Zn. The set of basic functions comprises $2 \cdot (A_N + 13)$ functions with the following properties:

$$\left.\begin{aligned} &\varphi(x)=\sum_{j=0}^{N}[A_j \cdot H_j(\alpha_1 \cdot x) \cdot \exp(-\tfrac{1}{2} \cdot \alpha_1^2 \cdot x^2)+B_j \cdot H_j(\alpha_2 \cdot x) \cdot \exp(-\tfrac{1}{2} \cdot \alpha_2^2 \cdot x^2)] \\ &\varphi(x,y,z)=\varphi(x) \cdot \varphi(y) \cdot \varphi(z) \\ &N = A_N + 12 \end{aligned}\right\} \quad (209)$$

Both $\alpha_1$-functions and $\alpha_2$-functions are chosen such that the number of functions is $A_N + 13$. The different range parameters $\alpha_1$ and $\alpha_2$ are useful, since different ranges can be accounted for. If $\alpha \gg \beta$, the related wavefunctions decrease much more rapidly (central part of the nucleus), whereas the β-contributions preferably describe the behavior in the domain $r \geq r_c$. With the help of this set of trial functions[2] (Ritz's variation principle), we obtain the best approximation of the total energy E by $E_{app}$ and the nuclear shell energies (for occupied and unoccupied shells). For bound states, $E_{app} > E$ is always fulfilled. It should be noted that for computational reasons it is useful to replace the set of functions (209) by the nonorthogonal set:

$$\begin{aligned} \varphi_{n1,n2,n3} = x^{n1} \cdot y^{n2} \cdot z^{n3} \cdot [&A_{n1,n2,n3}(\alpha_1) \cdot \exp(-\tfrac{1}{2}\alpha_1^2 \cdot r^2) + \\ &+ B_{n1,n2,n3}(\alpha_2) \cdot \exp(-\tfrac{1}{2}\alpha_2^2 \cdot r^2)] \end{aligned} \quad (210)$$

By forming arbitrary linear combinations depending on $\alpha_1$ and $\alpha_2$ we obtain the same results as by the expansion (260). The exploding coefficients of the Hermite polynomials are an obstacle in numerical calculations and can be avoided by the expansion (210). The minimal basis set for the calculation of deuteron would be one single trial function, i.e. a Gaussian without further polynomials. This is, however, a crude approximation and already far from the HF limit. Using this simple approximation, we obtain the result that the ground state $E_g$ depends solely on $\alpha_1$. The best approximation exceeds the HF limit by about 15 %. Various tasks, such as resonance scatter, nuclear reactions, and spin-orbit coupling cannot be described; the cross section of the pure potential scatter is also 12 % too low.

Using 13 $\alpha_1$-dependent and 13 $\alpha_2$-dependent functions, we have obtained the HF limit and virtually-excited states (a bound excited state does not exist). The HF wavefunction had to be subjected to virtually-excited configurations, i.e., all possible singlet and triplet states. This calculation had to be completed by introducing a further proton (interaction proton) and including all virtual configurations (besides a configuration with three independent nucleons, a configuration of a virtual $He^3$ state). Thus, for low proton energies (slightly above $E_{Th}$), the $He^3$ formation is possible. The exceeding energy can be transferred to the total system and/or to rotations/vibrations of $He^3$. In the same fashion, we have to proceed to the calculations for other nuclei: the configurations of all possible fragments have also to be taken into account. (The cases, corresponding to the O nucleus, are given in listing 52). In order to keep these considerations short, we now only give a skeleton of the calculation procedures, which are necessary to evaluate the cross sections. When – besides the ground state – all excited states (including virtually-excited states and configurations of fragments) are determined (wavefunctions and related energy levels), then Green's function is readily determined by taking the sum over all states. This function contains all coordinates in the configuration space (including the spin), quantum numbers of oscillations, and rotational bands:

$$G(r_n, r'_n) = \sum_{j=0}^{2\cdot N} \psi^*{}_j(r'_n) \cdot \psi_j(r_n) \quad (211)$$

The S-matrix is given by:

$$S_{kl} = \int \psi^*{}_k(r'_n) \cdot G(r_n, r'_n) \cdot \psi_l(r_n) d^3r_1 \ldots d^3r_{2N} \cdot d^3r'_1 \ldots d^3r'_{2N} \quad (212)$$

The transition matrix $T_{kl}$ is defined by all transitions with $k \neq l$:

$$T_{kl} = S_{kl} - \delta_{kl} \quad (213)$$

In order to determine the differential cross section, we need the transition probability. For this purpose, we assume that, before the interaction of the proton with the nucleus, this nucleus is in the ground state. Thus, it might be possible that a proton produces excited states of the nucleus by resonance scatter (inelastic), and a second proton hits the excited nucleus before the transition to the ground state (by emission of a $\gamma$ quantum) has occurred. The second proton would require a lower energy to release either a nucleon or to induce a much higher excited state of the nucleus. However, due to the nuclear cross section, the probability for an inelastic nuclear reaction is very small and would require a very high proton density to yield a noteworthy effect. Therefore, we have calculated the transition probability using the assumption that the occupation probability of the ground state $P_0$ is 1, i.e., $P_0 = 1$ and $P_k = 0$ ($k > 1$). (This is very special case of the Pauli master equation). The differential cross section is obtained by the transition probability divided by the incoming proton flux:

$$\mathbf{dq/d\,\Omega} = \frac{\textbf{Transition probabilit y}}{\textbf{Incoming proton current}} \quad \textbf{(214)}$$

At lower energies, this flux could be calculated by the current given by the Schrödinger equation. To be consistent, we have always used the Dirac equation, since proton energies E > 200 MeV show a significant relativistic effect. With regard to the incoming proton current, we have to point out an important feature:

- The Breit-Wigner formula only considers S states and the incoming current is along the z direction.
- The generalization of this formula by Flügge (1948) includes P states, but the incoming beam is also restricted to the z direction.

Since for our purpose it is necessary to take account for the x/y/z direction by $k_x$, $k_y$, $k_z$ in the Dirac equation, we have not yet succeeded in obtaining a compact and simple analytical form.

*A2.3 Application to reaction protons of the inelastic cross section $S_{sp,r}$*

We have already pointed out that the main purpose for calculations with the extended nuclear shell theory incorporate nuclear reaction contributions of protons, neutrons and further small nuclei to the total nuclear cross sections of nuclei discussed in this presentation. We should also mention that the default calculation procedure of nuclear reactions in GEANT4 is an evaporation/cascade model, which has been developed on the basis of statistical thermodynamics.

Figs. 15 - 18 do not yet provide final information about the contributions $S_{sp,n}$ and $S_{sp,r}$. The first case of nonreaction protons has already treated. According to Fig. 17 the contribution of reaction protons is particular important for E > 150 MeV with increasing energy. We now present the calculation formulas for this case. Thus $S_{sp,r}$ is proportional to $\Phi_0 \cdot 2 \cdot \upsilon \cdot C_{heavy}$ and a function $F_r$, depending on some further parameters. It should be mentioned that the parameters of Eq. (215) exclusively refer to the oxygen nucleus. However, from Fig. 10 the corresponding parameters of some further nuclei can be verified, e.g., $E_{Th}$, $E_{res}$, and some necessary information on the total cross section. We use the following definitions and abbreviations:

$$\left.\begin{array}{l} z_R = R_{CSDA}/\pi \\ \tau_s = 0.55411;\ \tau_f = \tau_s - 0.000585437 \cdot (E - E_{res}) \\ \arg 1 = z/\sqrt{\tau_s^2 + \tau_{in}^2 + (R_{CSDA}/4\pi)^2} \\ \arg 2 = (R_{CSDA} - z - \sqrt{2} \cdot \pi \cdot z_{shift})/\sqrt{\tau_f^2 + (\tau_{straggle}/7.07)^2 + (R_{CSDA}/\sqrt{3} \cdot 4\pi)^2} \\ \arg = (R_{CSDA} - 0.5 \cdot z_{shift} \cdot \sqrt{\pi} - z)/(\sqrt{\pi} \cdot z_{shift}) \end{array}\right\} (215)$$

Formulas (215 – 216) can only be partially derived, and the adaptation to computed data with the help of the extended nuclear-shell theory is also needed. This can be seen best via computation model M3, of which the main contribution consists of the term $I_2$ indicating a proportionality to [erf(z/τ) + erf(($R_{CSDA}$ – z)/τ)], if the particles are emerging at surface (i.e., erf(z/τ) at z = 0, whereas the integration boundary z ⟶ - ∞ implies the term [1 + erf(($R_{CSDA}$ – z)/τ)]). The transport of secondary reaction protons resulting from the spectral distribution of these protons has to be taken into account; and the spectral distributions rather obey a Landau than a Gaussian distribution (Fig. 39). The result is the following connection:

$$\left.\begin{array}{l} S_{sp,r} = (E_0/N_{abs}) \cdot \Phi_0 \cdot [2 \cdot \upsilon \cdot C_{heavy} \cdot F_r + G] \cdot Q^{tot}{}_{as}(medium) \cdot A_N{}^{1/3} / (Q^{tot}{}_{as}(oxygen) \cdot A_{oxygen}{}^{1/3}) \\ F_r = \varphi \cdot [\varphi_1 + \varphi_2 \cdot \theta] \\ \varphi = 0.5 \cdot (1 + erf(\arg));\quad \varphi_1 = erf(\arg 1);\quad \varphi_2 = erf(\arg 2) \\ \theta = \begin{pmatrix} e^{-1} - \exp(-(z - z_R)^2 / z_R{}^2)/(e^{-1} - 1) & (if\ z \le z_R) \\ 1 & (else) \end{pmatrix} \\ G = (c_1 \cdot z_{shift} \cdot \sqrt{\pi} / R_{CSDA}) \cdot \exp(-(\frac{1}{\sqrt{\pi}} R_{CSDA} - z)^2 / z_R{}^2) \end{array}\right\} (216)$$

The tails at z ≥ $R_{CSDA}$ result from tertiary protons induced by neutrons and the resonance interaction via meson exchange as pointed out in a previous section. Some consequences of these contributions with regard to buildup have been thoroughly discussed in this work, since the role skew symmetric energy transfer (Landau distributions) and energy transfer from reaction protons along the proton track represents a principal question in understanding the physical foundation of Bragg curves. It appears that the interpretation of Carlsson et al. (1977) is too simplified with regard to the contributions of the so-called secondary protons.

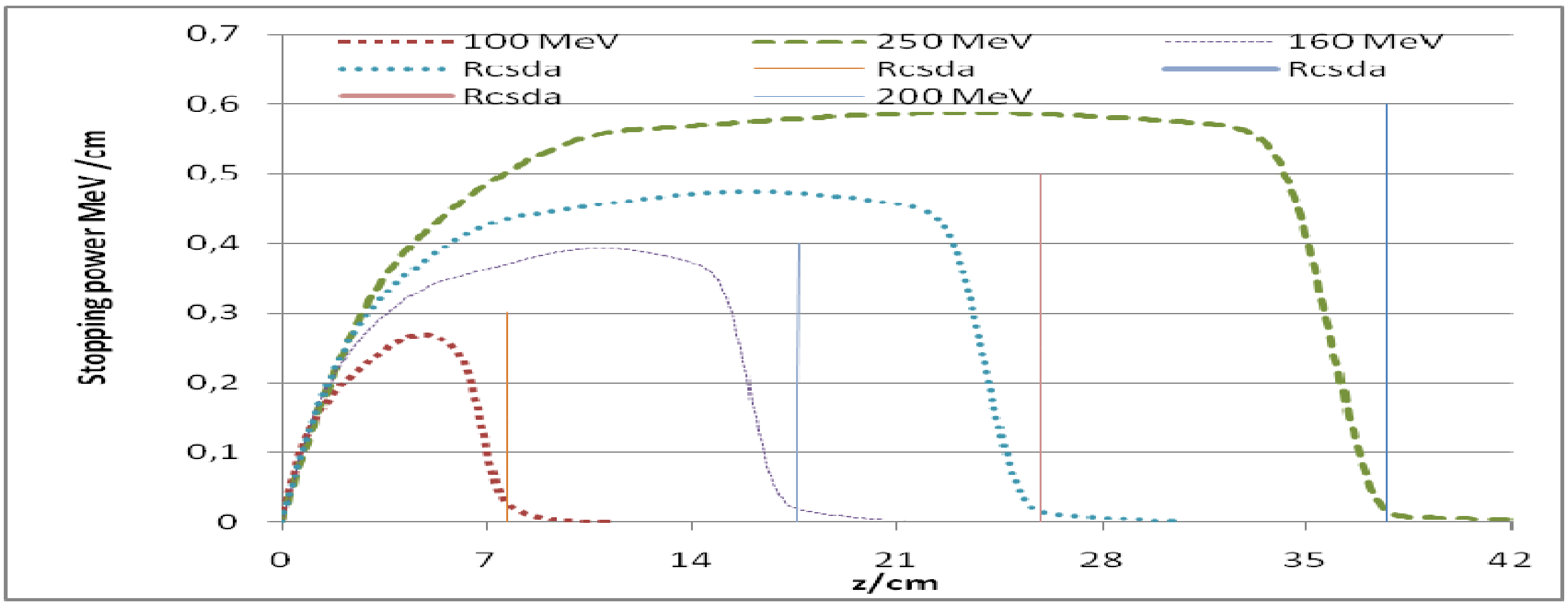

**Fig. 39:** Stopping power of secondary/tertiary protons (+ further charged particles) induced by nuclear reactions of protons with oxygen.

With regard to Fig. 39 and to the qualitative listing (52) valid for the proton - oxygen nucleus interaction, we should finally point out that by a suitable modification a similar listing for nuclear reactions will be obtained, if the oxygen nucleus is replaced by anotherone such as calcium or copper. All formulas necessary for the calculation of $Q^{tot}$ are applicable, as long as the rotational symmetry of the nuclei approximately holds.